\documentclass[amsmath,aps,showpacs,floats,amssymb,preprintnumbers,floatfix,superscriptaddress,nofootinbib]{revtex4}

\usepackage{graphicx}
\usepackage{dcolumn}
\usepackage{bm}
\usepackage{hyperref} 
\usepackage{color}
\usepackage[pdftex]{epsfig} 
\usepackage{amsmath}
\usepackage{feynmf}
\usepackage{feyn}

\def\square{\kern1pt\vbox{\hrule height 1.2pt\hbox{\vrule width 1.2pt\hskip 3pt
   \vbox{\vskip 6pt}\hskip 3pt\vrule width 0.6pt}\hrule height 0.6pt}\kern1pt}

\newcommand{\be}{\begin{equation}}
\newcommand{\ee}{\end{equation}}
\newcommand{\bea}{\begin{eqnarray}}
\newcommand{\eea}{\end{eqnarray}}
\newcommand{\nn}{\nonumber}

\begin{document}

\title{Quantum gravity corrections to the conformally coupled scalar self-mass-squared 
\\ on de Sitter II: kinetic-conformal cross terms}

\author{Sibel Boran}
\email{borans@itu.edu.tr}
\affiliation{Department of Physics, Istanbul Technical University, Maslak 34469 Istanbul, Turkey}

\author{Emre Onur Kahya}
\email{eokahya@itu.edu.tr}
\affiliation{Department of Physics, Istanbul Technical University, Maslak 34469 Istanbul, Turkey}

\author{Sohyun Park}
\email{spark1@kasi.re.kr}
\affiliation{Korea Astronomy and Space Science Institute, Daejeon 34055, Korea}

\date{\today}
%%%%%%%%%%%%%%%%%%%%%%%%%%%%%%%%%%%%%%%%%%%%%%%%%%%%%%%%%%%%%%%%
\begin{abstract}
The present work is the second part of a series of computations for the self-mass-squared of the conformally coupled (CC) scalar interacting with gravitons. This work includes the kinetic-kinetic and kinetic-conformal parts, and thus completes the full scalar self-mass squared at one loop order in de Sitter background when combined with the conformal-conformal part previously evaluated. We use dimensional regularization and renormalize the results by subtracting appropriate counterterms. The self-mass squared is finally ready to quantum-correct the CC scalar field equation so that one can study the effect of inflationary produced gravitons on the CC scalar and its observational consequences.
\end{abstract}


\maketitle

%%%%%%%%%%%%%%%%%%%%%%%%%%%%%%%%%%%%%%%%%%%%%%%%%%%%%%%%%%%%
\section{Introduction}\label{introduction}
%%%%%%%%%%%%%%%%%%%%%%%%%%%%%%%%%%%%%%%%%%%%%%%%%%%%%%%%%%%%
It has been fifty years since Leonard Parker's breakthrough discovery in late 1960's  that particles can be created out of the vacuum in an expanding universe \cite{Parker1}. 
See Refs. \cite{Parker-history-2015, Parker-history-2017} for the historical notes on the discovery.
Parker also 
pointed out that the particle production is maximized 
if the expansion is exponential and the particles are not conformally invariant \cite{Parker1}. 
The Poincar\'e patch of de Sitter space we consider is such an exponentially expanding background.
The particle creation in an exponentially expanding universe had not been emphasized much until the idea of inflation \cite{inflation} 
was proposed in early 1980's \cite{Parker-history-2017}. However, the inflationary paradigm drastically changed the situation and particle production during inflation has been attracting a great deal of attention since then 
\cite{phi4, SQED, PP, Yukawa, DW-0505, mw, SPM, db, KW1, kahya, kahya2, kahya3, PW, LPPW, MTW, WW-1508, PPW-1510, BW-1606,
CV, FRV1, FPRV, FRV2, Frob, Frob-1601, Frob-1603, Frob-1701,
Sloth-0604, Sloth-0612, Seery-Lidsey-Sloth-0610, Seery-0707-1, Seery-0707-2, Riotto-Sloth-0801, Giddings-Sloth-1005, Seery-1005,  
UM, MM, ABG, YK, APS, AK, YU, LBH, Akh, GRZ,
LPW-1210, LPW-1211, LW, GMPW-1308, WW-1408, GMPW-1504, GMPW-1609, BTW, SW, W-QGage, WY,
Holman-1985, BHLS-0912, BHLS-1005, BH-1103, Holman-1410, Kitamoto-1204, Kitamoto-1204-2, Kitamoto-1211, Kitamoto-1402, Kitamoto-1609}. 
The unique particles having no conformal invariance are the massless, minimally coupled (MMC) scalars and gravitons,  
and these particles comprise the scalar \cite{MC} and tensor \cite{S} perturbations predicted by inflationary theories.

The pioneering computations on the scalar and tensor perturbations were at tree order \cite{MC, S}, and it is a natural next step to include quantum loop corrections from MMC scalars and gravitons. A number of computations regarding 
quantum corrections in an inflationary background have been carried out 
\cite{
phi4, SQED, PP, Yukawa, DW-0505, mw, SPM, db, KW1, kahya, kahya2, kahya3, PW, LPPW, MTW, WW-1508, PPW-1510, BW-1606,
CV, FRV1, FPRV, FRV2, Frob, Frob-1601, Frob-1603, Frob-1701,
Sloth-0604, Sloth-0612, Seery-Lidsey-Sloth-0610, Seery-0707-1, Seery-0707-2, Riotto-Sloth-0801, Giddings-Sloth-1005,
Seery-1005,  
UM, MM, ABG, YK, APS, AK, YU, LBH, Akh, GRZ,
LPW-1210, LPW-1211, LW, GMPW-1308, WW-1408, GMPW-1504, GMPW-1609, BTW, SW, W-QGage, WY,
Holman-1985, BHLS-0912, BHLS-1005, BH-1103, Holman-1410, Kitamoto-1204, Kitamoto-1204-2, Kitamoto-1211, Kitamoto-1402, Kitamoto-1609}. 
Besides the cause of inflation, which still has no consensus, studying 
quantum loop effects during inflation can be well estimated by taking the locally de Sitter background as an inflationary universe.
A generic procedure to study loop effects is first to compute the one-particle-irreducible (1PI) 2-point function, and second to use it to quantum-correct the field equation of the particle in question. 
 For the case of a scalar, the 1PI 2-point function is called the self-mass-squared and denoted by $-iM^2(x;x')$. In the present paper, we compute a part of $-iM^2(x;x')$ for the CC scalar, which is indeed the third and last part followed by the first done in Ref. \cite{kahya} and the second in Ref. \cite{kahya3}. In a subsequent paper, we will use the full result of $-iM^2(x;x')$ to quantum-correct the CC scalar field equation
\be
\partial_{\mu}\Bigl(\sqrt{-g}g^{\mu\nu}\partial_{\mu}\phi(x)\Bigr) -\frac{1}{6}R\phi(x) - \int d^4x' M^2(x;x')\phi(x') = 0\;,
\label{linear_eq}
\ee
and solve it to study how gravitons produced by inflation affect the dynamics of the CC scalar.    

The motivation why we consider the CC scalar in particular is multifold. 
First, the tree order CC scalar mode functions
\be
\phi_0(t,\vec{x}) = u_0(t,k)e^{i\vec{k}\cdot\vec{x}}\;, \quad
\mbox{where} \quad
u_0(t,k) = \sqrt{\frac1{2k}}\frac{\exp\Big[-ik\int_{t_i}^{t}\frac{dt'}{a(t')}\Big]}{a(t)}\;,
\ee
vanish as $a(t)$ exponentially grows. Hence, any finite result can be attributed to pure loop corrections. 
Second, we note that the scalar is not differentiated in the conformal coupling by recalling 
derivative interactions redshift away during inflation. 
In fact, the authors previously studied the interaction between MMC scalars and gravitons, which are the two non-conformally invariant particles vastly produced by inflation, and verified that both loop effects from gravitons \cite{kahya, kahya2} and from MMC scalars \cite{PW, LPPW} redshift to zero. The reason for this is that their interactions are only through their kinetic energies, or in other words derivatives. 
Another observation is that the non-conformal and conformal combination seems to have more interesting effects than the combination of non-conformal particles.
In MMC scalar quantum electrodynamics, a MMC scalar loop leads to photons having an increasing effective mass \cite{SQED}. When a MMC scalar is Yukawa-coupled to a massless fermion, fermions acquire an effectively growing mass \cite{Yukawa}.
Graviton loop corrections to a massless fermion cause the fermion field strength to grow with time \cite{mw}.
Graviton loops also induce secular corrections to the electromagnetic field \cite{LPW-1210, LPW-1211, LW, GMPW-1308, WW-1408, GMPW-1504, GMPW-1609}. If our case does induce a secular loop effect, it will be added to this list of the non-conformal and conformal combinations having a significant effect. Furthermore, it might give an insight why this particular combination tends to give significant loop corrections.

The rest of this paper is organized as follows: In Section \ref{setup}, we recast the definitions and notations employed in our previous paper \cite{kahya3}. 
We provide a formal expression for the CC scalar self-mass-squared which consists of 4-point and 3-point interactions at one-loop order. The 4-point interaction is computed in Section \ref{comp4ptint}.
The 3-point interaction is further divided by the kinetic-kinetic, conformal-conformal and kinetic-conformal parts.
The first two parts were evaluated in Refs. \cite{kahya} and \cite{kahya3}, respectively and the kinetic-conformal part is computed in Section \ref{comp3ptint}. We also compute the kinetic-kinetic part using the CC scalar propagator in \ref{comp3ptint}.
The results are fully renormalized and the unregulated limit is taken in Section \ref{renormalization}. 
We summarize and discuss our results in Section \ref{dis}. 

%%%%%%%%%%%%%%%%%%%%%%%%%%%%%%%%%%%%%%%%%%%%%%%%%%%%%%%%%%%%%%%%
\section{The Self-Mass-Squared}\label{setup}
%%%%%%%%%%%%%%%%%%%%%%%%%%%%%%%%%%%%%%%%%%%%%%%%%%%%%%%%%%%%%%%%

We start with the Lagrangian describing pure gravity plus the interaction between gravitons and the massless CC scalar
\bea
\label{Lagrangian}
\mathcal{L} = 
- \frac12 \partial_{\mu} \phi \partial_{\nu} \phi g^{\mu\nu} \sqrt{-g} - \frac{D-2}{8(D-1)}\phi^2 R \sqrt{-g} 
+ \frac{1}{16 \pi G} (R-(D-2)\Lambda )\sqrt{-g}\;,
\eea
where $G$ is Newton's constant, $R$ is the Ricci scalar and $\Lambda = (D-1) H^2$ is the cosmological constant with the Hubble constant $H$. We work in $D$ spacetime dimensions to facilitate dimensional regularization.  
Our background geometry is the open conformal coordinate patch of de Sitter space
\be
ds^2 = \hat{g}_{\mu\nu}dx^{\mu}dx^{\nu} 
= a^2(\eta) \Bigl[ -d\eta^2 + d\vec{x} \!\cdot\! d\vec{x}
\Bigr] 
\;, \quad \mbox{where} \quad a(\eta) = -\frac1{H \eta} \;,
\label{bkgd_geometry}
\ee
with the coordinate ranges
\be
-\infty < x^0 \equiv \eta < 0 \quad , \quad -\infty < x^i <
+\infty \quad, \quad i = 1\;,\; 2\;,\; \cdots , D\!-\!1 \; .
\ee
The full metric is expressed as 
\be
g_{\mu\nu}(x) \equiv a^2 \Bigl[\eta_{\mu\nu} + \kappa h_{\mu\nu}(x)\Bigr]\;, 
\quad \mbox{where} \quad \kappa^2 \equiv 16\pi G\;,
\ee  
where $\hat{g}_{\mu\nu}$ is the background metric and $h_{\mu\nu}$  the conformally rescaled graviton field. 
The conformal coupling makes the matter sector of the Lagrangian conformally invariant:
\bea
g_{\mu\nu} &\equiv&  \Omega^2 \tilde{g}_{\mu\nu}\;, \quad
\phi \equiv  \Omega^{\frac{2-D}{2}} \tilde{\phi}\;.
\label{rescaling} \\
\Rightarrow
\mathcal{L}_{\rm Mat} &=& 
- \frac12 \partial_{\mu} \phi \partial_{\nu} \phi g^{\mu\nu} \sqrt{-g} 
- \frac{D-2}{8(D-1)}\phi^2 R \sqrt{-g} 
= - \frac12 \partial_{\mu} \tilde{\phi} \partial_{\nu} \tilde{\phi} \tilde{g}^{\mu\nu} \sqrt{-\tilde{g}} 
- \frac{D-2}{8(D-1)}\tilde{\phi}^2 \tilde{R} \sqrt{-\tilde{g}} \;.
\label{L_matter}
\eea
We take $\Omega = a$ and work with the conformally rescaled metric
\be
\tilde{g}_{\mu\nu} = \eta_{\mu\nu} + \kappa h_{\mu\nu}\;.
\ee
The expansions of the inverse and the volume element of the conformally rescaled metric $\tilde{g}_{\mu\nu}$ are 
\bea
\tilde{g}^{\mu\nu} &=& 
\eta ^{\mu\nu} - \kappa h^{\mu\nu} + \kappa^2 h^{\mu}_{\phantom{\rho}\rho}h^{\rho\nu} 
+ \mathcal{O}(\kappa^3)\;, 
\\
\sqrt{-\tilde{g}} &=& 
1 + \frac{1}{2}\kappa h + \frac{1}{8}\kappa^2 h^2 - \frac{1}{4}\kappa^2 h^{\rho\sigma}h_{\rho\sigma}
+ \mathcal{O}(\kappa^3)\;.
\label{sqrt_g}
\eea
Also the conformally rescaled Ricci scalar is
\bea
\tilde{R} &=& \kappa \Bigl(-h^{,\mu}_{\;\mu} + h^{\mu\nu}_{\phantom{\rho\rho},\mu\nu}\Bigr)
+\kappa^2 \Bigl( -2 h^{\mu\nu} h_{\lambda\nu ,\mu}^{\phantom{\rho\rho\rho\rho}\lambda} 
+ h^{\mu\nu} h_{\mu\nu,\lambda}^{\phantom{\rho\rho\rho\rho}\lambda}  
+ h^{\mu\nu} h_{,\mu\nu} +\frac34 h^{\mu\nu}_{\phantom{\rho\rho},\lambda} h_{\mu\nu}^{\phantom{\rho\rho},\lambda} 
\nonumber\\
& &\hspace{2.0cm} +h^{\mu\nu}_{\phantom{\rho\rho},\mu} h_{,\nu} 
-h^{\mu\nu}_{\phantom{\rho\rho},\mu} h_{\lambda\nu}^{\phantom{\rho\rho},\lambda}
-\frac12 h^{\mu\nu}_{\phantom{\rho\rho},\lambda} h^{\lambda}_{\phantom{\rho}\nu ,\mu} 
-\frac14 h_{,\mu}h^{,\mu} 
-\frac12 h h^{,\mu}_{\;\mu} 
+\frac12 h h^{\mu\nu}_{\phantom{\rho\rho},\mu\nu}
\Bigr)\;+ \mathcal{O}(\kappa^3) \;.
\eea

Using the perturbative expansion, we derive the self-mass-squared $-iM^2(x;x')$ at one loop order, which consists of the following three Feynman diagrams corresponding to the analytic expressions (written next to the diagrams, respectively):
\bea\label{3diagrams}
& &
\feyn{f%fsx\;
\vertexlabel_{x\phantom{'}
}fglBffs\vertexlabel_{x'}%x 
f} = -iM^2_{\rm 3pt}(x;x') \equiv  
\Bigl\langle i\frac{\delta S[\phi,h]}{\delta\phi(x)}i\frac{\delta S[\phi,h]}{\delta\phi(x')}\Bigr\rangle_0
\;+
\nonumber\\
\phantom{s}
\nonumber\\
& &
\phantom{ssssssss}\feyn{g1g2g3g4}
\nonumber\\
& &
\atop{\feyn{ff
\vertexlabel_{x\phantom{'}}
ff
}} = -iM^2_{\rm 4pt}(x;x') \equiv 
\Bigl\langle i\frac{\delta^2 S[\phi,h]}{\delta\phi(x)\delta\phi(x')} \Bigr\rangle_0
\;+
\nonumber\\
\phantom{s}
\nonumber\\ 
& &
\feyn{ffx\vertexlabel_{x\phantom{'}}ff} = {\rm{counterterm}}\;.
\eea
Here the subscript $0$ means that the expectation value is taken in the free theory.

The 3-point and 4-point interactions (terms proportional to $\tilde{\phi}^2 h$ and $\tilde{\phi}^2 h^2$) 
can be read off from the expansion of the matter part of the Lagrangian \eqref{L_matter}.
The kinetic term of the Lagrangian is expanded as \cite{kahya}
\bea
\mathcal{L}_{\rm K} &\!\equiv\! & 
- \frac12 \partial_{\mu} \tilde{\phi} \partial_{\nu} \tilde{\phi} \tilde{g}^{\mu\nu} \sqrt{-\tilde{g}} \;,
\nonumber\\
&\!=\!&-\frac12 \partial_{\mu} \tilde{\phi} \partial_{\nu} \tilde{\phi} 
\; \Biggr\{ \eta^{\mu\nu} \;+\;\kappa (\frac12 \eta^{\mu\nu}  h - h^{\mu\nu}) 
+\; \kappa^2( \frac18 \eta^{\mu\nu}  h^2 \;-\; \frac14 \eta^{\mu\nu}  h^{\rho\sigma} h_{\rho\sigma}
- \frac12  h h^{\mu\nu} +  h^{\mu\rho} h_{\rho}^{\phantom{\eta}\nu} )\;+\; \mathcal{O}(\kappa^3) \Biggl\}\;.
\label{expandedLk}
\eea
and the conformal coupling term is expanded as \cite{kahya3}
\begin{eqnarray}
\mathcal{L}_{\rm CC} &\equiv& -\frac{D-2}{8(D-1)}\tilde{\phi}^2 \tilde{R} \sqrt{-\tilde{g}}\;,
\nonumber\\
&=& -\frac{D-2}{8(D-1)} \kappa\tilde{\phi}^2 \Bigl(-h^{,\mu}_{\;\mu} + h^{\mu\nu}_{\phantom{\rho\rho},\mu\nu}\Bigr)
-\frac{D-2}{8(D-1)}\kappa^2 \tilde{\phi}^2 \Bigl(
-2 h^{\mu\nu} h_{\lambda\nu ,\mu}^{\phantom{\rho\rho\rho\rho}\lambda} 
+ h^{\mu\nu} h_{\mu\nu,\lambda}^{\phantom{\rho\rho\rho\rho}\lambda}  
+ h^{\mu\nu} h_{,\mu\nu} 
\nonumber\\
& & \hspace{1.5cm}
+\frac34 h^{\mu\nu}_{\phantom{\rho\rho},\lambda} h_{\mu\nu}^{\phantom{\rho\rho},\lambda} 
+h^{\mu\nu}_{\phantom{\rho\rho},\mu} h_{,\nu} 
-h^{\mu\nu}_{\phantom{\rho\rho},\mu} h_{\lambda\nu}^{\phantom{\rho\rho},\lambda} 
-\frac12 h^{\mu\nu}_{\phantom{\rho\rho},\lambda} h^{\lambda}_{\phantom{\rho}\nu ,\mu} 
-\frac14 h_{,\mu}h^{,\mu} 
-\frac12 h h^{,\mu}_{\mu}
+\frac12 h h^{\mu\nu}_{\phantom{\rho\rho},\mu\nu}
\Bigr)\;.
\label{expandedLcc}
\end{eqnarray}

In the following subsection we demonstrate the formal expressions of the 3-point and 4-point interactions for the self-mass-squared and drop tilde for notational simplicity. However, we re-emphasize here that our metric, the scalar field $\phi$ and graviton field $h_{\mu\nu}$ are conformally rescaled ones:
\bea
\tilde{g}_{\mu\nu} = \eta_{\mu\nu} + \kappa h_{\mu\nu}\;, \qquad
\tilde{\phi} 
\equiv \phi\;.
\label{rescaling_notation}
\eea

%%%%%%%%%%%%%%%%%%%%%%%%%%%%%%%%%%%%%%%%%%%%%%%%%%%%%%%%%%%%%%%%
\subsection{Formal expressions for the one loop self-mass-squared}
%%%%%%%%%%%%%%%%%%%%%%%%%%%%%%%%%%%%%%%%%%%%%%%%%%%%%%%%%%%%%%%%

We first recast the 4-point and 3-point contributions from Refs. \cite{kahya} and \cite{kahya3} and derive the kinetic-conformal cross part of the 3-point interaction.
%%%%%%%%%%%%%%%%%%%%%%%%%%%%%%%%%%%%%%%%%%%%%%%%%%%%%%%%%
\subsubsection{4-point contributions}
%%%%%%%%%%%%%%%%%%%%%%%%%%%%%%%%%%%%%%%%%%%%%%%%%%%%%%%%%

The 4-point contribution from the kinetic term \eqref{expandedLk} is similar to \cite{kahya} where the background metric for this case is flat. Therefore the expression becomes
\bea\label{K4-pt}
-iM^2_{\rm K4pt}(x;x') 
&\!\!\equiv\!\!& 
\Bigl\langle i\frac{\delta^2 S_{\rm K4}}{\delta\phi(x')\delta\phi(x)} \Bigr\rangle_0\;,
\nonumber\\
&\!\!=\!\!& 
\frac{i}{8} \kappa^2 \partial^{\mu} [\, i[^{\alpha}_{\phantom{\eta}\alpha} \Delta^{\rho}_{\phantom{\eta}\rho} ](x;x') \partial_{\mu} \delta^{D}(x-x')] \;-\; \frac{i}{4} \kappa^2 \partial^{\mu} [\, i[^{\alpha \beta} \Delta_{\alpha\beta} ](x;x') \partial_{\sigma} \delta^{D} (x-x')] 
\nonumber\\
&\!\!\!\! & - \frac{i}{4} \kappa^2 \partial^{\rho} [\, i[^{\alpha}_{\phantom{\eta}\alpha} \Delta^{\rho\sigma} ](x;x') \partial_{\mu} \delta^{D}(x-x')] \;+\; i \kappa^2 \partial^{\mu} [\, i[_{\alpha \beta} \Delta^{\rho\sigma} ](x;x') \partial_{\sigma} \delta^{D} (x-x')]\;,
\eea  
where the expression
\be\label{grprop}
i[^{\mu\nu}\Delta_{\lambda\nu}](x;x') = \Bigl\langle h^{\mu\nu}(x) h_{\lambda\nu}(x') \Bigr\rangle_0\;,
\ee
is the graviton propagator which will be given in the next subsection.
The 4-point contribution from the conformal coupling term \eqref{expandedLcc} was calculated in \cite{kahya3}.

%%%%%%%%%%%%%%%%%%%%%%%%%%%%%%%%%%%%%%%%%%%%%%%%%%%%%%%%%%%%%%%
\subsubsection{3-point contributions}
%%%%%%%%%%%%%%%%%%%%%%%%%%%%%%%%%%%%%%%%%%%%%%%%%%%%%%%%%%%%%%%%
The first Feynman diagram that appears in \eqref{3diagrams} can be read off as two 3-point interaction vertices connected by a graviton propagator and a conformally coupled scalar propagator. There are two, conformal and kinetic, types of 3-point interactions for a scalar coupled to gravity conformally as one can see from \eqref{Lagrangian}.

First, we label the 3-point interactions from the conformal coupling term \eqref{expandedLcc} as
\begin{eqnarray}
\mathcal{L}_{\rm 3pt} = 
-\frac{D-2}{8(D-1)} \kappa\phi^2 (-h^{,\mu}_{\;\mu} + h^{\mu\nu}_{\phantom{\rho\rho},\mu\nu})  = 
- \tilde{\kappa} \phi^2 (-h^{,\mu}_{\;\mu} + h^{\mu\nu}_{\phantom{\rho\rho},\mu\nu})
\equiv \mathcal{L}_{3a} + \mathcal{L}_{3b}\;,
\label{3pntlagr}
\end{eqnarray}
where we defined $\tilde{\kappa} \equiv \frac{D-2}{8(D-1)}\times \kappa$. 
This leads to the {\it conformal-conformal} part of the 3-point interaction \cite{kahya3}
\bea
-iM^2_{\rm 3pt}(x;x') &\!\!=\!\!& 
\Biggl\langle
i\frac{\delta S_{3a}}{\delta\phi(x)}i\frac{\delta S_{3a}}{\delta\phi(x')}
+i\frac{\delta S_{3a}}{\delta\phi(x)}i\frac{\delta S_{3b}}{\delta\phi(x')}
+i\frac{\delta S_{3b}}{\delta\phi(x)}i\frac{\delta S_{3a}}{\delta\phi(x')}
+i\frac{\delta S_{3b}}{\delta\phi(x)}i\frac{\delta S_{3b}}{\delta\phi(x')}
\Biggr\rangle_0\;,
\nn \\
&\!\!=\!\!& \!- \tilde{\kappa}^2 i\Delta_{\rm cf} \left\{\!
\partial^2 \partial'^2 \; i[^{\alpha}_{\phantom{\rho}\alpha}\Delta^{\beta}_{\phantom{\rho}\beta}](x;x')
\!-\![\partial_{\mu} \partial_{\nu} \partial'^2 
\!+\!\partial'_{\mu} \partial'_{\nu} \partial^2 ] i[^{\alpha}_{\phantom{\rho}\alpha}\Delta^{\mu\nu}](x;x')
\!+\!\partial_{\mu} \partial_{\nu} \partial'_{\alpha} \partial'_{\beta} i[^{\mu\nu}\Delta^{\alpha\beta}](x;x')\!\right\}\;.\nonumber\\
\label{3-pt}
\eea
This contribution was calculated in a previous work \cite{kahya3}.

Second, we label the 3-point interactions from the kinetic term of the Lagrangian \eqref{expandedLk} as
\be
\mathcal{L}_{\rm K 3pt} = -\frac{1}{2}\kappa\partial_{\mu}\phi\partial_{\nu}\phi
\Bigl(- h^{\mu\nu} +\frac{1}{2}\eta^{\mu\nu} h\Bigr) 
\equiv \mathcal{L}_{K3c} + \mathcal{L}_{K3d}\;.
\ee
Taking $c$ and $d$ at each vertex point we obtain the  {\it kinetic-kinetic} part of 3-point interaction of the self-mass-squared
\bea
-iM^2_{\rm 3pt K}(x;x') 
&=& \Biggl\langle
i\frac{\delta S_{K3c}}{\delta\phi(x)}i\frac{\delta S_{K3c}}{\delta\phi(x')}
+i\frac{\delta S_{K3c}}{\delta\phi(x)}i\frac{\delta S_{K3d}}{\delta\phi(x')}
+i\frac{\delta S_{K3d}}{\delta\phi(x)}i\frac{\delta S_{K3c}}{\delta\phi(x')}
+i\frac{\delta S_{K3d}}{\delta\phi(x)}i\frac{\delta S_{K3d}}{\delta\phi(x')}
\Biggr\rangle_0\;,
\nn \\
&=& 
\kappa^2 \biggl\{ 
\frac12 \partial_{\alpha} \partial'^{\nu} [ i[^{\alpha \beta} \Delta ^{\rho}_{\phantom{\eta}\rho}] (x;x')\; \partial_{\beta} \partial'_{\nu} i\Delta_{\rm{cf}} (x;x')]
\;-\;
\partial_{\alpha} \partial'_{\rho} [ i[^{\alpha \beta} \Delta ^{\rho\sigma}](x;x') \;\partial_{\beta} \partial'_{\sigma} i\Delta_{\rm{cf}}(x;x')]
\nonumber\\
\;\;\;\;\;\;\;\;\; & & -\;
\frac14 \partial^{\mu} \partial'^{\nu} [ i[^{\alpha}_{\phantom{\eta}\alpha} \Delta ^{\rho}_{\phantom{\eta}\rho}] (x;x')\; \partial_{\mu} \partial'_{\nu} i\Delta_{\rm{cf}}(x;x')]
\;+\;
\frac12 \partial^{\mu} \partial'_{\rho} [ i[^{\alpha}_{\phantom{\eta}\alpha} \Delta ^{\rho \sigma}] (x;x')\; \partial_{\mu} \partial'_{\sigma} i\Delta_{\rm{cf}}(x;x')]
\biggr\}\;.
\label{M^2_3point_K_propagator}
\eea   
Note that this part can be read off from the 3-point contribution of Ref. \cite{kahya} by replacing the minimally conformally coupled propagator term $i\Delta_{A}(x;x')$ by the conformal propagator 
$i\Delta_{\rm{cf}}(x;x')  \equiv \Bigl\langle \phi(x)\phi(x') \Bigr\rangle_0 = (aa')^{\frac{D}{2}-1}F(y)$ and dividing by $(aa')^{D-2}$ (extracting $(aa')^{\frac{D}{2}-1}$ from each vertex point). Although the structure of the kinetic part of our calculation is similar to Ref. \cite{kahya} the powers of the scale factor gives quite a different form, and thus one has to redo all of the calculation using new identities.

Finally, the cross terms, which we name as {\it kinetic-conformal} part can be obtained by 
taking one kinetic term at one vertex and one conformal term at the other vertex:
\bea
\feyn{f\vertexlabel^{a}\vertexlabel_{b\atop{CC}}fglBf\vertexlabel^{c}\vertexlabel_{d\atop{KC}}f} 
&+& \feyn{f\vertexlabel^{a}\vertexlabel_{b\atop{KC}}fglBf\vertexlabel^{c}\vertexlabel_{d\atop{CC}}f}=
-iM^2_{\rm 3pt cross}(x;x') 
\nonumber\\
\phantom{\eta}
\nonumber\\
&=& 
\Biggl\langle
i\frac{\delta S_{3a}}{\delta\phi(x)}i\frac{\delta S_{K3c}}{\delta\phi(x')}
+i\frac{\delta S_{3a}}{\delta\phi(x)}i\frac{\delta S_{K3d}}{\delta\phi(x')}
+i\frac{\delta S_{3b}}{\delta\phi(x)}i\frac{\delta S_{K3c}}{\delta\phi(x')}
+i\frac{\delta S_{3b}}{\delta\phi(x)}i\frac{\delta S_{K3d}}{\delta\phi(x')}
\nonumber \\
& &+i\frac{\delta S_{3c}}{\delta\phi(x')}i\frac{\delta S_{K3a}}{\delta\phi(x)}
+i\frac{\delta S_{3d}}{\delta\phi(x')}i\frac{\delta S_{K3a}}{\delta\phi(x)}
+i\frac{\delta S_{3c}}{\delta\phi(x')}i\frac{\delta S_{K3b}}{\delta\phi(x)}
+i\frac{\delta S_{3d}}{\delta\phi(x')}i\frac{\delta S_{K3b}}{\delta\phi(x)}
\Biggr\rangle_0\;.
\label{M^2_3point_cross}
\eea   
Taking the variation of each term with respect to the scalar,
\bea
\frac{\delta S_{3a}}{\delta\phi(x)} 
&=&-\tilde{\kappa} \frac{\delta}{\delta\phi(x)} \int d^Dy
\Bigl[-\phi^2(y)\partial_{\mu}\partial^{\mu} h(y)\Bigr] 
= - \tilde{\kappa} \Bigl[-2\phi(x)\partial_{\mu}\partial^{\mu} h(x)\Bigr]\;,
\label{S_3a}
\\
\frac{\delta S_{3b}}{\delta\phi(x)} 
&=& - \tilde{\kappa} \frac{\delta}{\delta\phi(x)} \int d^Dy
\Bigl[\phi^2(y)\partial_{\mu}\partial_{\nu}h^{\mu\nu}(y)\Bigr] 
= - \tilde{\kappa} \Bigl[2\phi(x)\partial_{\mu}\partial_{\nu}h^{\mu\nu}(x)\Bigr]\;,
\label{S_3b}
\\
\frac{\delta S_{K3c}}{\delta\phi(x')} 
&=& \tilde{\kappa} \frac{\delta}{\delta\phi(x')} \int d^Dy
\Bigl[\frac12 \partial_{\mu} \phi(y) \partial_{\nu} \phi(y) h^{\mu\nu}(y)\Bigr] 
= - \tilde{\kappa} \partial'_{\mu} \Bigl[\partial'_{\nu} \phi(x') h^{\mu\nu}(x')\Bigr]\;,
\label{S_K3c}
\\
\frac{\delta S_{K3d}}{\delta\phi(x')} 
&=& \tilde{\kappa} \frac{\delta}{\delta\phi(x')} \int d^Dy
\Bigl[-\frac14 \partial_{\mu} \phi(y) \partial_{\nu} \phi(y) \eta^{\mu\nu}h(y)\Bigr] 
= \frac{\tilde{\kappa}}{2} \partial'_{\mu}\Bigl[\partial'^{\mu} \phi(x') h(x')\Bigr]\;,
\label{S_K3d}
\eea
and plugging these into \eqref{M^2_3point_cross} leads to 
\begin{eqnarray} 
\lefteqn{-iM^2_{\rm 3ptcross}(x;x')
= \! \tilde{\kappa}^2 \biggl\{\!
2\partial'_{\mu}\Bigl[\partial'_{\nu} i\Delta_{\rm cf} (x;x') \partial_{\rho}\partial^{\rho} i[^{\mu\nu}\Delta^{\gamma}_{\phantom{\eta}\gamma}](x;x')\Bigr]
-\partial'_{\mu}\Bigl[\partial'^{\mu} i\Delta_{\rm cf}(x;x') \partial_{\rho}\partial^{\rho} i[^{\alpha}_{\phantom{\eta}\alpha}\Delta^{\gamma}_{\phantom{\eta}\gamma}](x;x')\Bigr]
}
\nonumber\\
& & \hspace{3.3cm}
-2\partial'_{\mu}\Bigl[\partial'_{\nu} i\Delta_{\rm cf}(x;x') \partial_{\rho}\partial_{\sigma} i[^{\mu\nu}\Delta^{\rho\sigma}](x;x')\Bigr]
+\partial'_{\mu}\Bigl[\partial'^{\mu} i\Delta_{\rm cf}(x;x') \partial_{\rho}\partial_{\sigma} i[^{\alpha}_{\phantom{\eta}\alpha}\Delta^{\rho\sigma}](x;x')\Bigr]
\nonumber\\
& & \hspace{3.3cm}
+2\partial_{\mu}\Bigl[\partial_{\nu} i\Delta_{\rm cf}(x;x') \partial'_{\rho}\partial'^{\rho} i[^{\mu\nu}\Delta^{\gamma}_{\phantom{\eta}\gamma}](x;x')\Bigr]
-\partial_{\mu}\Bigl[\partial^{\mu} i\Delta_{\rm cf}(x;x') \partial'_{\rho}\partial'^{\rho} i[^{\alpha}_{\phantom{\eta}\alpha}\Delta^{\gamma}_{\phantom{\eta}\gamma}](x;x')\Bigr]
\nonumber\\
& & \hspace{3.3cm}
-2\partial_{\mu}\Bigl[\partial_{\nu} i\Delta_{\rm cf}(x;x') \partial'_{\rho}\partial'_{\sigma} i[^{\mu\nu}\Delta^{\rho\sigma}](x;x')\Bigr]
+\partial_{\mu}\Bigl[\partial^{\mu} i\Delta_{\rm cf}(x;x') \partial'_{\rho}\partial'_{\sigma} i[^{\alpha}_{\phantom{\eta}\alpha}\Delta^{\rho\sigma}](x;x')\Bigr]
\! \biggr\}\;.
\label{3-ptcross}
\end{eqnarray}

%%%%%%%%%%%%%%%%%%%%%%%%%%%%%%%%%%%%%%%%%%%%%%%%%%%%%%%%%%%%%%%%
\subsection{Propagators}
%%%%%%%%%%%%%%%%%%%%%%%%%%%%%%%%%%%%%%%%%%%%%%%%%%%%%%%%%%%%%%%%

To express the propagators in the same notation as that of the previous work, we recall the three notational conventions employed in Refs. \cite{kahya, kahya3}. First the background metric is denoted with a hat,
\be
\hat{g}_{\mu\nu} = \eta_{\mu\nu} \quad \mbox{accordingly} \quad \hat{R} = 0\;.
\ee
Second, the spatial parts of the Lorentz metric and the Kronecker delta are defined with a bar, 
\bea\label{lormetkrodel}
\overline{\eta}_{\mu\nu} \equiv \eta_{\mu\nu} + \delta^0_{\mu} \delta^0_{\nu} \qquad {\rm and} \qquad \overline{\delta}^{\mu}_{\nu} \equiv \delta^{\mu}_{\nu} - \delta_0^{\mu} \delta^0_{\nu}\;.
\eea
Third, we define the de Sitter length function $y(x;x')$ which is related to the de Sitter invariant length $\ell(x;x')$ from $x^{\mu}$ to $x^{\prime \mu}$ as
\bea\label{ydef}
y(x;x')
=4 \sin^2\Bigl( \frac12 H \ell(x;x')\Bigr) 
= a a' H^2 \Bigl\{ \Vert \vec{x} - \vec{x}' \Vert^2 - \Bigl(\vert \eta - \eta'\vert - i \delta\Bigr)^2 \Bigr\}\;,
\eea
where $a \equiv a(\eta)$ and $a' \equiv a(\eta')$.

The propagator for a massless conformally coupled scalar for the flat background is \cite{BD},
\bea
i\Delta_{\rm cf} (x;x') & = & \frac{\Gamma\Bigl( \frac{D}2 \!-\! 1\Bigr)}{(4\pi)^{\frac{D}2}}
\Bigl(\frac4{\Delta x^2}\Bigr)^{\frac{D}2-1} = (aa')^{\frac{D}{2}-1} \frac{H^{D-2}}{(4\pi)^{\frac{D}2}}
\Gamma\Bigl( \frac{D}2 \!-\! 1\Bigr)
\Bigl(\frac4{y}\Bigr)^{\frac{D}2-1} \equiv 
(aa')^{\frac{D}{2}-1} F(y) \;. \label{CCP}
\eea
Here we defined $F(y)$ above, in order to work with de Sitter invariant functions.  

To obtain the graviton propagator, we add the gauge fixing term (first derived in Ref. \cite{tw}) to the Lagrangian,
\bea
\label{gf}
\mathcal{L}_{\rm GF} = -\frac12 a^{D-2} \eta^{\mu\nu} F_{\mu} F_{\nu}\;, 
\quad
F_{\mu} & \equiv & \eta^{\rho\sigma} \Bigl(h_{\mu\rho ,\sigma} 
- \frac12 h_{\rho \sigma ,\mu} + (D-2) H a h_{\mu \rho} \delta^0_{\sigma} \Bigr)\;.
\eea
We partially integrate the quadratic part of the gauge fixed Lagrangian 
to extract the kinetic operator $D_{\mu\nu}^{~~\rho \sigma}$ as follows 
\be
\frac12 h^{\mu\nu} D_{\mu\nu}^{~~\rho \sigma} h_{\rho\sigma}\;,
\ee
where
\bea
D_{\mu\nu}^{\phantom{\rho\rho}\rho\sigma}&\equiv &\Biggl\{ \frac12 \overline{\delta}_{\mu}^{\phantom{\rho}(\rho} \overline{\delta}_{\nu}^{\phantom{\rho}\sigma)} - \frac14 \eta_{\mu\nu} \eta^{\rho\sigma}-\frac1{2(D-3)} \delta_{\mu}^0 \delta_{\nu}^0 \delta_0^{\rho} \delta_0^{\sigma} \Biggr\} D_A 
+\delta^0_{(\mu} \overline{\delta}_{\nu )}^{( \rho} \delta_0^{\sigma )} D_B + \frac12 \Bigl(\frac{D-2}{D-3}\Bigr) \delta_{\mu}^0 \delta_{\nu}^0 \delta_0^{\rho} \delta_0^{\sigma} D_C\;,
\eea
and the three scalar differential operators are 
\bea
D_A & \equiv & \partial_{\mu} \Bigl(\sqrt{-\hat{g}} \,
\hat{g}^{\mu\nu} \partial_{\nu}\Bigr) \;, \\
D_B & \equiv & \partial_{\mu} \Bigl(\sqrt{-\hat{g}} \,
\hat{g}^{\mu\nu} \partial_{\nu}\Bigr) - \frac1{D} 
\Bigl(\frac{D\!-\!2}{D\!-\!1}\Bigr) \hat{R} 
\sqrt{-\hat{g}} \; , \\
D_C & \equiv & \partial_{\mu} \Bigl(\sqrt{-\hat{g}} \,
\hat{g}^{\mu\nu} \partial_{\nu}\Bigr) - \frac2{D}
\Bigl(\frac{D\!-\!3}{D\!-\!1}\Bigr) \hat{R} 
\sqrt{-\hat{g}} \; .
\eea
The graviton propagator obeys the defining equation,
\bea
D_{\mu\nu}^{~~\rho\sigma} \times i\Bigl[{}_{\rho\sigma}
\Delta^{\alpha\beta} \Bigr](x;x') = \delta_{\mu}^{(\alpha}
\delta_{\nu}^{\beta)} i \delta^D(x-x') \;,
\eea 
and inverting the kinetic operator leads to the graviton propagator in the gauge of \eqref{gf} 
being a sum of constant tensor factors times scalar propagators, 
\bea
i[_{\mu\nu}\Delta_{\rho\sigma}](x;x') = \underset{I =A, B, C}{\Sigma} [_{\mu\nu} T^I_{\rho\sigma}] i \Delta_I(x;x')\;.
\eea
Here each of the scalar propagators obeys
\bea
D_I \times i\Delta_I(x;x') = i \delta^D(x - x') \qquad {\rm for}
\qquad I = A,B,C \;, \label{sprops}
\eea
and the tensor factors are
\bea
\Bigl[{}_{\mu\nu} T^A_{\rho\sigma}\Bigr] &=&  2 \overline{\eta}_{\mu (\rho} \overline{\eta}_{\sigma) \nu} -
\frac{2}{D-3} \overline{\eta}_{\mu\nu} \overline{\eta}_{\rho \sigma}\;,\label{T^A} \\
\Bigl[{}_{\mu\nu} T^B_{\rho\sigma}\Bigr] &=& -4 \delta^0_{(\mu} \overline{\eta}_{\nu) (\rho} \delta^0_{\sigma)}\;, \label{T^B} \\
\Bigl[{}_{\mu\nu} T^C_{\rho\sigma}\Bigr] &=&  \frac{2}{(D-3)(D-2)} 
[(D-3) \delta^0_{\mu} \delta^0_{\nu} + \overline{\eta}_{\mu\nu}] 
[(D-3)\delta^{0}_{\rho}\delta^{0}_{\sigma} + \overline{\eta}_{\rho\sigma}]\;. \label{T^C}
\eea

The $A$-type propagator is the same as the MMC scalar propagator, which consists of the de Sitter invariant and breaking parts \cite{ow1,ow2}.
\bea\label{compactA}
i \Delta_A (x;x') &=& A(y) + k \ln(aa')\;, 
\eea
where $k \equiv \frac{H^{D-2}}{(4 \pi )^{D/2}} \frac{\Gamma(D-1)}{\Gamma(\frac{D}2)}$ and 
the de Sitter invariant part $A(y)$ has the following solution \cite{ow2},
\bea\label{A}
\lefteqn{A(y)\equiv \frac{H^{D-2}}{(4\pi)^{\frac{D}2}} \Biggl\{ 
\Gamma(\frac{D}2 -1)\Bigl(\frac{4}{y}\Bigr)^{\frac{D}2 -1} 
+\frac{\Gamma(\frac{D}2 +1)}{\frac{D}2 -2} \Bigl(\frac{4}{y} \Bigr)^{\frac{D}2 -2}
-\pi \cot\Bigl(\frac{\pi D}2\Bigr) \frac{\Gamma(D-1)}{\Gamma(\frac{D}2)}
}
\nonumber \\
&& \hspace{2.5cm} +\sum_{n=1}^{\infty}\biggl[ 
\frac1{n} \frac{\Gamma(n+D-1)}{\Gamma(n+\frac{D}2)} \Bigl(\frac{y}4 \Bigr)^n 
-\frac1{n -\frac{D}2 +2} \frac{\Gamma(n +\frac{D}2 +1)}{ \Gamma(n+2)}\Bigl(\frac{y}4 \Bigr)^{n -\frac{D}2 +2} \biggr] \Biggr\}\;.
\eea
The $B$-type and $C$-type propagators are de Sitter invariant and have the following solutions 
\bea
i \Delta_B (x;x') &\!\!\!=\!\!\!& 
i \Delta_{\rm cf}(x;x') 
-\frac{H^{D-2}}{(4\pi )^{\frac{D}2}} \sum_{n=0}^{\infty} \Biggl\{ 
\frac{\Gamma(n+D\!-\!2)}{\Gamma (n+\frac{D}2)}\Bigl(\frac{y}4 \Bigr)^n 
-\frac{\Gamma (n\!+\!\frac{D}2)}{\Gamma (n\!+\!2)} \Bigl(\frac{y}4 \Bigr)^{n \!-\!\frac{D}2 \!+\!2}
\Biggr\} \;, 
\label{DeltaB} 
\\
i \Delta_C (x;x') &\!\!\!=\!\!\!& 
i \Delta_{\rm cf}(x;x')
+\frac{H^{D-2}}{(4\pi )^{\frac{D}2}} \sum_{n=0}^{\infty} \Biggl\{ 
(n\!+\!1)\frac{\Gamma(n\!+\!D\!-\!3)}{\Gamma (n\!+\!\frac{D}2)}\Bigl(\frac{y}4 \Bigr)^n 
-\Bigl(n\!-\!\frac{D}2\!+\!3\Bigr)\frac{\Gamma(n\!+\!\frac{D}2 \!-\!1)}{\Gamma(n+2)}
\Bigl(\frac{y}4 \Bigr)^{n\!-\!\frac{D}2\!+\!2}
 \Biggr\}\;. 
\label{DeltaC} 
\eea
The infinite sums vanish in $D=4$ and each term is a positive power of $y$, therefore we only need to retain a small number of terms when multiplied by a potentially divergent quantity.

%%%%%%%%%%%%%%%%%%%%%%%%%%%%%%%%%%%%%%%%%%%%%%%%%%%%%%%%%%%%%%%%
\section{Computation of the 4-point interactions}\label{comp4ptint}
%%%%%%%%%%%%%%%%%%%%%%%%%%%%%%%%%%%%%%%%%%%%%%%%%%%%%%%%%%%%%%%%

In this section we evaluate the contribution from the kinetic 4-point interactions \eqref{K4-pt} and bring the conformal 4-point contribution from our previous work \cite{kahya3}.  
First note that in the coincidence limit the three scalar propagators become \cite{kahya}
\bea
\lim_{x' \rightarrow x} i\Delta _{A} (x;x') &=& \frac{H^{D-2}}{(4\pi)^{\frac{D}{2}}} \frac{\Gamma(D-1)}{\Gamma(\frac{D}{2})} \biggl[-\pi \cot(\frac{\pi D}{2}) \;+\; 2\ln(a) \biggl]\;,
\label{coa}
\\
\lim_{x' \rightarrow x} i\Delta _{B} (x;x') &=& \frac{H^{D-2}}{(4\pi)^{\frac{D}{2}}} \frac{\Gamma(D-1)}{\Gamma(\frac{D}{2})} \times -\frac{1}{D-2} \rightarrow -\frac{H^2}{(4\pi)^2}\;,
\label{cob}
\\
\lim_{x' \rightarrow x} i\Delta _{C} (x;x') &=& \frac{H^{D-2}}{(4\pi)^{\frac{D}{2}}} \frac{\Gamma(D-1)}{\Gamma(\frac{D}{2})} \times \frac{1}{(D-2)(D-3)} \rightarrow \frac{H^2}{(4\pi)^2}\;.
\label{coc}
\eea
This leads the four contractions of the coincident limit graviton propagator in \eqref{K4-pt} to 
\bea
& &  i[ ^{\alpha}_{\phantom{\eta}\alpha} \Delta^{\rho}_{\phantom{\eta}\rho} ](x;x') \rightarrow  -4 (\frac{D-1}{D-3}) i\Delta_{A} (x;x') \;+\;  \frac{H^2}{4\pi^2}\;,
\nonumber\\
& &  i[ ^{\alpha\beta} \Delta_{\alpha\beta} ](x;x') \rightarrow  \frac{(D-1)(D^2-3D-2)}{(D-3)} i\Delta_{A} (x;x') \;-\;  \frac{H^2}{8\pi^2}\;,
\nonumber\\
& &  i[ ^{\alpha}_{\phantom{\eta}\alpha} \Delta^{\rho\sigma} ](x;x') \rightarrow  -\frac{4}{(D-3)} i\Delta_{A} (x;x') \;+\; [2 \delta_{0}^{\rho} \delta_{0}^{\sigma} \;+\; \bar{\eta}^{\rho\sigma} ] \frac{H^2}{16\pi^2}\;,
\nonumber\\
& &  i[ _{\alpha \beta} \Delta^{\rho\sigma} ](x;x')  \rightarrow  (\frac{D^2-3D-2}{D-3}) \bar{\delta}^{\sigma}_{\alpha} i\Delta_{A} (x;x') \;-\; \delta_{0}^{\sigma} \delta_{\alpha}^{0} \frac{H^2}{8\pi^2}\;.
\eea
Next, substituting these relations into the expression \eqref{K4-pt}, we obtain 
\bea\label{4-ptlim}
-iM^2_{\rm K4pt}(x;x') &=& 
i\kappa^2 \frac{H^{D-2}}{(4\pi )^{\frac{D}{2}}} \frac{\Gamma{(D-1)}}{\Gamma{(\frac{D}{2})}} \pi \cot(\frac{\pi D}{2}) 
 \; \biggl\{\Bigl[ \frac14 D(D-1) - D \Bigr]\; \partial^{\mu} \partial_{\mu} \; \delta^{D}(x-x') \biggr\}
 \nonumber\\
& & \;-\; i\kappa^2 \frac{H^{2}}{4\pi^{\frac{D}{2}}}  \biggl\{\Bigl[ \frac14 D(D-1) - D \Bigr]\; \partial^{\mu} [\; \ln(a) \; \partial_{\mu} ]\; \delta^{D}(x-x') \;+\;\frac14 \partial^{2}_{0} \;\delta^{D}(x-x')  \biggr\}
\nonumber\\
& & \;+\; \mathcal{O}(D-4)\;.
\eea 
To evaluate the action of derivatives on the Dirac delta function in \eqref{4-ptlim}, we note the following identities,
\bea\label{idfor4}
\partial_{\mu} \partial^{\mu} \delta^{D}(x-x') &\!=\!& \Bigl[-\partial^{2}_{0} + \nabla^2 \Bigr] \delta^{D}(x-x')\;,
\nonumber\\
\partial^{\mu} \Bigl[ \ln(a) \partial_{\mu} \Bigr] \delta^{D}(x-x') &\!=\!& -\left[\;Ha\partial_{0}  +\frac12 \ln(aa') a^2 \square +\frac{(D-2)}2\ln(aa') Ha\partial_{0} -\ln(aa') \nabla^2 \right] \delta^{D}(x-x')\;.
\eea 
The covariant d' Alembertian operator in our background is
\bea\label{dsda}
\square \equiv \frac1{\sqrt{-\hat{g}}} \partial_{\mu} \Bigl(\sqrt{-\hat{g}} \hat{g}^{\mu\nu} \partial_{\nu}\Bigr) = -\frac1{a^2} \partial_{0}^2 - \frac{(D-2)H}{a} \partial_{0}+\frac1{a^2}\nabla^2.
\eea
The following properties of the differential operators are also useful.
\bea\label{siden} 
\partial'_i = -\partial_i\;, \quad
\nabla'^2 = \nabla^2\;, \quad 
\partial^2 = -\partial_{0}^{2} + \nabla^2\;.\label{h}
\eea
By plugging \eqref{idfor4}, \eqref{dsda}, and \eqref{siden} into the expression \eqref{4-ptlim}, we obtain 
\bea
\lefteqn{-iM^2_{\rm K4pt}(x;x') = i\kappa^2 \frac{H^{D-4}}{(4\pi)^{\frac{D}{2}}} \frac{\Gamma(D-1)}{\Gamma(\frac{D}{2})} 
\Biggl\{ \biggl[ \pi \cot\Bigl(\frac{\pi D}2\Bigr) \frac14 D(D-5)  - \frac14 \ln(aa') \biggr] H^2 a^{\frac{D}{2}} \square}
\nonumber\\
& & \hspace{2.5cm} + \biggl[ \pi \cot\Bigl(\frac{\pi D}2\Bigr) (D-2) - \frac12 - \frac12 \ln(aa') \biggr] H^3 a^{\frac{D}{2}-1} \partial_0
\nonumber\\ 
& & \hspace{2.5cm} 
-\biggl[ \pi \cot\Bigl(\frac{\pi D}2\Bigr) \frac14 D(D-5)  + \frac12 \ln(aa') \biggr] H^2  \nabla^2  \Biggr\} \delta^{D} (x-x')
+\mathcal{O}(D-4)\;.
\label{M^2_4point_KWinteraction}
\eea  
This coincidence limit expression \eqref{M^2_4point_KWinteraction} will be used in Section \ref{renormalization}.

%%%%%%%%%%%%%%%%%%%%%%%%%%%%%%%%%%%%%%%%%%%%%%%%%%%%%%%%%%%%%%%%%%%%%%%%%%%%
\section{Computation of the 3-point interactions}\label{comp3ptint}
%%%%%%%%%%%%%%%%%%%%%%%%%%%%%%%%%%%%%%%%%%%%%%%%%%%%%%%%%%%%%%%%%%%%%%%%%%%%

In this section we evaluate the contributions from the 3-point kinetic and cross (kinetic-conformal) interactions.  
Let us remember the form of all the propagators in terms of the de Sitter invariant function $y(x;x')$ in (\ref{ydef}),
\bea
i\Delta_{\rm cf} (x;x') & = & \frac{\Gamma\Bigl( \frac{D}2 \!-\! 1\Bigr)}{(4\pi)^{\frac{D}2}}
\Bigl(\frac4{\Delta x^2}\Bigr)^{\frac{D}2-1} = (aa')^{\frac{D}{2}-1} \frac{H^{D-2}}{(4\pi)^{\frac{D}2}}\label{newF}
\Gamma\Bigl( \frac{D}2 \!-\! 1\Bigr)
\Bigl(\frac4{y}\Bigr)^{\frac{D}2-1} \equiv 
(aa')^{\frac{D}{2}-1} F(y)\;,   \\ 
i\Delta_{A, B, C} (x;x') & \equiv & A(y),\; B(y),\; C(y)\;.
\label{newABC}
\eea
Following the organization of  \cite{kahya}, we will split the 3-point  interactions into three parts:

(i) {\it Local contributions} including the delta function; 
\bea\label{dfordsb}
\partial_{\mu} \partial'_{\nu} i\Delta_{A}(x;x')&=& \delta_{\mu}^{0} \delta_{\nu}^{0} \frac{i}{a^{D-2}} \delta^D (x-x') 
\;+\; A''(y) \partial_{\mu} y \partial'_{\nu} y
\;+\; A'(y) \partial_{\mu} \partial'_{\nu} y \;,
\eea

(ii) {\it logarithm contributions} originated from the factor of $k\ln(aa')$ in the $A$-type propagator 
\bea
\kappa^2\;\partial_i \partial'_i \Bigl[ i\Delta_{A}(x;x') \; \partial_i \partial'_i i\Delta_{\rm{cf}}(x;x')\Bigr]\;,
\eea

(iii) {\it normal contributions} not having either delta function or $\ln(aa')$.  

At this point, it should be noted that the 3-point conformal interactions (that we evaluated in Ref. \cite{kahya3}) do not contain the local and logarithm contributions. 
%%%%%%%%%%%%%%%%%%%%%%%%%%%%%%%%%%%%%%%%%%%%%%%%%%%%%%%%%%%%%%%%%%%%%%%%%%%%
\subsection{3-point kinetic interactions}\label{3ptk}
%%%%%%%%%%%%%%%%%%%%%%%%%%%%%%%%%%%%%%%%%%%%%%%%%%%%%%%%%%%%%%%%%%%%%%%%%%%%

Contracting the tensor indices in the graviton propagator, 
the contribution from the kinetic 3-point interaction vertices  (\ref{M^2_3point_K_propagator}) becomes
\bea
-iM^2_{\rm 3pt K}(x;x')& =& \kappa^2 \;
\biggl\{ 
-\partial_i \partial'_i [ i\Delta_{A} (x;x') \; \partial_i \partial'_i \; i\Delta_{\rm{cf}}(x;x') ]
\;+\; C_{A} \; \partial_0 \partial'_0 [ i\Delta_{A}(x;x') \; \partial_0 \partial'_0 \; i\Delta_{\rm{cf}}(x;x') ]
\nonumber\\
& &\hspace{0.8cm}-\;\partial_i \partial'_0 [ i\Delta_{A}(x;x') \; \partial_i \partial'_0 \; i\Delta_{\rm{cf}}(x;x')]
\;-\;\partial_0 \partial'_i [ i\Delta_{A}(x;x') \; \partial_0 \partial'_i \; i\Delta_{\rm{cf}}(x;x')]
\nonumber\\
& &\hspace{0.8cm}+\;\partial_0 \partial'_0 [ i\Delta_{B} (x;x') \; \partial_i \partial'_i \;i\Delta_{\rm{cf}}(x;x')]
\;+\;\partial_i \partial'_0 [ i\Delta_{B} (x;x')\; \partial_0 \partial'_i \;i\Delta_{\rm{cf}}(x;x')]
\nonumber\\
& &\hspace{0.8cm}+\;\partial_0 \partial'_i [ i\Delta_{B} (x;x')\; \partial_i \partial'_0 \;i\Delta_{\rm{cf}}(x;x')]
\;+\;\partial_i \partial'_i [ i\Delta_{B} (x;x')\; \partial_0 \partial'_0 \;i\Delta_{\rm{cf}}(x;x')]
\nonumber\\
& &\hspace{0.8cm}-\;C_{C} \; \partial_0 \partial'_0 [ i\Delta_{C} (x;x')\; \partial_0 \partial'_0 \;i\Delta_{\rm{cf}}(x;x')]
\biggr\}\;.
\label{M^2_3point_K_interaction}
\eea 
Here the coefficients $C_{A}$ and $C_{C}$ are
\bea
C_{A} \equiv (\frac{D-1}{D-3}) \quad \mbox{and} \quad C_{C} \equiv 2(\frac{D-2}{D-3})\;.
\eea
We will break up \eqref{M^2_3point_K_interaction} into the three (local, logarithm and normal) parts and examine each of them in the next three subsections. 

%%%%%%%%%%%%%%%%%%%%%%%%%%%%%%%%%%%%%%%%%%%%%%%%%%%%%%%%%%%%%%%%
\subsubsection{Local contributions for 3-point kinetic part}\label{3ptklocal}
%%%%%%%%%%%%%%%%%%%%%%%%%%%%%%%%%%%%%%%%%%%%%%%%%%%%%%%%%%%%%%%%

The local contributions in \eqref{M^2_3point_K_interaction} come from the second, eighth, and ninth terms.
We apply the relation \eqref{dfordsb} to these terms. 

The second term becomes
\bea
\lefteqn{
\kappa^2 C_{A} \partial_0 \partial'_0 [ i\Delta_{A}(x;x') \partial_0 \partial'_0  i\Delta_{\rm{cf}}(x;x')]
= \kappa^2 (\frac{D-1}{D-3}) \partial_0 \partial'_0 [ i\Delta_{A} (x;x') \partial_0 \partial'_0 (aa')^{\frac{D}2 -1}F(y)]\;,
}
\nonumber\\
& &\hspace{0cm} =
 \kappa^2 \; \Bigl(\frac{D-1}{D-3}\Bigr) \biggl\{ \partial_0 \partial'_0 [ i\Delta_{A} (x;x') (\partial_0 \partial'_0 (aa')^{\frac{D}2 -1}) F(y)]
+ \partial_0 \partial'_0 [ i\Delta_{A}(x;x')  (\partial'_0 (aa')^{\frac{D}2 -1}) (\partial_0  F(y))]
\nn \\
& & \hspace{2.5cm} + \partial_0 \partial'_0 [ i\Delta_{A}(x;x')  (\partial_0 (aa')^{\frac{D}2 -1}) (\partial'_0  F(y)) ]
+ \partial_0 \partial'_0 [ (aa')^{\frac{D}2 -1}  i\Delta_{A}(x;x')  (\partial_0 \partial'_0  F(y))]
\biggr\}\;,
\nonumber\\
& &\hspace{0cm}=\kappa^2 \; \Bigl(\frac{D-1}{D-3}\Bigr) \biggl\{ \partial_0 \partial'_0 [ i\Delta_{A}(x;x') (\frac{D}{2}-1)^2 H^2 (aa')^{\frac{D}{2}} F(y)]
+ \partial_0 \partial'_0 [ i\Delta_{A}(x;x')  (\frac{D}{2}-1) H (aa')^{\frac{D}{2}-1}  [a'\partial_0 + a\partial'_0]  F(y)]
\nonumber\\
& &\hspace{2.5cm}+\partial_0 \partial'_0 [ (aa')^{\frac{D}2 -1}  i\Delta_{A}(x;x') \partial_0 \partial'_0   F(y) ]
\biggr\}\;,
\nonumber\\
& &\hspace{0cm}\rightarrow  -i \kappa^2  \frac{H^{D-4}}{(4\pi )^{\frac{D}{2}}} \frac{\Gamma (D-1)}{\Gamma(\frac{D}{2})}  \pi \cot\Bigl(\frac{\pi D}{2}\Bigr) \Bigl(\frac{D-1}{D-3}\Bigr) 
\biggl\{ H^2 a^{\frac{D}{2}} \square +(D-2) \Bigl[H^3 a^{\frac{D}{2}-1} \partial_{0} - \frac{(3D-8)}{4} H^4 a^{\frac{D}{2}} \Bigr]- H^2\nabla^2 \biggr\} \delta^{D}(x-x')
\nonumber\\
& &\hspace{0.5cm}
+\;\mathcal{O}(D-4)\;. 
\eea
Here we used the coincidence limit of the $A$-type propagator  \eqref{coa} and covariant d'Alembertian \eqref{dsda}. 

The eighth term becomes
\bea
\lefteqn{\kappa^2 \partial_i \partial'_i [ i\Delta_{B}(x;x')  \partial_0 \partial'_0  i\Delta_{\rm{cf}}(x;x')]
= - \kappa^2  \nabla^2 [ i\Delta_{B}(x;x') \partial_0 \partial'_0 (aa')^{\frac{D}2 -1} F(y) ]\;,
}
\nonumber\\
& &\hspace{0cm}= - \kappa^2  \biggl\{ \nabla^2\Bigl[ i\Delta_{B}(x;x')  (\partial_0 \partial'_0 (aa')^{\frac{D}2 -1}) F(y) \Bigr]
+ \nabla^2\Bigl[ i\Delta_{B} (x;x')  (\partial'_0 (aa')^{\frac{D}2 -1}) (\partial_0  F(y)) \Bigr]
\nonumber\\
& &\hspace{1.5cm}
+ \nabla^2\Bigl[ i\Delta_{B} (x;x')  (\partial_0 (aa')^{\frac{D}2 -1}) (\partial'_0  F(y)) \Bigr]
+ \nabla^2\Bigl[ (aa')^{\frac{D}2 -1}  i\Delta_{B}(x;x')  (\partial_0 \partial'_0  F(y)) \Bigr]
\biggr\}\;,
\nonumber\\
& &\hspace{0cm}\rightarrow i \kappa^2  \frac{H^2}{(4\pi)^{2}} \nabla^2 \delta^4 (x-x') + \mathcal{O}(D-4)\;.
\eea

The ninth term becomes
\bea
\lefteqn{-\kappa^2 C_{C} \partial_0 \partial'_0 [ i\Delta_{C}(x;x')  \partial_0 \partial'_0 i\Delta_{\rm{cf}}(x;x')]
}
\nonumber\\
& &\hspace{0cm}
= -\kappa^2 \; 2(\frac{D-2}{D-3})  \biggl\{ \partial_0 \partial'_0 [ i\Delta_{C}(x;x') (\partial_0 \partial'_0 (aa')^{\frac{D}2 -1}) F(y) ]
+ \partial_0 \partial'_0 [ i\Delta_{C}(x;x')  (\partial'_0 (aa')^{\frac{D}2 -1}) (\partial_0  F(y)) ]
\nonumber\\& &\hspace{3cm}
+ \partial_0 \partial'_0 [ i\Delta_{C}(x;x') (\partial_0 (aa')^{\frac{D}2 -1}) (\partial'_0  F(y)) ]
+ \partial_0 \partial'_0 [ (aa')^{\frac{D}2 -1} i\Delta_{C}(x;x') (\partial_0 \partial'_0 F(y)) ]
\biggr\}\;,
\nonumber\\
& &\hspace{0cm}= - \kappa^2 2(\frac{D-2}{D-3}) \biggl\{ \partial_0 \partial'_0 [ i\Delta_{C}(x;x') (\frac{D}{2}-1)^2 H^2 (aa')^{\frac{D}{2}} F(y) ]
+ \partial_0 \partial'_0 [ i\Delta_{C}(x;x')   (\frac{D}{2}-1) H (aa')^{\frac{D}{2}-1} [a'\partial_0 + a\partial'_0]  F(y) ]
\nonumber\\& &\hspace{3cm}
+ \partial_0 \partial'_0 [ (aa')^{\frac{D}2 -1} i\Delta_{C}(x;x') \partial_0 \partial'_0  F(y) ]
\biggr\}\;,
\nonumber\\
& &\hspace{0cm}\rightarrow \; i \kappa^2  \frac{H^2}{(4\pi )^{2}} 
\biggl\{ -4H^2a^2\square -8 H a \partial_{0} + 8 H^2a^2 + 4 \nabla^2 \biggr\} \delta^4 (x-x')+\mathcal{O}(D-4)\;.
\eea

Summing the three local contributions leads to
\bea\label{3ptlock}
\lefteqn{-iM^2_{\rm{3pt K} \atop {\rm loc}}(x;x') = -i\kappa^2 \;\frac{H^{D-4}}{(4\pi)^{\frac{D}{2}}} \frac{\Gamma (D-1)}{\Gamma(\frac{D}{2})}  
\biggl\{ [ \pi \cot(\frac{\pi D}{2}) (\frac{D-1}{D-3})\;+\;2 ] \;H^2 a^{\frac{D}{2}} \square 
}
\nonumber\\
& &\hspace{0.0cm}+\; [ \pi \cot(\frac{\pi D}{2}) (\frac{D-1}{D-3})(D-2)\;+\;4 ] \;H^3 a^{\frac{D}{2}-1} \partial_0 
-\; [ \pi \cot(\frac{\pi D}{2}) (\frac{D-1}{D-3})(D-2)\frac{(3D-8)}{4}\;+\;4 ] \;H^4 a^{\frac{D}{2}} 
\nonumber\\
& &\hspace{0.0cm}-\; [ \pi \cot(\frac{\pi D}{2}) (\frac{D-1}{D-3})\;+\;\frac52 ] \;H^2 \nabla^2 \biggr\} \delta^D (x-x') 
\;+\;\mathcal{O} (D-4)\;. 
\eea

%%%%%%%%%%%%%%%%%%%%%%%%%%%%%%%%%%%%%%%%%%%%%%%%%%%%%%%%%%%%%%%%
\subsubsection{Logarithm contributions for 3-point kinetic part}\label{3ptklog}
%%%%%%%%%%%%%%%%%%%%%%%%%%%%%%%%%%%%%%%%%%%%%%%%%%%%%%%%%%%%%%%%

The logarithm contributions come from the first to fourth terms in \eqref{M^2_3point_K_interaction}, which contain the $k\ln(aa')$ term in the A-type propagator \eqref{compactA}.
Because all the logarithm contributions are finite we can take the limit $D=4$. 

The first term is 
\bea
-\kappa^2  \partial_i \partial'_i [ i\Delta_{A}(x;x')  \partial_i \partial'_i i\Delta_{\rm{cf}}(x;x')] 
&\rightarrow& -\kappa^2  \partial_i \partial'_i [ k  \ln (aa') \partial_i \partial'_i (aa')^{\frac{D}{2}-1} F(y)] 
\nonumber\\
&=& -\kappa^2  k \ln (aa') (aa')^{\frac{D}{2}-1} \nabla^4 F(y)\;,
\nonumber\\
&=& -\kappa^2  \frac{H^2}{8 \pi^2} \ln (aa')  (aa') \nabla^4 F(y) +\mathcal{O}(D-4)\;.
\eea
Here the $D=4$ limit was taken for $k \equiv \frac{H^{D-2}}{(4 \pi )^{D/2}} \frac{\Gamma(D-1)}{\Gamma(\frac{D}2)}\rightarrow \frac{H^2}{8 \pi^2}$. 

The second term becomes
\bea
 \lefteqn{\kappa^2 C_{A} \partial_0 \partial'_0 [ i\Delta_{A}(x;x') \partial_0 \partial'_0 i\Delta_{\rm{cf}}(x;x')]
 \rightarrow \kappa^2 (\frac{D-1}{D-3}) \partial_0 \partial'_0 [ k \ln (aa') \partial_0 \partial'_0 (aa')^{\frac{D}2 -1} F(y)]
}
\nonumber\\
& &\hspace{0cm} = \kappa^2 (\frac{D-1}{D-3}) \biggl\{ \partial_0 \partial'_0 [ k  \ln (aa')  (\partial_0 \partial'_0 (aa')^{\frac{D}2 -1}) F(y) ]
+ \partial_0 \partial'_0 [ k  \ln (aa')  (\partial'_0 (aa')^{\frac{D}2 -1}) (\partial_0  F(y)) ]
\nonumber\\& &\hspace{2.5cm}
+ \partial_0 \partial'_0 [ k  \ln (aa') (\partial_0 (aa')^{\frac{D}2 -1}) (\partial'_0  F(y)) ]
+ \partial_0 \partial'_0 [ (aa')^{\frac{D}2 -1}  k  \ln (aa') \partial_0 \partial'_0 F(y) ]
\biggr\}\;,
\nonumber\\
& &\hspace{0cm}=\; \kappa^2 \frac{3H^2}{8\pi^2} \biggl\{ H^2 \partial_0 \partial'_0 [  \ln (aa') \; (aa')^{2} F(y) ] 
\;+\; H \partial_0 \partial'_0 [ \ln (aa') \; (aa')\;[a' \partial_{0} + a \partial'_{0} ] F(y) 
\nonumber\\
& &\hspace{2cm}+\; \partial_0 \partial'_0 [ \ln (aa') \; i\delta^{4} (x-x') ] \biggr\}\;+\; \mathcal{O}(D-4)\;,
\nonumber\\
& &\hspace{0cm}=\; \kappa^2 \frac{H^2}{8\pi^2} \biggl\{ -[24+15\ln(aa')] H^4 (aa')^{3} F(y) \;+\; [6+\frac92 \ln(aa') ] H^2 (aa')^2 \nabla^2 I[F](y) \biggr\}
\nonumber\\
& &\hspace{2cm}+\; i\kappa^2 \frac{H^2}{8\pi^2}  \ln (aa')\biggl\{ 3 \; a^2\square + 6\; H a \partial_{0}- 6\;H^2 a^2 - 3 \;\nabla^2  \biggr\} \; \delta^{4} (x-x') \;+\; \mathcal{O}(D-4)\;.
\eea

The third term gives
\bea
\lefteqn{-\kappa^2  \partial_i \partial'_0 [ i\Delta_{A}(x;x')  \partial_i \partial'_0 i\Delta_{\rm{cf}}(x;x')]
 \rightarrow  -\kappa^2  \partial_i \partial'_0 [ k \ln (aa') \partial_i \partial'_0 (aa')^{\frac{D}{2}-1} F(y)] 
}
\nonumber\\
& &\hspace{0cm}= 
-\kappa^2 \biggl\{ k \ln (aa') \partial_i \partial'_0 [ \partial_i \partial'_0  (aa')^{\frac{D}{2}-1} F(y)]
+ (\partial_i k \ln (aa'))  \partial'_0 [\partial_i \partial'_0 (aa')^{\frac{D}{2}-1} F(y) ] 
+ (\partial'_0 k \ln (aa')) \partial_i[\partial_i \partial'_0 (aa')^{\frac{D}{2}-1}  F(y) ] \biggr\}\;,
\nonumber\\
& &\hspace{0cm}=
\kappa^2 \frac{H^2}{8\pi^2} \biggl\{ [1-2\ln(aa') ] [H^2 (aa') a^{'2} \nabla^2 + H(aa') a'\partial'_{0} ]\biggr\} F(y)
+ i\kappa^2 \frac{H^2}{8\pi^2} \ln(aa') \nabla^2  \delta^4(x-x') +\mathcal{O}(D-4) \;.
\eea

Similarily, the fourth term gives
\bea
\lefteqn{-\kappa^2  \partial_0 \partial'_i [ i\Delta_{A}(x;x')  \partial_0 \partial'_i i\Delta_{\rm{cf}}(x;x')]
\rightarrow  -\kappa^2  \partial_0 \partial'_i [k \ln (aa') \partial_0 \partial'_i (aa')^{\frac{D}{2}-1} F(y)]
}
\nonumber\\
& &\hspace{0cm}=-\kappa^2 \biggl\{ k \ln (aa') \partial_0 \partial'_i [ \partial_0 \partial'_i  (aa')^{\frac{D}{2}-1} F(y)]
+ (\partial_0 k \ln (aa'))  \partial'_i [\partial_0 \partial'_i (aa')^{\frac{D}{2}-1} F(y) ] 
+ (\partial'_i k \ln (aa')) \partial_0[\partial_0 \partial'_i (aa')^{\frac{D}{2}-1}  F(y) ] \biggr\}\;,
\nonumber\\
& &\hspace{0cm}=\kappa^2 \frac{H^2}{8\pi^2} \Biggl\{ [1-2\ln(aa') ]  [H^2 (aa') a^{2} \nabla^2 + H(aa') a\partial_{0} ]\Biggr\} F(y) 
+ i\kappa^2 \frac{H^2}{8\pi^2}  \ln(aa') \nabla^2  \delta^4(x-x') +\mathcal{O}(D-4) \;.
\eea

Collecting all the four logarithm contributions, we have  
\bea\label{3ptlogk}
 \lefteqn{-iM^2_{\rm{3pt K} \atop {\rm log}}(x;x') = \kappa^2 \frac{H^2}{8\pi^2} \Biggl\{ -15\;[\frac85 + \ln(aa') ] H^4 (aa')^3 F(y)
+ \frac92 [\frac43 + \ln(aa') ] H^2 (aa')^2 \nabla^2 I[F](y) }
\nonumber\\
& &\hspace{0cm}
- [-\frac12 + \ln(aa') ] H^2(aa')^2 \nabla^2 [ 6 F(y) + 4 y F'(y) ]
- [-\frac12 + \ln(aa') ] H^2 (aa') (a^2 + a^{'2}) \nabla^2 [ 2 F(y) + 2 y F'(y) ]
\nonumber\\
& &\hspace{0cm}+ [-\frac12 + \ln(aa') ]  (aa')  \nabla^4 I[F](y) - \ln(aa') (aa') \nabla^4 F(y)
+\ln(aa') \left[3a^2 \square + 6H a \partial_{0} - 6 H^2 a^2 - \nabla^2\right]  i\delta^4 (x-x') \biggr\}
\nonumber\\& &\hspace{0cm}
+\mathcal{O}(D-4)\;.
\eea
Here we used the following identities derived in Refs. \cite{kahya, kahya3} 
\bea
\bigl[ a' \partial_{0} \;+\; a \partial'_{0} \bigr]  f(y) & = & -(D-1) H a a' f(y) \;+\; \frac{\nabla^2}{2 H} I[f](y)\;,\label{ides1}
\\
\bigl[a \partial_{0} \;+\; a' \partial'_{0} \bigr] f(y) & = & H (a + a')^2 y f'(y) \;+\; (D-1) H a a' f(y) \;-\; \frac{\nabla^2}{2 H} I[f](y)\;.\label{ides2}
\eea

The terms not multiplied by the delta function in Eq. \eqref{3ptlogk} (namely,  the nonlocal logarithm contributions) are summarized in TABLE ~\ref{knloclog}. Later this will be added to TABLE ~\ref{kt4}. In the tables, we write the function $F(y)$ defined in Eq.  \eqref{newF} in terms of $x$, i.e., by rescaling $x\equiv \frac{y}{4}$ and take the $D=4$ limit:
\bea\label{defF(x)}
F(y) &\rightarrow& \frac{H^2}{(4 \pi)^{2}} \frac{1}{x}\;.
\eea
The last term multiplied by the delta function will give finite local terms after renormalization.
\begin{table}[htbp]
\begin{center}
\caption{Nonlocal logarithm contributions from the relation \eqref{3ptlogk} with $x \equiv \frac{y}{4}$.}
\begin{ruledtabular}
\begin{tabular}{lcl}
${\phantom{ssssssssss}\rm External\; operators}$&&${\rm Coef.\;of\;} \frac{\kappa^2 H^{4}}{(4\pi)^4}$ \\[1ex]
\hline\\
$\phantom{ssssssssss}(aa')^3 H^4$&&$  -\frac{48}{x} - \ln(aa') \frac{30}{x}\phantom{ssssssssssssssssssssss}$ \\[2ex]
$\phantom{ssssssssss}(aa')^2 H^2 \nabla^2$&& $ 12\ln(x) - \frac{10}{x} + \ln(aa')[ 9\ln(x) +  \frac{20}{x}]$ \\[2ex]
$\phantom{ssssssssss}(aa') (a^2+a'^2) H^2 \nabla^2$&& $ -\frac{6}{x} + \ln(aa') \frac{12}{x} $ \\[1ex]
$\phantom{ssssssssss}(aa')  \nabla^4$&& $ -\ln(x)  + \ln(aa') [ 2\ln(x) - \frac{2}{x}] $ \\[2ex]
\end{tabular}\label{knloclog}
\end{ruledtabular}
\end{center}
\end{table}

%%%%%%%%%%%%%%%%%%%%%%%%%%%%%%%%%%%%%%%%%%%%%%%%%%%%%%%%%%%%%%%%
\subsubsection{Normal contributions for 3-point kinetic part}\label{3ptknormal}
%%%%%%%%%%%%%%%%%%%%%%%%%%%%%%%%%%%%%%%%%%%%%%%%%%%%%%%%%%%%%%%%

The remaining terms in the 3-point kinetic part (\ref{M^2_3point_K_interaction}) are normal contributions: The terms having the $A$-type propagator are
\bea
& & -\kappa^2 \partial_i \partial'_i [ i\Delta_{A}(x;x')  \partial_i \partial'_i i\Delta_{\rm{cf}}(x;x')] 
= - \kappa^2 (aa')^{\frac{D}{2}-1} \partial_i \partial'_i [ A \; \partial_i \partial'_i F ] \;,
\\
& & \kappa^2 C_{A}  \partial_0 \partial'_0 [ i\Delta_{A}(x;x')  \partial_0 \partial'_0 i\Delta_{\rm{cf}}(x;x')]
= \kappa^2 C_{A}  \biggl\{ (\frac{D}{2}-1)^2 H^2 \partial_0\partial'_0[\;(aa')^{\frac{D}{2}} [AF] ] 
\nn \\ 
& & \hspace{1cm} + (\frac{D}{2}-1) H \partial_0\partial'_0[(aa')^{\frac{D}{2}-1} [A (a'\partial_0 + a\partial'_0 )F] ]
+\partial_0\partial'_0[\;(aa')^{\frac{D}{2}-1} [A \partial_0\partial'_0 F] ]\biggr\}\;,
\\
& & -\kappa^2 \biggl\{\partial_i \partial'_0 [ i\Delta_{A} (x;x')\; \partial_i \partial'_0 \;i\Delta_{\rm{cf}}(x;x')]
\;+\;\partial_0 \partial'_i [ i\Delta_{A} (x;x')\; \partial_0 \partial'_i \;i\Delta_{\rm{cf}}(x;x')] \biggr\}  
\nn\\
& & \hspace{1cm}= -\kappa^2 \biggl\{ (\frac{D}{2}-1) a^{\frac{D}{2}-1} H \partial_i \partial'_0 [A a'^{\frac{D}{2}} \partial_i F ] + \partial_i \partial'_0 [(aa')^{\frac{D}{2}-1} A \partial_i \partial'_0 F]
\nonumber\\
& & \hspace{2.5cm} +(\frac{D}{2}-1) a'^{\frac{D}{2}-1} H \partial'_i \partial_0 [A a^{\frac{D}{2}} \partial'_i F ] + \partial'_i \partial_0 [(aa')^{\frac{D}{2}-1} A \partial'_i \partial_0 F]
\biggr\}\;.
\eea
The terms including the $B$-type propagator are
\bea
& & 
\kappa^2 \partial_0 \partial'_0 [ i\Delta_{B}(x;x')  \partial_i \partial'_i \;i\Delta_{\rm{cf}}(x;x')]=- \kappa^2 \partial_0 \partial'_0 [(aa')^{\frac{D}{2}-1} B \partial_i^2 F]\;,
\\
& & 
\kappa^2 \biggl\{ \partial_i \partial'_0 [ i\Delta_{B} (x;x') \; \partial_0 \partial'_i \;i\Delta_{\rm{cf}}(x;x')]
\;+\;\partial_0 \partial'_i [ i\Delta_{B} (x;x') \; \partial_i \partial'_0 \;i\Delta_{\rm{cf}}(x;x')] \biggr\} 
\nonumber\\
& &  \hspace{1cm} = \kappa^2 \biggl\{
\partial_i\partial'_0[(aa')^{\frac{D}{2}-1}(\frac{D}{2}-1)Ha B \partial'_i F] 
+ \partial_i\partial'_0[(aa')^{\frac{D}{2}-1} B\partial_0\partial'_i F] 
\nonumber\\
& & \hspace{2.5cm} +\partial'_i\partial_0[(aa')^{\frac{D}{2}-1}(\frac{D}{2}-1)Ha' B \partial'_i F] 
+\partial'_i\partial_0[(aa')^{\frac{D}{2}-1} B\partial'_0\partial_i F] \biggr\} \;, 
\\
& & 
\kappa^2 \partial_i \partial'_i [ i\Delta_{B} (x;x') \partial_0 \partial'_0 i\Delta_{\rm{cf}}(x;x')] = -\kappa^2 \partial_i^2 [B \partial_0 \partial'_0 ((aa')^{\frac{D}{2}-1} F)]\;.
\eea
Finally the terms containing the $C$-type propagator are
\bea
& & 
\kappa^2 C_{C}  \partial_0 \partial'_0 [ i\Delta_{C}(x;x')  \partial_0 \partial'_0 i\Delta_{\rm{cf}}(x;x')]
=-\kappa^2 C_{C} \; \biggl\{ (\frac{D}{2}-1)^2 H^2 \partial_0\partial'_0[\;(aa')^{\frac{D}{2}} [CF] ]
\nonumber\\
& & \hspace{1cm} +(\frac{D}{2}-1) H \partial_0\partial'_0[\;(aa')^{\frac{D}{2}-1} [C (a'\partial_0 + a\partial'_0 )F]]
+\partial_0\partial'_0[(aa')^{\frac{D}{2}-1} [C \partial_0\partial'_0 F] ]
\biggr\}\;.
\eea
To reduce these normal contributions we will apply the following identity
\bea\label{ui}
\lefteqn{
\partial_0\partial'_0[(aa')^n f] = (aa')^n  \biggl\{ \frac12 (a+a')^2 [(D-1)yf' +y^2 f''] + \frac12 (a^2+a'^2) [\square f + H^2yf']
}
\nonumber\\
& & \hspace{3.5cm} + [n^2 -(D-1) [n-\frac{D-2}{2}]] H^2aa' f +\frac12 [n-\frac{D-2}{2}] \nabla^2 I[f] -\nabla^2 f
\biggr\}\;.
\eea

Keeping in mind that the self-mass-squared will eventually be integrated in the quantum-corrected field equation \eqref{linear_eq}, our goal is to make the integral finite by pulling the derivatives outside the integral.  
The strategy is to convert primed derivatives into unprimed ones so that we can freely pull them outside the integral. Once this procedure is completed, the self-mass-squared becomes in the form of eleven unprimed external operators acting on the functions of $y$. Extracting derivatives involves indefinite integrations and we denote this operation by   
\bea
I[f](y) \equiv \int^y dy' f(y') \;.
\eea
The following example demonstrates the operation \cite{kahya}:
\bea
f(y) \Biggl\{A''(y) \frac{\partial y}{\partial x^{\rho}} \frac{\partial y}{
\partial x^{\prime \sigma}} \!+\! A'(y) \frac{\partial^2 y}{\partial x^{\rho}
\partial x^{\prime \sigma}} \Biggr\} \!=\!
\partial_{\rho} \partial_{\sigma}' I^2[f A''](y) \!+\! \frac{\partial^2 y}{
\partial x^{\rho} \partial x^{\prime \sigma}} I[f' A'](y) \; . \label{keyI}
\eea
The terms in the left-hand-side are converted to the desired form of derivatives acting on a function of $y$ plus an extra term as a function of $y$ in the right-hand-side.  
Similar identities were also derived in Refs. \cite{kahya, kahya3}
\bea
f(y)\nabla^2 A(y) &=& \nabla^2 I^2[f A''](y) + 2(D-1) H^2 a a' I[f'A'](y)\;,
\label{ID017} \qquad\\
f(y)\partial'_0 \partial'_0 A(y) &=& \partial'_0 \partial'_0 I^2[f A''](y) + (\partial'_0 \partial'_0 y) I[F'A'](y)\;,
\label{ID018} \qquad\\
f(y)\partial'_i \partial'_0 A(y) &=& \partial'_i \partial'_0 I^2[f A''](y) + H a'\; \partial'_i I^2[f' A'](y)\;,
\label{ID019} \qquad\\
f(y)\partial'_i \partial'_j A(y) &=& \partial'_i \partial'_j I^2[f A''](y) + 2 H^2 a a' \eta_{ij} I[f' A'](y)\;.
\label{ID020}
\eea
Using these identities, all the remaining terms in (\ref{M^2_3point_K_interaction}) can be converted to functions of $y$ which are acted upon by one of the following eleven external (unprimed) operators
\bea
\alpha & \equiv & (a a')^{\frac{D}{2}+1} \square^2 \label{alpha} \; , 
\label{newa}\\ 
\beta_1 & \equiv & (a a')^{\frac{D}{2}+1}  H^2 \square
 \label{beta1}\; , \\
\beta_2 & \equiv & (a a')^{\frac{D}{2}} (a^2 + a^{\prime 2}) H^2 \square \; , \\
\gamma_1 & \equiv &  (a a')^{\frac{D}{2}+1} H^4 \; , \\
\gamma_2 & \equiv &  (a a')^{\frac{D}{2}} (a^2 + a^{\prime 2}) H^4 \; , \\
\gamma_3 & \equiv &  (a a')^{\frac{D}{2}} (a + a')^2 H^4 = 2 \gamma_1 + \gamma_2 \; , \\
\delta & \equiv &  (a a')^{\frac{D}{2}-1}(a^2 + a^{\prime 2}) \nabla^2 \square \; , \\
\epsilon_1 & \equiv &  (a a')^{\frac{D}{2}} H^2 \nabla^2 \; , \\
\epsilon_2 & \equiv &  (a a')^{\frac{D}{2}-1}(a^2 + a^{\prime 2}) H^2 \nabla^2 \; , \\
\epsilon_3 & \equiv &  (a a')^{\frac{D}{2}-1}(a + a')^2 H^2 \nabla^2 = 2\epsilon_1 + \epsilon_2 \; ,\\
\zeta & \equiv &  (a a')^{\frac{D}{2}-1}\nabla^4 \label{zeta}\;.
\label{newz} 
\eea
For instance, applying the identity (\ref{ID017}) to the first term (which is the simplest) in (\ref{M^2_3point_K_interaction}) gives
\bea
-\partial_i \partial'_i [ i\Delta_{A}(x;x') \; \partial_i \partial'_i \;i\Delta_{\rm{cf}}(x;x')] &=& - (aa')^{\frac{D}{2}-1} \partial_i \partial'_i [ A \; \partial_i \partial'_i F ] = - (aa')^{\frac{D}{2}-1} \nabla^2 [ A \; \nabla^2 \; F ] \;,
\nonumber\\
&=& - (aa')^{\frac{D}{2}-1} \nabla^2 \;[\nabla^2 I^2[AF''] \;+\;2(D-1)H^2aa' I[A'F'] ]\;,
\nonumber\\
&=& - (aa')^{\frac{D}{2}-1} [\nabla^4 I^2[AF''] \;+\;2(D-1)H^2aa' \nabla^2 I[A'F'] ]\;,
\nonumber\\
&=& - ( \zeta I^2[AF''] \;+\; 2(D-1)\epsilon_1 I[A'F'])\;.
\eea
To convert the fifth term (which is the most complicated) in (\ref{M^2_3point_K_interaction}) we use the identity (\ref{ui}),
\bea
\lefteqn{
\kappa^2 \partial_0 \partial'_0 [ i\Delta_{B} (x;x') \partial_i \partial'_i i\Delta_{\rm{cf}}(x;x')] 
= - \kappa^2 \partial_0 \partial'_0 [(aa')^{\frac{D}{2}-1} B \nabla^2 F]\;,
}
\nonumber\\
& & = - \kappa^2 \partial_0 \partial'_0 [(aa')^{\frac{D}{2}-1} \;(\nabla^2 I^2[BF''] +2(D-1)H^2aa' I[B'F'] )]
\nonumber\\
& & = - \kappa^2 \nabla^2 \partial_0 \partial'_0 [(aa')^{\frac{D}{2}-1}  I^2[BF''] ] - \kappa^2  2(D-1)H^2  \partial_0 \partial'_0 [ (aa')^{\frac{D}{2}} I[B'F'] ]\;,
\nonumber\\
& & =- \kappa^2 \biggl\{ \frac12  \epsilon_3 [(D-1)y I[BF''] +y^2 [BF'']] 
+  \frac12 [\delta I^2[BF''] + \epsilon_2 y I[BF'']] + (\frac{D}{2}-1)^2 \; \epsilon_1 I^2[BF'']
\nonumber\\
& &\hspace{1.5cm} - \zeta I^2[BF'']
+ (D-1)\gamma_3 [(D-1)y [B'F']+y^2 [B'F']'] 
+ (D-1) [\beta_2 I[B'F'] + \gamma_2 y [B'F']']
\nonumber\\
& & \hspace{1.5cm} +\frac12 (D-1)(D^2-4D+4) \gamma_1 I[B'F'] + (D-1) \epsilon_1 I^2[B'F'] 
- 2(D-1)\;\epsilon_1 I[B'F']
\biggr\}\;.
\eea
The result is simply written as
\bea
\lefteqn{-iM^2_{\rm 3ptK}(x;x')
= -\kappa^2  \Bigl\{\alpha f_{\alpha}(y) 
+ \beta_2 f_{\beta_2}(y)
+ \gamma_1 f_{\gamma_1}(y) + \gamma_2 f_{\gamma_2}(y) + \gamma_3 f_{\gamma_3}(y) 
} \nonumber\\
& & \hspace{4.5cm}+ \delta f_{\delta}(y) 
+ \epsilon_1 f_{\epsilon_1}(y) + \epsilon_2 f_{\epsilon_2}(y) + \epsilon_3 f_{\epsilon_3}(y) 
+ \zeta f_{\zeta}(y) \Bigl \}\;,
\label{3ptK-alphazeta}
\eea
where we give the functions $f_i$ (on which the external operators are acting) in TABLEs \ref{kta} through \ref{ktz}.

%%%%%%%%%%%%%%%%%%%%%%%%%%%%%%%%%%%%%%%%%%%%%%%%%%%%%%%%%%%%%%%%
\subsection{3-point cross interactions}\label{3ptcross}
%%%%%%%%%%%%%%%%%%%%%%%%%%%%%%%%%%%%%%%%%%%%%%%%%%%%%%%%%%%%%%%%

By contracting the tensor indices in (\ref{3-ptcross}), the contribution from the 3-point cross interactions (namely kinetic-conformal terms of the self-mass-squared) becomes
\bea\label{M^2_3point_cross_interaction}
-iM^2_{\rm 3ptcross}(x;x')
& =&  \tilde{\kappa}^2 \Biggl\{
C_{1} \partial_i[\partial_i \; i\Delta_{\rm{cf}}(x;x') \Bigl(\nabla^{2}  i\Delta_{A}(x;x') + \nabla^{'2} i\Delta_{A}(x;x') \Bigr)]
\nonumber\\
& &\hspace{1.0cm} 
- C_{1} \partial_i[\partial_i \;i\Delta_{\rm{cf}}(x;x') \Bigl(\partial_{0}^{2} i\Delta_{A}(x;x') + \partial_{0}^{'2} i\Delta_{A} (x;x')\Bigr)] 
\nonumber\\
& &\hspace{1.0cm} 
+\partial_0[\partial_0 i\Delta_{\rm{cf}}(x;x') \nabla^{'2} \Bigl( - C_{2} i\Delta_{A}(x;x') + C_{3} i\Delta_{C} (x;x')\Bigr)]
\nonumber\\
& &\hspace{1.0cm} 
+\partial'_0[\partial'_0 \;i\Delta_{\rm{cf}}(x;x') \nabla^{2} \Bigl(- C_{2} i\Delta_{A} (x;x') + C_{3} i\Delta_{C} (x;x')\Bigr)]
\nonumber\\ 
& &\hspace{1.0cm} 
+ C_{4}\partial_0[\partial_0 \;i\Delta_{\rm{cf}}(x;x') \partial_{0}^{'2} \Bigl( i\Delta_{A} (x;x') - i\Delta_{C} (x;x') \Bigr) ] 
\nonumber\\ 
& &\hspace{1.0cm} 
+C_{4}\partial'_0[\partial'_0 \;i\Delta_{\rm{cf}}(x;x') \partial_{0}^{2} \Bigl( i\Delta_{A}(x;x') - i\Delta_{C} (x;x') \Bigr) ] 
\nonumber\\
& &\hspace{1.0cm} 
+C_{1}\partial_0[\partial_i \;i\Delta_{\rm{cf}}(x;x') \partial'_i \partial'_0 \;i\Delta_{B} (x;x')] 
+ C_{1} \partial'_0[\partial'_i \; i\Delta_{\rm{cf}}(x;x') \partial_i \partial_0  \;i\Delta_{B}(x;x')]
\nonumber\\
& &\hspace{1.0cm} 
+\;C_{1}\partial_i[\partial_0 \;i\Delta_{\rm{cf}}(x;x') \partial'_i \partial'_0\; i\Delta_{B}(x;x')] 
+ C_{1} \partial'_i[\partial'_0 \;i\Delta_{\rm{cf}}(x;x') \partial_i \partial_0\; i\Delta_{B}(x;x')]
\nonumber\\
& &\hspace{1.0cm} 
-C_{1}\partial_i[\partial_j \;i\Delta_{\rm{cf}}(x;x') \partial'_i \partial'_j \;i\Delta_{A}(x;x')] 
- C_{1} \partial'_i[\partial'_j \;i\Delta_{\rm{cf}}(x;x') \partial_i \partial_j \;i\Delta_{A}(x;x')]\Biggr\}\;.\label{3-ptcrosscoeff}
\eea
Here the coefficients are 
\bea\label{coeffsABC}
C_{1} \equiv 4 \;, \;  
C_{2} \equiv 4(\frac{D-2}{D-3}) \;, \;
C_{3} \equiv \frac{4}{(D-3)} \;, \; 
C_{4} \equiv 4(\frac{D-1}{D-3}) \;. 
\eea
In the following subsections we split this 3-point cross part into the local, logarithm, and normal contributions  as we have done for the 3-point kinetic part.

%%%%%%%%%%%%%%%%%%%%%%%%%%%%%%%%%%%%%%%%%%%%%%%%%%%%%%%%%%%%%%%%
\subsubsection{Local contributions for 3-point cross part}\label{3ptcrosslocal}
%%%%%%%%%%%%%%%%%%%%%%%%%%%%%%%%%%%%%%%%%%%%%%%%%%%%%%%%%%%%%%%%

The fifth and sixth terms of  \eqref{M^2_3point_cross_interaction} produce the local contributions. 
We will again apply the equation \eqref{dfordsb}. The fifth term becomes
\bea
\tilde{\kappa}^2  C_{4} \partial_0[ \partial_0 \;i\Delta_{\rm{cf}}(x;x') \partial_{0}^{'2} \Bigl( i\Delta_{A}(x;x') - i\Delta_{C}(x;x') \Bigr) ] 
&\!\!=\!\!& \tilde{\kappa}^2  4(\frac{D-1}{D-3}) \partial_0[\partial_0 (aa')^{\frac{D}{2}-1}  F(y) \partial_{0}^{'2} \Bigl( i\Delta_{A}(x;x') - i\Delta_{C}(x;x')\Bigr)]\;,
\nonumber\\
&\!\!=\!\!& \tilde{\kappa}^2  4(\frac{D-1}{D-3}) \partial_0[\partial_0 (aa')^{\frac{D}{2}-1} F(y) \partial_{0}^{'2} ( 0 )]
= 0 \;.
\eea
The sixth term becomes
\bea
\tilde{\kappa}^2  C_{4} \partial'_0[ \partial'_0 \;i\Delta_{\rm{cf}}(x;x') \partial_{0}^{2} \Bigl( i\Delta_{A}(x;x') - i\Delta_{C}(x;x') \Bigr) ] 
&=& 0 \;.
\eea
That is, the local contribution coming from the 3-point cross part is zero.
\bea\label{3-ptcrossloc}
-iM^2_{\rm{3pt cross} \atop {\rm loc}}(x;x') = 0\;.
\eea

%%%%%%%%%%%%%%%%%%%%%%%%%%%%%%%%%%%%%%%%%%%%%%%%%%%%%%%%%%%%%%%%
\subsubsection{Logarithm contributions for 3-point cross part}\label{3ptcrosslog}
%%%%%%%%%%%%%%%%%%%%%%%%%%%%%%%%%%%%%%%%%%%%%%%%%%%%%%%%%%%%%%%%

The logarithm contributions come from the second, fifth and sixth terms in \eqref{M^2_3point_cross_interaction}. Since all the logarithm contributions are finite, we again take the limit $D=4$. 
The second term is
\bea
-\tilde{\kappa}^2 \;C_{1} \; \partial_i[\partial_i \;i\Delta_{\rm{cf}}(x;x') \Bigl(\partial_{0}^{2} i\Delta_{A}(x;x') + \partial_{0}^{'2} i\Delta_{A}(x;x') \Bigr)] 
&\rightarrow  & -\tilde{\kappa}^2 \; 4 \; \partial_i[\partial_i (aa')^{\frac{D}{2}-1} F(y) \Bigl(\partial_{0}^{2} k \ln (aa') + \partial_{0}^{'2} k \ln(aa') \Bigr)] 
\nonumber\\
&=& -\tilde{\kappa}^2 \;4\;  (aa')^{\frac{D}{2}-1} k \partial_i  [ (\partial_i F(y) ) \;[\partial_0 \;Ha \;+\; \partial'_0 \;Ha'] ]\;,
\nonumber\\
&=& -\tilde{\kappa}^2  \frac{H^2}{8\pi^2} 4 H^2 (a^2+a'^2) (aa') \nabla^2 F(y)\;.
\eea
The fifth term is
\bea
\lefteqn{
\tilde{\kappa}^2 \;C_{4}\; \partial_0[\partial_0  i\Delta_{\rm{cf}}(x;x') \partial_{0}^{'2} \Bigl( i\Delta_{A}(x;x') - i\Delta_{C}(x;x') \Bigr) ] 
} 
\nn \\
&& \rightarrow   \tilde{\kappa}^2 \;4\Bigl(\frac{D-1}{D-3}\Bigr) \; \partial_0\left[\partial_0  (aa')^{\frac{D}{2}-1} F(y) \partial_{0}^{'2} \Bigl( i\Delta_{A}(x;x') - i\Delta_{C}(x;x')\Bigr) \right] 
\nonumber\\
& & = \tilde{\kappa}^2 4\Bigl(\frac{D-1}{D-3}\Bigr)  \partial_0\left[ \Bigl[(\frac{D}{2}-1) Ha(aa')^{\frac{D}{2}-1} F(y) 
+\;(aa')^{\frac{D}{2}-1} \partial_{0} F(y) \Bigr] \partial_{0}^{'2} (k\ln(aa') ) \right]\;,
\nonumber\\
& & = \tilde{\kappa}^2 \frac{3H^2}{2\pi^2}\; [2 H^4 (aa')^{3} \;+\; 2(aa')^{2} H^3 a \partial_0 \;+\; (aa') H^2 a^2] F(y)\;.
\eea
Here note that only $i\Delta_{A}(x;x')$ has the logarithm piece, thus we can take only $i\Delta_{A}(x;x')$.
Similarly, the sixth term becomes
\bea
\tilde{\kappa}^2 C_{4} \partial'_0[\partial'_0 i\Delta_{\rm{cf}}(x;x') \partial_{0}^{2} \Bigl( i\Delta_{A}(x;x') - i\Delta_{C}(x;x') \Bigr) ] 
= \tilde{\kappa}^2 \frac{3H^2}{2\pi^2}[2 H^4 (aa')^{3} + 2 (aa')^{2} H^3 a' \partial'_0 +(aa') H^2 a'^2] F(y)\;.
\eea
Summing these three logarithm contributions results in
\bea\label{3-ptcrosslog}
\lefteqn{
-iM^2_{\rm{3pt cross} \atop {\rm log}}(x;x') = \tilde{\kappa}^2 \frac{H^2}{8\pi^2} \biggl\{ 24 H^4(aa')^3 [5F(y) + 2 y F'(y)] + 24 H^4 (aa')^2 (a^2 + a^{'2}) y F'(y)
} 
\nonumber\\
& & \hspace{4.0cm} - 12 H^2 (aa')^2 \nabla^2 I[F](y) - 4 H^2 (aa')(a^2+a^{'2}) \nabla^2 F(y)  \biggr\}\;. 
\qquad \qquad \qquad \qquad 
\eea
Note that this result includes only nonlocal (not having delta function) logarithm pieces. 
TABLE~\ref{cnloclog} summarizes the result and the nonlocal logarithm contributions will be added to TABLE ~\ref{ct4} later.
\begin{table}[htbp]
\begin{center}
\caption{Nonlocal logarithm contributions from the relation \eqref{3-ptcrosslog} with $x \equiv \frac{y}{4}$.}
\begin{ruledtabular}
\begin{tabular}{lcl}
${\phantom{ssssssssss}\rm External\; operators}$&&${\rm Coef.\;of\;} \frac{\tilde{\kappa}^2 H^{4}}{(4\pi)^4}$ \\[1ex]
\hline\\
$\phantom{ssssssssss}(aa')^3 H^4$&&$ -\frac{144}{x} \phantom{ssssssssssssssssssssss}$ \\[2ex]
$\phantom{ssssssssss}(aa')^2 (a^2+a^{'2}) H^4 $&&$ -\frac{192}{x} $ \\[2ex]
$\phantom{ssssssssss}(aa')^2 H^2 \nabla^2$&& $ -24 \;\ln(x)$ \\[2ex]
$\phantom{ssssssssss}(aa') (a^2+a'^2) H^2 \nabla^2$&& $ -\frac{8}{x}$ \\[1ex]
\end{tabular}\label{cnloclog}
\end{ruledtabular}
\end{center}
\end{table}

%%%%%%%%%%%%%%%%%%%%%%%%%%%%%%%%%%%%%%%%%%%%%%%%%%%%%%%%%%%%%%%%
\subsubsection{Normal contributions for 3-point cross part}\label{3ptcrossnormal}
%%%%%%%%%%%%%%%%%%%%%%%%%%%%%%%%%%%%%%%%%%%%%%%%%%%%%%%%%%%%%%%%

Every term in the 3-point cross part in \eqref{3-ptcross} generates normal contributions.
The first term in \eqref{3-ptcross} is
\bea
\lefteqn{2 \tilde{\kappa}^2 \partial'_{\mu} \Bigl[\partial'_{\nu} \; i\Delta_{\rm cf}(x;x') \partial_{\rho}\partial^{\rho} i[^{\mu\nu}\Delta^{\gamma}_{\phantom{\eta}\gamma}](x;x')\Bigr]
}
\nonumber\\
& &=\; \tilde{\kappa}^2 \Biggl\{-\frac{8}{(D-3)} \partial'_i[\partial'_i \;i\Delta_{\rm cf}(x;x') \partial^2 i\Delta_{A}(x;x')]
+\frac{8}{(D-2)} \partial'_0[\partial'_0 \;i\Delta_{\rm cf}(x;x') \partial^2 i\Delta_{C}(x;x')] 
\nonumber\\
& &\hspace{0.5cm} + \frac{8}{(D-3)(D-2)} \partial'_i[\partial'_i \;i\Delta_{\rm cf}(x;x') \partial^2 i\Delta_{C}(x;x')]\Biggr\}\;,
\nonumber\\
& & =\; \tilde{\kappa}^2 \Biggl\{\frac{8}{(D-3)} \partial'_i[\partial'_i \;i\Delta_{\rm cf}(x;x') \partial_0^2 \;i\Delta_{A}(x;x')]
-\frac{8}{(D-3)} \partial'_i[\partial'_i \;i\Delta_{\rm cf}(x;x') \nabla^2 i\Delta_{A}(x;x')] 
\nonumber\\
& &\hspace{0.5cm} - \frac{8}{(D-2)} \partial'_0[\partial'_0 \;i\Delta_{\rm cf}(x;x') \partial_0^2 \;i\Delta_{C}(x;x')] 
+ \frac{8}{(D-2)} \partial'_0[\partial'_0 \;i\Delta_{\rm cf}(x;x') \nabla^2 i\Delta_{C} (x;x')]
\nonumber\\
& &\hspace{0.5cm} -\frac{8}{(D-3)(D-2)} \partial'_i[\partial'_i \;i\Delta_{\rm cf}(x;x') \partial_0^2 \;i\Delta_{C}(x;x')]
+\frac{8}{(D-3)(D-2)} \partial'_i[\partial'_i \;i\Delta_{\rm cf}(x;x') \nabla^2 i\Delta_{C}(x;x')]\Biggr\}\;.
\eea
The second term in \eqref{3-ptcross} is
\bea
\lefteqn{-\tilde{\kappa}^2 \partial'_{\mu}\Bigl[\partial'^{\mu} \;i\Delta_{\rm cf}(x;x') \partial_{\rho}\partial^{\rho} i[^{\alpha}_{\phantom{\eta}\alpha}\Delta^{\gamma}_{\phantom{\eta}\gamma}](x;x')\Bigr]
}
\nonumber\\
& & = \tilde{\kappa}^2 \Biggl\{4(\frac{D-1}{D-3}) \partial'_i[\partial'_i \;i\Delta_{\rm cf}(x;x') \partial^2 i\Delta_{A}(x;x')]
-4(\frac{D-1}{D-3}) \partial'_0[\partial'_0 \;i\Delta_{\rm cf}(x;x') \partial^2 i\Delta_{A}(x;x')]
\nonumber\\
& &\hspace{0.5cm} -(\frac{8}{(D-3)(D-2)}) \partial'_i[\partial'_i \;i\Delta_{\rm cf}(x;x') \partial^2 i\Delta_{C}(x;x')]
+(\frac{8}{(D-3)(D-2)}) \partial'_0[\partial'_0 \;i\Delta_{\rm cf}(x;x') \partial^2 i\Delta_{C}(x;x')]\Biggr\}\;,
\nonumber\\
& & = \tilde{\kappa}^2 \Biggl\{-4(\frac{D-1}{D-3}) \partial'_i[\partial'_i \;i\Delta_{\rm cf}(x;x') \partial_0^2 i\Delta_{A}(x;x')]
+4(\frac{D-1}{D-3}) \partial'_i[\partial'_i \;i\Delta_{\rm cf}(x;x') \nabla^2 i\Delta_{A}(x;x')] 
\nonumber\\
& &\hspace{0.5cm} +4(\frac{D-1}{D-3}) \partial'_0[\partial'_0 \;i\Delta_{\rm cf}(x;x') \partial_0^2 i\Delta_{A}(x;x')]
-4(\frac{D-1}{D-3}) \partial'_0[\partial'_0 \;i\Delta_{\rm cf}(x;x') \nabla^2 i\Delta_{A}(x;x')] 
\nonumber\\
& &\hspace{0.5cm} +(\frac{8}{(D-3)(D-2)}) \partial'_i[\partial'_i \;i\Delta_{\rm cf}(x;x') \partial_0^2 i\Delta_{C}(x;x')]
-(\frac{8}{(D-3)(D-2)}) \partial'_i[\partial'_i \;i\Delta_{\rm cf}(x;x') \nabla^2 i\Delta_{C}(x;x')]
\nonumber\\
& &\hspace{0.5cm} -(\frac{8}{(D-3)(D-2)}) \partial'_0[\partial'_0 \;i\Delta_{\rm cf}(x;x') \partial_0^2 i\Delta_{C}(x;x')]
+(\frac{8}{(D-3)(D-2)}) \partial'_0[\partial'_0 \;i\Delta_{\rm cf}(x;x')) \nabla^2 i\Delta_{C}(x;x')]\Biggr\}\;.
\eea
The third term in \eqref{3-ptcross} is
\bea
\lefteqn{-2\tilde{\kappa}^2 \partial'_{\mu}\Bigl[\partial'_{\nu} \;i\Delta_{\rm cf}(x;x') \partial_{\rho}\partial_{\sigma} i[^{\mu\nu}\Delta^{\rho\sigma}](x;x')\Bigr]}
\nonumber\\
& & = \tilde{\kappa}^2 \Biggl\{- 4 \partial'_i[\partial'_j \;i\Delta_{\rm cf}(x;x') \partial_i \partial_j i\Delta_{A}(x;x')] 
+\frac{4}{(D-3)} \partial'_i[\partial'_i \;i\Delta_{\rm cf}(x;x') \nabla^2 i\Delta_{A}(x;x')] 
\nonumber\\
& &\hspace{0.5cm} +4 \partial'_i[\partial'_0 \;i\Delta_{\rm cf}(x;x') \partial_0 \partial_i\; i\Delta_{B}(x;x')] 
+4 \partial'_0[\partial'_i \;i\Delta_{\rm cf}(x;x') \partial_0 \partial_i\; i\Delta_{B}(x;x')]
\nonumber\\
& &\hspace{0.5cm} -\frac{4}{(D-2)} \partial'_i[\partial'_i \;i\Delta_{\rm cf}(x;x') \partial_0^2 i\Delta_{C}(x;x')] 
- \frac{4}{(D-3)(D-2)} \partial'_i[\partial'_i \;i\Delta_{\rm cf}(x;x') \nabla^2 i\Delta_{C}(x;x')]
\nonumber\\
& &\hspace{0.5cm} -4(\frac{D-3}{D-2}) \partial'_0[\partial'_0 \;i\Delta_{\rm cf}(x;x') \partial_0^2 i\Delta_{C}(x;x')] 
-\frac{4}{(D-2)} \partial'_0[\partial'_0 \;i\Delta_{\rm cf}(x;x') \nabla^2 i\Delta_{C}(x;x')]\Biggr\}\;.
\eea
The fourth term in \eqref{3-ptcross} is
\bea
\lefteqn{\tilde{\kappa}^2 \partial'_{\mu}\Bigl[\partial'^{\mu} \;i\Delta_{\rm cf}(x;x') \partial_{\rho}\partial_{\sigma} i[^{\alpha}_{\phantom{\eta}\alpha}\Delta^{\rho\sigma}](x;x')\Bigr]}
\nonumber\\
& & = \tilde{\kappa}^2 \Biggl\{-\frac{4}{(D-3)}\partial'_i[\partial'_i \;i\Delta_{\rm cf}(x;x') \nabla^2 i\Delta_{A}(x;x')] 
+ \frac{4}{(D-3)} \partial'_0[\partial'_0 \;i\Delta_{\rm cf}(x;x') \nabla^2 i\Delta_{A}(x;x')]
\nonumber\\
& &\hspace{0.5cm} +\frac{4}{(D-3)(D-2)} \partial'_i[\partial'_i \;i\Delta_{\rm cf}(x;x') \nabla^2 i\Delta_{C}(x;x')] 
+ \frac{4}{(D-2)} \partial'_i[\partial'_i \;i\Delta_{\rm cf}(x;x') \partial_0^2 i\Delta_{C}(x;x')]
\nonumber\\
& &\hspace{0.5cm} -\frac{4}{(D-3)(D-2)} \partial'_0[\partial'_0 \;i\Delta_{\rm cf}(x;x') \nabla^2 i\Delta_{C}(x;x')] 
- \frac{4}{(D-2)} \partial'_0[\partial'_0 \;i\Delta_{\rm cf}(x;x') \partial_0^2 i\Delta_{C}(x;x')]\Biggr\}\;.
\eea
The fifth term in \eqref{3-ptcross} is,
\bea
\lefteqn{2 \tilde{\kappa}^2 \partial_{\mu} \Bigl[\partial_{\nu} \;i\Delta_{\rm cf}(x;x') \partial'_{\rho}\partial'^{\rho} i[^{\mu\nu}\Delta^{\gamma}_{\phantom{\eta}\gamma}](x;x')\Bigr]}
\nonumber\\
& & = \tilde{\kappa}^2 \Biggl\{\frac{8}{(D-3)} \partial_i[\partial_i \;i\Delta_{\rm cf}(x;x') \partial_{0}^{'2} i\Delta_{A}(x;x')] 
- \frac{8}{(D-3)} \partial_i[\partial_i \;i\Delta_{\rm cf}(x;x') \nabla'^2 i\Delta_{A}(x;x')]  
\nonumber\\
& &\hspace{0.5cm} - \frac{8}{(D-2)} \partial_0[\partial_0 \;i\Delta_{\rm cf}(x;x') \partial_{0}^{'2} i\Delta_{C}(x;x')]
+ \frac{8}{(D-2)} \partial_0[\partial_0 \;i\Delta_{\rm cf}(x;x') \nabla'^2 i\Delta_{C}(x;x')]
\nonumber\\
& &\hspace{0.5cm} -\frac{8}{(D-3)(D-2)} \partial_i[\partial_i \;i\Delta_{\rm cf}(x;x') \partial_{0}^{'2} i\Delta_{C}(x;x')]
+\frac{8}{(D-3)(D-2)} \partial_i[\partial_i \;i\Delta_{\rm cf}(x;x') \nabla'^2 i\Delta_{C}(x;x')]\Biggr\}\;.
\eea
The sixth term in \eqref{3-ptcross} is
\bea
\lefteqn{-\tilde{\kappa}^2 \partial_{\mu}\Bigl[\partial^{\mu} \;i\Delta_{\rm cf}(x;x') \partial'_{\rho}\partial'^{\rho} i[^{\alpha}_{\phantom{\eta}\alpha}\Delta^{\gamma}_{\phantom{\eta}\gamma}](x;x')\Bigr]}
\nonumber\\
& & = \tilde{\kappa}^2 \Biggl\{-4(\frac{D-1}{D-3}) \partial_i[\partial_i \;i\Delta_{\rm cf}(x;x')) \partial_{0}^{'2} i\Delta_{A}(x;x')]
+4(\frac{D-1}{D-3}) \partial_i[\partial_i \;i\Delta_{\rm cf}(x;x') \nabla'^2 i\Delta_{A}(x;x')] 
\nonumber\\
& &\hspace{0.5cm} +4(\frac{D-1}{D-3}) \partial_0[\partial_0 \;i\Delta_{\rm cf}(x;x') \partial_{0}^{'2} i\Delta_{A}(x;x')]
-4(\frac{D-1}{D-3}) \partial_0[\partial_0 \;i\Delta_{\rm cf}(x;x') \nabla'^2 i\Delta_{A}(x;x')] 
\nonumber\\
& &\hspace{0.5cm} +(\frac{8}{(D-3)(D-2)}) \partial_i[\partial_i \;i\Delta_{\rm cf}(x;x') \partial_{0}^{'2} i\Delta_{C}(x;x')]
-(\frac{8}{(D-3)(D-2)}) \partial_i[\partial_i \;i\Delta_{\rm cf}(x;x') \nabla'^2 i\Delta_{C}(x;x')]
\nonumber\\
& &\hspace{0.5cm} -(\frac{8}{(D-3)(D-2)}) \partial_0[\partial_0 \;i\Delta_{\rm cf}(x;x') \partial_{0}^{'2} i\Delta_{C}(x;x')]
+(\frac{8}{(D-3)(D-2)}) \partial_0[\partial_0 \;i\Delta_{\rm cf}(x;x') \nabla'^2 i\Delta_{C}(x;x')]\Biggr\}\;.
\eea
The seventh term in \eqref{3-ptcross} is
\bea
\lefteqn{-2\tilde{\kappa}^2 \partial_{\mu}\Bigl[\partial_{\nu} \;i\Delta_{\rm cf}(x;x') \partial'_{\rho}\partial'_{\sigma} i[^{\mu\nu}\Delta^{\rho\sigma}](x;x')\Bigr]}
\nonumber\\
& & = \tilde{\kappa}^2 \Biggl\{- 4 \partial_i[\partial_j \;i\Delta_{\rm cf}(x;x') \partial'_i \partial'_j i\Delta_{A}(x;x')] 
+\frac{4}{(D-3)} \partial_i[\partial_i \;i\Delta_{\rm cf}(x;x') \nabla'^2 i\Delta_{A}(x;x')]
\nonumber\\ 
& &\hspace{0.5cm} +4 \partial_i[\partial_0 \;i\Delta_{\rm cf}(x;x') \partial'_0 \partial'_i \;i\Delta_{B}(x;x')] 
+ 4 \partial_0[\partial_i \;i\Delta_{\rm cf}(x;x') \partial'_0 \partial'_i \;i\Delta_{B}(x;x')]
\nonumber\\
& &\hspace{0.5cm} -\frac{4}{(D-2)} \partial_i[\partial_i \;i\Delta_{\rm cf}(x;x') \partial_{0}^{'2} i\Delta_{C}(x;x')] 
- \frac{4}{(D-3)(D-2)} \partial_i[\partial_i \;i\Delta_{\rm cf}(x;x') \nabla'^2 i\Delta_{C}(x;x')]
\nonumber\\
& &\hspace{0.5cm} -4(\frac{D-3}{D-2}) \partial_0[\partial_0 \;i\Delta_{\rm cf}(x;x') \partial_{0}^{'2} i\Delta_{C}(x;x')] 
-\frac{4}{(D-2)} \partial_0[\partial_0 \;i\Delta_{\rm cf}(x;x') \nabla'^2 i\Delta_{C}(x;x')]\Biggr\}\;.
\eea
Finally, the eighth term in \eqref{3-ptcross} is
\bea
\lefteqn{\tilde{\kappa}^2 \partial_{\mu}\Bigl[\partial^{\mu} \;i\Delta_{\rm cf}(x;x') \partial'_{\rho}\partial'_{\sigma} i[^{\alpha}_{\phantom{\eta}\alpha}\Delta^{\rho\sigma}](x;x')\Bigr]}
\nonumber\\
& & = \tilde{\kappa}^2 \Biggl\{-\frac{4}{(D-3)}\partial_i[\partial_i \;i\Delta_{\rm cf}(x;x') \nabla'^2 i\Delta_{A}(x;x')] 
+ \frac{4}{(D-3)} \partial_0[\partial_0 \;i\Delta_{\rm cf}(x;x') \nabla'^2 i\Delta_{A}(x;x')]
\nonumber\\
& &\hspace{0.5cm} +\frac{4}{(D-3)(D-2)} \partial_i[\partial_i \;i\Delta_{\rm cf}(x;x') \nabla'^2 i\Delta_{C}(x;x')] 
+ \frac{4}{(D-2)} \partial_i[\partial_i \;i\Delta_{\rm cf}(x;x') \partial_{0}^{'2} i\Delta_{C}(x;x')]
\nonumber\\
& &\hspace{0.5cm} -\frac{4}{(D-3)(D-2)} \partial_0[\partial_0 \;i\Delta_{\rm cf}(x;x') \nabla'^2 i\Delta_{C}(x;x')] 
- \frac{4}{(D-2)} \partial_0[\partial_0 \;i\Delta_{\rm cf}(x;x') \partial_{0}^{'2} i\Delta_{C}(x;x')]\Biggr\}\;.
\eea

We aim at putting these normal contributions in the form of unprimed derivatives acting on the functions of $y$. In our computation three derivatives, spatial and temporal, acting on functions $A(y), B(y), C(y)$ arise, which requires an additional identity as follows, 
\bea\label{threederivi}
F(y)\partial_{\alpha} \partial'_{\rho} \partial'_{\sigma} A(y) &=& \partial_{\alpha} \partial'_{\rho} \partial'_{\sigma} \bigg\{I^3[FA'''](y) \bigg\} 
+ \frac{\partial^3 y}{\partial x^{\alpha} \partial x^{'\rho} \partial x^{'\sigma}}
\bigg\{I[F'A'](y) + I^2[F'A''](y)\bigg\}
\nonumber\\
&& +\bigg(\!
\frac{\partial y}{\partial x^{\alpha}} \frac{\partial^2 y}{\partial x^{'\rho} \partial x^{'\sigma}}
+2 \frac{\partial y}{\partial x^{'(\rho}} \frac{\partial^2 y}{\partial x^{'\sigma)} \partial x^{\alpha}}\!\bigg)\bigg\{\!I[F'A''](y)\bigg\}\;.
\eea
Applying this identity to the first term of \eqref{3-ptcrosscoeff}, which is the simplest because all of the derivatives are spatial, i.e., in the form of $F(y)\partial_i \nabla'^2 A(y)$,
\bea\label{inst}
2 C_{1}\;\tilde{\kappa}^2 \partial_i [\partial_i \;i\Delta_{\rm{cf}}(x;x') \nabla^{2} i\Delta_{A}(x;x')] &=& 2 C_{1}\;\tilde{\kappa}^2 \partial_i [\partial_i (aa')^{\frac{D}{2}-1} F(y) \nabla^{2} A(y)]\;,
\nonumber\\
&=& 8\; \tilde{\kappa}^2 (aa')^{\frac{D}{2}-1}\bigg( \partial_i \partial_i[F(y) \nabla^2 A(y)] - \partial_i [F(y) \partial_i \nabla^2 A(y)] \!\bigg)\;.
\eea
Next, applying the identity \eqref{threederivi} to the second term of the last expression \eqref{inst} gives
\bea
F(y) \partial_i \nabla'^2 A(y) 
&=& \partial_{i} \nabla'^2 \bigg\{I^3[F A'''](y) \bigg\} + \frac{\partial^{3} y}{\partial x^{i} \partial x^{'j} \partial x^{'j}} \bigg\{I[F'A'](y)+ I^2[F'A''](y)\bigg\}
\nonumber\\
& & + \bigg(\!
\frac{\partial y}{\partial x^{i}} \frac{\partial^{2} y}{\partial x^{'j} \partial x^{'j}}
+2 \frac{\partial y}{\partial x^{'(j}} \frac{\partial^{2} y}{\partial x^{'j)} \partial x^{i}}\!\bigg)\bigg\{\! I[F'A''](y)\bigg\}\;.\label{deriv1}
\eea   
Further using the following derivatives of $y$ with respect to $x^i$ or $x'^{j}$,
\bea
\frac{\partial y}{\partial x^i} = 2 H^2 a a' \Delta x_i \; ,\; \frac{\partial y}{\partial x'^{j}} = - 2 H^2 a a' \Delta x_j \; , \; \frac{\partial^2 y}{\partial x^{i} \partial x^{j}} = 2 H^2 a a' \eta_{i j} \; , \; \frac{\partial^2 y}{\partial x^i \partial x'^{j}} = - 2 H^2 a a' \eta_{i j} \; \nonumber \\ \frac{\partial^2 y}{\partial x^i \partial x^{i}} = \frac{\partial^2 y}{\partial x'^j \partial x'^{j}} = 2 H^2 a a' (D-1) \; , \; \frac{\partial^3 y}{\partial x^i \partial x^i \partial x'^j} = \frac{\partial^3 y}{\partial x^i \partial x'^j \partial x'^j} = \frac{\partial^4 y}{\partial x^{i} \partial x^{i} \partial x^{'j} \partial x^{'j}} = 0\;, 
\eea
leads Eq. \eqref{deriv1} to
\bea
F(y) \partial_i \nabla'^2 A(y) = \partial_{i} \nabla'^2 \bigg\{I^3[F A'''](y) \bigg\} + \bigg(\! 4 H^4 (a a')^2 \Delta x_i(D-1) + 8 H^4 (a a')^2 \Delta x_j \eta_{ij}  \!\bigg)\bigg\{\! I[F'A''](y)\bigg\}\;.\label{deriv1fin}
\eea
Next, plugging the identity \eqref{ID017} and the expression \eqref{deriv1fin} into Eq. \eqref{inst} gives
\bea
8\;\tilde{\kappa}^2 \partial_i[\partial_i (aa')^{\frac{D}{2}-1} F(y) \nabla^{2} A(y)] = 8\; \tilde{\kappa}^2 (aa')^{\frac{D}{2}-1}\bigg( \partial_i \partial_i [\nabla^2 I^2[F A''](y) + 2(D-1) H^2 a a' I[F'A'](y)]
\nonumber\\
- \partial_i \bigg( \partial_{i} \nabla'^2 \bigg\{I^3[F A'''](y) \bigg\} + \bigg(\! 4 H^4 (a a')^2 \Delta x_i(D-1) + 8 H^4 (a a')^2 \Delta x_j \eta_{ij}  \!\bigg)\bigg\{\! I[F'A''](y)\bigg\}\bigg)\;\bigg)\;,\nonumber\\
= 8\; \tilde{\kappa}^2 (aa')^{\frac{D}{2}-1}\bigg( 
\nabla^4 I^2[F A''](y) + 2 (D-1) H^2 a a' \nabla^2 I[F'A'](y)] - \nabla^4 I^3[F A'''](y) 
\nonumber\\
- 4 (D-1)(D+1) H^4 (a a')^2 I[F'A''](y) -8 (D+1) H^6 (a a')^3 \Vert \Delta \vec{x} \Vert^2 [F' A''](y) \bigg)\;.
\eea
We use the following identity
\bea
a a' H^2 \Vert \Delta \vec{x} \Vert^2 f(y) = -\frac12 (D-1) I[f](y) - \frac{\nabla \cdot \nabla'}{4 a a' H^2}  I^2[f](y)\;.
\eea
to obtain 
\bea
\lefteqn{8\;\tilde{\kappa}^2 \partial_i[\partial_i (aa')^{\frac{D}{2}-1} F(y) \nabla^{2} A(y)] 
= 8 \tilde{\kappa}^2 (aa')^{\frac{D}{2}-1}\bigg( \nabla^4 I^2[F A''](y) + 2 (D-1) H^2 a a' \nabla^2 I[F'A'](y)] 
}
\nonumber\\
& & \hspace{1cm} -  \nabla^4 I^3[F A'''](y) -2 (D+1) H^2 a a' \nabla^2 I^2[F' A''](y) \bigg)\;,
\\
& & = -\tilde{\kappa}^2 \biggl\{ \epsilon_1 16\Bigl( (D+1) I^2[F' A''](y)-(D-1) I[F'A'](y)\bigg)
+ \zeta 8\bigg(I^3[F A'''](y)-I^2[F A''](y)\Bigr)\biggl\}\;.
\eea

The results including this pure spatial derivative, and also the pure temporal and temporal spatial mixed derivatives in equation \eqref{3-ptcrosscoeff} are tabulated in TABLEs ~\ref{ct1}, ~\ref{ct2}, ~\ref{ct3}, ~\ref{ct4}, in the form of the eleven operators (\ref{newa}-\ref{newz}) acting on functions of $y$.
Finally, by adding all the 3-point cross contributions coming from $A(y)$, $B(y)$ and $C(y)$ in \eqref{3-ptcrosscoeff}, we have
\bea
\lefteqn{-iM^2_{\rm 3ptcross}(x;x')
= -\tilde{\kappa}^2  \Bigl\{\alpha f_{\alpha}(y) 
+ \beta_1 f_{\beta_1}(y) + \beta_2 f_{\beta_2}(y)
+ \gamma_1 f_{\gamma_1}(y) + \gamma_2 f_{\gamma_2}(y) + \gamma_3 f_{\gamma_3}(y) 
} \nonumber\\
& & \hspace{4.5cm}+ \delta f_{\delta}(y) 
+ \epsilon_1 f_{\epsilon_1}(y) + \epsilon_2 f_{\epsilon_2}(y) + \epsilon_3 f_{\epsilon_3}(y) 
+ \zeta f_{\zeta}(y) \Bigl \}\;. 
\label{3pt-alphazeta}
\eea
Here we give the functions $f_i$ (on which the eleven external operators are acting) in TABLEs \ref{cta} through \ref{ctz} in Appendix \ref{app2}.

%%%%%%%%%%%%%%%%%%%%%%%%%%%%%%%%%%%%%%%%%%%%%%%%%%%%%%%%%%%
\section{Renormalization}\label{renormalization}
%%%%%%%%%%%%%%%%%%%%%%%%%%%%%%%%%%%%%%%%%%%%%%%%%%%%%%%%%%%

In this section we absorb the divergences occurred in the primitive diagrams (the first two terms in Eq. \eqref{3diagrams}) by counterterms (the third term in Eq. \eqref{3diagrams}) and obtain a finite result for the scalar self-mass-squared.   
First we recall the appropriate counterterms for our Lagrangian \eqref{Lagrangian}, which we constructed by applying the theorem of Bogoliubov, Parasiuk, Hepp and Zimmermann (BPHZ) \cite{bogo} in our previous work \cite{kahya3}. 
Since these BPHZ counterterm vertices are local (in the form of $\delta$-function), the next task is to segregate all the divergent terms into a $\delta$-function. 
The localized divergences can then be subtracted by the local counterterms. We perform this procedure of renormalization in the following subsections.

%%%%%%%%%%%%%%%%%%%%%%%%%%%%%%%%%%%%%%%%%%%%%%%%%%%%%%%%%%%%%%%%
\subsection{Allowed counterterms and the associated vertices} \label{allow}
%%%%%%%%%%%%%%%%%%%%%%%%%%%%%%%%%%%%%%%%%%%%%%%%%%%%%%%%%%%%%%%%

First of all, the structure of our Lagrangian \eqref{Lagrangian} guides us to identify three de Sitter invariant counterterms at one loop order as
\bea
\Delta \mathcal{L}_1 = \frac12 c_1 \kappa^2 \square \phi \square \phi a^2 \;,\quad
\Delta \mathcal{L}_2 = -\frac12 c_2 \kappa^2 H^2 \partial_{\mu} \phi \partial^{\mu} \phi\;, \quad
\Delta \mathcal{L}_3 = \frac12 c_3 \kappa^2 H^4 \phi^2 a^2\;. 
\label{invar}
\eea
See Ref. \cite{kahya3} for the detailed derivation of these counterterms.
The associated vertices for the counterterms 
are obtained by taking the same variation performed for the primitive terms: 
\bea
\frac{i \delta \Delta S_1}{\delta \phi(x) \delta \phi(x')}\Biggl\vert_{\phi = 0} &=&
i c_1 \kappa^2 \Bigl[a^2 \square^2 + (D-2)H^2a^2\square + 2(D-2)Ha\partial_0\square \Bigr]\delta^D (x-x')\;,\label{alp1}
\\
\frac{i \delta \Delta S_2}{\delta \phi(x) \delta \phi(x')}\Biggl\vert_{\phi = 0} &=&
i c_2 \kappa^2 H^2 a^2 \square \delta^D (x-x')\;, \label{alp2} 
\\
\frac{i \delta \Delta S_3}{\delta \phi(x) \delta \phi(x')}\Biggl\vert_{\phi = 0} &=&
i c_3 \kappa^2 H^4 a^2 \delta^D (x-x')\;. \label{alp3} 
\eea
In the first vertex \eqref{alp1} we correct the mistake in Eq. (94) of Ref. \cite{kahya3} (also  in Eq. (130) of Ref. \cite{kahya}),
where the last two terms are missing. This mistake did not affect our previous results
because the divergences in the forms of $a^2 \square^2\delta^D (x-x')$ and $Ha\partial_0\square \delta^D (x-x')$ did not occur in the cases of Refs. \cite{kahya, kahya3}. 
These two forms of divergences only arise in the 3-point kinetic-conformal interactions, which is the main computation of the present paper. 
Moreover, as will be shown in the next subsection, these two divergences exactly match the relative coefficients of the first and third terms of Eq.  \eqref{alp1} and thus get cancelled simultaneously,
which serves as a nontrivial check for the correctness of our computation.

On the other hand, our gauge fixing Lagrangian \eqref{gf} breaks the spatial special conformal symmetry among 
the full de Sitter symmetries and leads to one noninvariant counterterm \cite{kahya3}
\bea\Delta \mathcal{L}_4 = -\frac12 \kappa^2 H^2 
\nabla \phi \cdot \nabla \phi  \;.\quad
\label{one-noninv}
\eea
It should be noted that this noninvariant counterterm still respects the residual de Sitter symmetries (homogeneity, isotropy and dilation symmetry) and it also becomes Poincar\'e invariant in flat space \cite{kahya3}. These conditions indeed highly restrict the form of counterterms and leave only one possibility given in Eq. \eqref{one-noninv}.   
The associated vertex to this noninvariant counterterm is
\bea
\frac{i \delta \Delta S_4}{\delta \phi(x) \delta \phi(x')}
\Biggl\vert_{\phi = 0}
= i c_4 \kappa^2 H^2
\nabla^2 \delta^D(x\!-\!x') \; . 
\label{alp4}
\eea

These four counterterms are the only possible ones for the CC scalar self-mass-squared at one loop order allowed by the symmetries of the bare Lagrangian for a scalar conformally coupled to gravity \eqref{Lagrangian} and the gauge-fixing term \eqref{gf}. In fact, all the terms in the self-mass-squared must respect the same symmetries, which implies that any divergent term can be cancelled by one of these four counterterms. In other words, any divergence that cannot be cancelled by these counterterms means an error in our computation. Indeed this has served us as a crucial error-detecting method in this heavy calculation.

%%%%%%%%%%%%%%%%%%%%%%%%%%%%%%%%%%%%%%%%%%%%%%%%%%%%%%%%%%%%%%%%
\subsection{Localization of divergences}
%%%%%%%%%%%%%%%%%%%%%%%%%%%%%%%%%%%%%%%%%%%%%%%%%%%%%%%%%%%%%%%%

We now localize the divergent terms so as to put them in the form of the counterterm vertices. Keeping in mind that the terms in the self-mass-squared will eventually be integrated in the effective field equation \eqref{linear_eq}, the first step is to make the powers of $y$,
\bea
\Bigl(\frac{4}{y}\Bigr)^{D}\;, \; 
\Bigl(\frac{4}{y}\Bigr)^{D-1} \quad \mbox{and} \quad \Bigl(\frac{4}{y}\Bigr)^{D-2}\;,
\eea
integrable in $D=4$ dimensions.
This can be done by extracting d' Alembertian operators using the following identity,
\bea
\square f(y) = H^2 \Bigl[(4y - y^2) f''(y) + D (2 - y)f'(y)\Bigr]
+{\rm Res} \Bigl[y^{\frac{D}2 -2} f\Bigr] 
\frac{4 \pi^{D/2} H^{2-D}}{\Gamma(\frac{D}2 - 1)} \frac{i}{\sqrt{-g}} \delta^D (x-x').
\label{key}
\eea
Here ${\rm Res}[F]$ means the residue of $F(y)$. 
By applying this identity (\ref{key}) and rearranging the terms, we can write these powers of $y$ as 
\bea
\Bigl(\frac{4}{y}\Bigr)^{D}\phantom{\mu\nu}
&=&\frac2{(D-1)D} \frac{\square}{H^2} \Bigl(\frac{4}{y}\Bigr)^{D-1}\;,
\label{0th}\\
\Bigl(\frac{4}{y}\Bigr)^{D-1}
&=&\frac2{(D-2)^2} \frac{\square}{H^2} \Bigl(\frac{4}{y}\Bigr)^{D-2}-\frac2{D-2} \Bigl(\frac{4}{y}\Bigr)^{D-2}\;,
\label{1st}\\
\Bigl(\frac{4}{y}\Bigr)^{D-2}&=&\frac2{(D-4)(D-3)} \frac{\square}{ H^2} \Bigl(\frac{4}{y}\Bigr)^{D-3}-\frac4{D-4} \Bigl(\frac{4}{y}\Bigr)^{D-3}\;.
\label{2nd} 
\eea
Note that extracting d'Alembertians subsequently transfers a quartically divergent power $1/y^D$ into a quadratically divergent $1/y^{D-1}$,  a logarithmically divergent $1/y^{D-2}$ and finally into an integrable power of $1/y^{D-3}$. A new problem is that its coefficient has a divergent factor $1/(D-4)$.  
To take care of this divergent factor, we apply the identity \eqref{key} to the power of  $1/y^{D/2-1}$ which leads to an expression of zero:

\bea
0=\frac{\square}{H^2} \Bigl(\frac{4}{y}\Bigr)^{\frac{D}2 -1} 
- \frac{D}2 \Bigl(\frac{D}2 -1 \Bigr) \Bigl(\frac{4}{y}\Bigr)^{\frac{D}2 -1} 
-\frac{(4\pi)^{D/2} H^{-D}}{\Gamma(\frac{D}2 -1)} \frac{i}{a^D} \delta^D (x-x') \;. 
\label{delt}
\eea
We add this particular form of zero to (\ref{2nd}) and segregate the divergent factor into the $\delta$-function,
\bea
\Bigl(\frac{4}{y}\Bigr)^{D-2}
&\!\!\!=\!\!\!&\frac{2}{(D-4)(D-3)}\Biggl\{ 
\frac{(4\pi)^{\frac{D}2} H^{-D}}{\Gamma(\frac{D}2 - 1)}\frac{i \delta^D( x-x')}{a^D} 
+\frac{\square}{H^2} \Biggl[\Bigl(\frac{4}{y} \Bigr)^{D-3} 
- \Bigl(\frac{4}{y}\Bigr)^{\frac{D}2 - 1} \Biggr] \Biggr\} 
\nonumber \\
& & \hspace{3cm} 
-\frac{4}{D-4} \Biggl\{\Bigl(\frac{4}{y}\Bigr)^{D-3} 
- \frac{D (D-2)}{8 (D-3)} \Bigl(\frac{4}{y}\Bigr)^{\frac{D}2 -1} \Biggr\}\;,
\\
&\!\!\!=\!\!\!&\frac{i H^{-D} (4\pi)^{\frac{D}2}}{(D-4)(D-3) \Gamma (\frac{D}2)} (D-2) \frac{\delta^D (x-x')}{a^D} 
-\frac{\square}{H^2} \Biggl\{\frac{4}{y} \ln\Bigl(\frac{y}4\Bigr) \Biggr\} + 2 \Bigl(\frac{4}{y}\Bigr) \ln\Bigl(\frac{y}{4}\Bigr) - \Bigl(\frac{4}{y} \Bigr) + \mathcal{O}(D-4)
\;,\label{D-2} 
\eea
For the rest of computations we take the powers of $1/y$ instead of $4/y$:
\bea
\lefteqn{
\Bigl(\frac{1}{y}\Bigr)^{D-2}
=\frac{i H^{-D} \pi^{\frac{D}{2}}}{(D-4)(D-3) \Gamma (\frac{D}2) 4^{\frac{D}{2}}} 16(D-2) \frac{\delta^D (x-x')}{a^D}} \nonumber \\ 
& & \hspace{2cm} -\frac{16}{4^D}\frac{\square}{H^2} \Biggl\{\frac{4}{y} \ln\Bigl(\frac{y}4\Bigr) \Biggr\}  
+\frac{32}{4^D} \Bigl(\frac{4}{y}\Bigr) \ln\Bigl(\frac{y}{4}\Bigr)
-\frac{16}{4^D}\Bigl(\frac{4}{y} \Bigr) + \mathcal{O}(D-4)\;.
\label{yup(-2+D)}
\eea
Subsequently replacing $1/y^{D-2}$ in \eqref{1st} by \eqref{yup(-2+D)} makes the quadratically divergent power $1/y^{D-1}$ become  
\bea
\lefteqn{
\Bigl(\frac{1}{y}\Bigr)^{D-1}
= \frac{i H^{-D} \pi^{\frac{D}{2}} }{(D-4)(D-3) \Gamma (\frac{D}{2}) 4^{\frac{D}{2}} } 
\Biggl\{ \frac{8}{D-2} \frac{\square}{H^2} - 8 \Biggr\}\frac{\delta^D (x-x')}{a^D} 
-\frac{2}{4^D} \frac{\square^2}{H^4} \Biggl\{\frac{4}{y} \ln \Bigl(\frac{y}{4}\Bigr) \Biggr\}
}
\nonumber\\
& & \hspace{1.5cm}
+\frac{\square}{H^2} \Biggl\{ \frac{8}{4^D}\Bigl(\frac{4}{y}\Bigr) \ln\Bigl(\frac{y}{4}\Bigr) 
-\frac{2}{4^D} \Bigl(\frac{4}{y} \Bigr)\Biggr\} 
-\frac{8}{4^D} \Bigl(\frac{4}{y} \Bigr)\ln\Bigl(\frac{y}{4}\Bigr) 
+ \frac{4}{4^D} \Bigl(\frac{4}{y} \Bigr) + \mathcal{O}(D-4)\;.
\label{yup(-1+D)}
\eea
and replacing $1/y^{D-1}$ in \eqref{0th} by \eqref{yup(-1+D)} makes the quartically divergent power $1/y^{D}$ become
\bea
\lefteqn{
\Bigl(\frac{1}{y}\Bigr)^{D}=\frac{i H^{-D} \pi^{\frac{D}2}}{(D-4) (D-3) \Gamma (\frac{D}2) 4^{\frac{D}2}}
\Biggl\{ \frac4{(D-2)(D-1)D} \frac{\square^2}{H^4} -\frac{4}{(D-1) D} \frac{\square}{H^2}\Biggr\} 
\frac{\delta^D (x-x')}{a^D}
}
\nonumber \\ 
& & \hspace{1cm} -\frac{1}{4^D}\Biggl\{ 
\frac{1}{12} \frac{\square^3}{H^6} \biggl[\frac{4}{y} \ln \Bigl(\frac{y}4 \Bigr) \bigg] 
+\frac{\square^{2}}{H^4} \biggl[\frac13 \Bigl( \frac{4}{y}\Bigr) \ln \Bigl(\frac{y}{4} \Bigr) 
-\frac1{12} \Bigl(\frac{4}{y} \Bigr)\biggr]  
-\frac1{3} \Bigl( \frac{4}{y}\Bigr) \ln\Bigl( \frac{y}{4}\Bigr) + \frac1{6} \Bigl( \frac{4}{y}\Bigr)
\Biggr\} + \mathcal{O}(D-4)\;.
\label{yup(-D)}
\eea
Here we note that the d'Alembertian operator $\square$ acts on both $a^{D}$ and  $\delta^D(x-x')$ while 
it directly acts on $\delta^D(x-x')$ in the counterterm vertices. The following identities allow us to move the factors of $a$ to the left of derivative operators:
\bea
\square a^n \delta^D (x-x')
&=&\Bigl( a^n \square - [n^2 + n(D-1)] H^2 a^n  -2 n H a^{n-1} \partial_{0} \Bigr) \delta^D (x-x')\;,
\label{ident1}
\\ 
\square ^2 a^n \delta^D (x-x')
&=&\Bigl(a^n \square ^2 - 2 n [3 n+(D-1)] H^2 a^n \square -4 n H a^{n-1} \partial_{0} \square  + 4 n^3 H^3 a^{n-1} \partial_{0} 
\nonumber \\
& & \hspace{0.5cm} +4 n(n-1) H^2 a^{n-2} \nabla^2 
+n^2 [n+(D-1)]^2 H^4 a^n \Bigr) \delta^D (x-x')\;,
\label{ident2} 
\\
\square  a^n \partial_{0}  \delta^D (x-x')
&=& a^n \Bigl(\square - n[n + (D-1)] H^2  -2 n H a^{-1} \partial_{0} \Bigr) \partial_{0} \delta^D (x-x')\;.
\label{ident3}
\eea

%%%%%%%%%%%%%%%%%%%%%%%%%%%%%%%%%%%%%%%%%%%%%%%%%%%%%%%%%%%%%%%%
\subsection{Regularization} 
%%%%%%%%%%%%%%%%%%%%%%%%%%%%%%%%%%%%%%%%%%%%%%%%%%%%%%%%%%%%%%%%

%%%%%%%%%%%%%%%%%%%%%%%%%%%%%%%%%%%%%%%%%%%%%%%%%%%%%%%%%%%%%%%%
\subsubsection{For 4-point part and 3-point kinetic part}\label{regk}
%%%%%%%%%%%%%%%%%%%%%%%%%%%%%%%%%%%%%%%%%%%%%%%%%%%%%%%%%%%%%%%%

To illustrate the process of putting the divergences exactly in the forms of the counterterms, we take the term with the external operator $\alpha$ in Eq. \eqref{newa} coming from the 3-point kinetic interactions as an example:
\bea\label{alphay^(2-D)}
\lefteqn{
\alpha \Biggl\{-\frac{D H^{2D-4}\Gamma(\frac{D}2 -1)^2}{64(D-1)\pi^D} \Bigl( \frac1{y}\Bigr)^{D-2} \Biggr\} 
= -\frac{D H^{2D-4}\Gamma(\frac{D}2 -1)^2}{64(D-1)\pi^D} (aa')^{\frac{D}{2}+1} \square ^2 \Bigl(\frac1{y}\Bigr)^{D-2}\;,
}
\nn\\
& & \hspace{0cm} = -\frac{D H^{2D-4}\Gamma(\frac{D}2 -1)^2}{64(D-1)\pi^D}\; \frac{i H^{-D} \pi^\frac{D}2}{(D-4)(D-3) \Gamma(\frac{D}2)4^{\frac{D}2}} a^{\frac{D}2+1}  16(D-2) \square ^2 a^{\frac{D}2+1} \frac{\delta^D (x-x')}{a^D}\;\nonumber\\
& & \hspace{0.5cm} -\frac{D H^{2D-4}\Gamma(\frac{D}2 -1)^2}{64(D-1)\pi^D} (aa')^{\frac{D}{2}+1} \square ^2 
( -\frac{16}{4^{D}} \frac{\square}{H^2} \Biggr\{ \frac{4}{y} \ln (\frac{y}{4}) \Biggl\} + \frac{32}{4^{D}} (\frac{4}{y}) \ln (\frac{y}{4}) 
-\frac{16}{4^{D}} (\frac{4}{y}) + \mathcal{O}(D-4) )\;,
\nonumber\\
& & \hspace{0cm} = \frac{i H^{D-4}}{(4\pi)^\frac{D}2} \frac{\Gamma(\frac{D}2)}{(D-4)(D-3)} 
\Biggl\{-\frac{D}{D^2-3D+2} H^2 a^2 \square^2 + \frac{D(D-4)}{2(D-1)} H^2 a^2 \square  
\nonumber\\
& & \hspace{0.5cm}  - \frac{2D}{(D-1)}Ha \partial_{0} \square + \frac{D}{(D-1)}H^2 \nabla^2 -\frac{D(D-2)}{(D-1)}H^3a  \partial_{0} + H^4a^2 \Biggr\} \delta^D (x-x')
\nonumber\\
& & \hspace{0.5cm} 
+ \frac{ H^{4}}{(4\pi)^{4}}  
\Biggl\{  \frac{1}{3} (aa')^{3} \frac{\square ^3}{H^2}  \frac{\ln (x)}{x}  
-  \frac{1}{3}  (aa')^{3}  \square ^2 \Bigl [\frac{2 \ln (x)}{x}\;-\;\frac{1}{x}\Bigr]  \Biggr\}\;.
\eea
Note that all the divergent pieces are exactly in the form of the counterterm vertices (\ref{alp1} - \ref{alp3}) and  \eqref{alp4}.
Following the same process, the terms with the remaining ten external operators (\ref{newa} - \ref{newz}) can be localized into the desired form of external operators acting on $\delta^D(x-x')$. 
We tabulate the results from TABLE~\ref{kt1} %, TABLE~\ref{kt2}  
to TABLE~\ref{kt3}.

\begin{table}[htbp]
\begin{center}
\caption{Contributions to 3-point kinetic counterterms. All terms are multiplied by $\frac{i \kappa^2 H^{D-4}}{(4\pi)^{D/2}} \frac{\Gamma(\frac{D}2)}{(D-4)(D-3)}$.}
\begin{ruledtabular}
\begin{tabular}{lcl}
${\rm External\; operators}$&&${\rm Coef.\;of\;}H^{4}a^{2}\delta^D(x-x')$ \\[1ex]
\hline\\
$\alpha$&& $-\frac{D^3(D-2)}{16(D-1)}$ \\[2ex]
$\beta_{2}$&& $\frac {D^6-9D^5+16D^4+36D^3-72D^2+16D+32}{16(D^2-3D+2)}$\\[2ex]
$\gamma_{1}$&& $\frac{15D^5-104D^4+284D^3-320D^2-32D+160}{16(D-1)}$ \\[2ex]
$\gamma_{2}$&& $\frac{-4D^4+22D^3-11D^2-90D+108}{4}$\\[2ex]
\hline\\
${\rm Total}$&&$\frac{-15D^6+130D^5-406D^4+434D^3+372D^2-1112D+608}{8(D^2-3D+2)}$ \\[1ex]
\end{tabular}\label{kt1}
\end{ruledtabular}
\end{center}
\end{table}

\begin{table}[htbp] 
\begin{center}
\caption{Contributions to 3-point kinetic counterterms. All terms are multiplied by $\frac{i \kappa^2 H^{D-4}}{(4\pi)^{D/2}}
\frac{\Gamma(\frac{D}2)}{(D-4)(D-3)}$.}
\begin{ruledtabular}
\begin{tabular}{lcc}
${\rm External}$&${\rm Coef.\;of}$&${\rm Coef.\;of}$\\[1ex]
${\rm operators}$&$a^{2}\square^2\delta^D(x-x')$&$H^{2}a^{2}\square\delta^D(x-x')$\\[1ex]
\hline\\
$\alpha$&    $-\frac{D}{D^2-3D+2}$& $\frac{D(D-4)}{2(D-1)}$ \\[2ex]
$\beta_{2}$& $\frac{D}{D^2-3D+2}$& $\frac{D^5-6D^4+5D^3+34D^2-12D+8}{4(D-2)(D-1)^2}$ \\[2ex]
$\gamma_{1}$&  & $\frac{D^3+4D-8}{4(D-2)(D-1)}$ \\[2ex]
$\gamma_{2}$&  & $-1$  \\[2ex]
\hline\\
${\rm Total}$&$0$&$\frac{D^5-3D^4-14D^3+82D^2-60D+24}{4(D-2)(D-1)^2}$ \\[1ex]
\end{tabular}\label{kt2}
\end{ruledtabular}
\end{center}
\end{table}

\begin{table}[htbp] 
\begin{center}
\caption{Contributions to 3-point kinetic counterterms. All terms are multiplied by $\frac{i \kappa^2 H^{D-4}}{(4\pi)^{D/2}}
\frac{\Gamma(\frac{D}2)}{(D-4)(D-3)}$.}
\begin{ruledtabular}
\begin{tabular}{lcccc}
${\rm External}$&${\rm Coef.\;of}$ &${\rm Coef.\;of}$ &${\rm Coef.\;of}$ &${\rm Coef.\;of}$\\[1ex]
${\rm operators}$&$H^{2}\nabla^2 \delta^D(x-x')$ &$Ha \partial_{0} \square \delta^D(x-x')$& $\nabla^2 \square \delta^D(x-x')$&$ Ha^{-1}\nabla^2\partial_0\delta^D(x-x')$\\[1ex]
\hline\\
$\alpha$ &$\frac{D}{(D-1)}$ & $-\frac{2D}{(D-1)}$& &\\[2ex]
$\beta_{2}$ &$\frac{D(D^2-2D+4)}{(D^2-3D+2)}$ & $\frac{2D}{(D-1)}$& &\\[2ex]
$\delta$&  $ \frac{D(D^2-2D-4)}{4(D^2-3D+2)}$ & &$-\frac{D}{D^2-3D+2}$ &$\frac{D^2}{D^2-3D+2}$\\[2ex]
$\epsilon_{1}$&  $-\frac{4(D^2-7D+4)}{4(D-1)}$ & &$\frac{D}{D^2-3D+2}$  &$-\frac{D^2}{D^2-3D+2}$\\[2ex]
$\epsilon_{2}$&  $-\frac{D}{(D-1)}$ & & &\\[2ex]
\hline\\
${\rm Total}$&$-\frac{D(D^3-14D^2+28D-20)}{4(D^2-3D+2)}$& $0$ &$0$ &$0$\\[1ex]
\end{tabular}\label{kt3}
\end{ruledtabular}
\end{center}
\end{table}
As we realized in section \ref{allow} the total coefficients for the contributions coming from  
the terms $\nabla^2 \partial_{0} \delta^D(x-x')$ and $\nabla^2 \square \delta^D(x-x')$ to 3-point kinetic counterterms add up to zero as tabulated in TABLE~\ref{kt3}. This is a very important check for the correctness of our calculation.

The terms in the last line of Eq. \eqref{alphay^(2-D)} are nonlocal and finite for $D=4$ dimension.
For notational simplicity, we define the function $y(x;x')$ as $x \equiv \frac{y}{4}$ and take the $D=4$ limit for these finite nonlocal pieces.
Again, applying the same procedure to the terms with the remaining ten external operators (\ref{newa} - \ref{newz}),
we obtain nonlocal finite terms. 
We add these newly found finite terms to the finite terms tabulated in TABLES ~\ref{kta} - ~\ref{ktz} and
give the summation of the two (i.e., all the finite terms from the 3-point kinetic interactions) in TABLE ~\ref{kt4}. 

%Table VI replaced  
\begin{table}[htbp]
\begin{center}
\caption{All finite nonlocal 3-point kinetic contributions with $x \equiv \frac{y}{4}$, where $y(x;x')$ is defined in the equation (\ref{ydef}).}
\begin{ruledtabular}
\begin{tabular*}{1.5\textwidth}{ll}
$\phantom{ssssssssss}\rm External \; operators$&$\rm Coef. \; of \; \frac{\kappa^2 H^4}{(4\pi)^4}$\\[1ex]
\hline\\
$\phantom{ssssssssss}(aa')^3 \square^3/H^2$&$\frac{\ln{x}}{3x}\phantom{ssssssssssssssssssssss}$\\[2ex]
$\phantom{ssssssssss}(aa')^3 \square^2$&$-\frac{3\ln{x}}{2x} $\\[2ex]
$\phantom{ssssssssss}(aa')^3 H^2\square$&$\frac{25\ln{x}}{x}-\frac{3}{2x} $\\[2ex]
$\phantom{ssssssssss}(aa')^3 H^4$&$-\frac{44\ln{x}}{x} - \frac{26}{x} - \ln(aa') \frac{30}{x}$\\[2ex]
$\phantom{ssssssssss}(aa')^2(a^2+a'^2)\square^3/H^2$&$-\frac{\ln{x}}{6x}$\\[2ex]
$\phantom{ssssssssss}(aa')^2(a^2+a'^2)\square^2$&$\frac{7\ln{x}}{6x} - \frac{1}{6x}$\\[2ex]
$\phantom{ssssssssss}(aa')^2(a^2+a'^2)H^2\square$&$\frac{\ln{x}}{3x} + \frac{5}{6x}$\\[2ex]
$\phantom{ssssssssss}(aa')^2(a^2+a'^2)H^4$&$-\frac{4\ln{x}}{x} - \frac{7}{2x}$\\[2ex]
$\phantom{ssssssssss}(aa')^2 H^2 \nabla^2$&$\frac{4\ln{x}}{x} - \frac{14}{x} + \ln(aa')[9\ln(x) + \frac{20}{x}]$\\[2ex]
$\phantom{ssssssssss}(aa')^2 \nabla^2 \square$&$\frac{8\ln{x}}{3x} + \frac{1}{3x}$\\[2ex]
$\phantom{ssssssssss}(aa')^2 H^2\nabla^2 \square^2$&$\frac{\ln{x}}{3x} $\\[2ex]
$\phantom{ssssssssss}(aa')(a^2+a'^2) H^2 \nabla^2$&$\frac{22\ln{x}}{3x} - \frac{67}{6x} + \ln(aa') \frac{12}{x}$\\[2ex]
$\phantom{ssssssssss}(aa')(a^2+a'^2)\nabla^2\square$&$-\frac{11\ln{x}}{3x} + \frac{4}{6x}$\\[2ex]
$\phantom{ssssssssss}(aa')\nabla^4$&$-\ln(x) + \frac{16}{3x} + \ln(aa') [2\ln(x) - \frac{2}{x}]$\\[2ex]
\end{tabular*}\label{kt4}
\end{ruledtabular}
\end{center}
\end{table}
 In summary, the regulated self-mass-squared from the kinetic interactions consists of three finite parts: 
\begin{itemize}
\item{the local 4-point contributions given in Eq. \eqref{M^2_4point_KWinteraction},}
\item{the local 3-point kinetic contributions arising from TABLES ~\ref{kt1} - ~\ref{kt3}, and from Eqs. \eqref{3ptlock} and \eqref{3ptlogk} (there are also the local 3-point kinetic contributions from our previous work, but they vanish),}
\item{the nonlocal terms from the 3-point kinetic interactions listed in TABLE ~\ref{kt4}.}
\end{itemize}
That is, we finally write it as
\bea
\lefteqn{-iM^2_{\rm regK}(x;x') \;=\; i \kappa^2 a^2 \Bigl( d_1 \square^2 + d_2 H^2 \square + d_3 H^4 + d_4 H^2 \frac{\nabla^2}{a^2} + d_5 H \frac{\partial_{0} \square}{a}   \Bigr) \delta^D(x-x')  + {\rm Table ~\ref{kt4}} 
}
\nonumber \\
& &\hspace{0cm}
+ i \kappa^2 \frac{1}{(4\pi)^2} \;a^2 \Bigl( \frac{11}{4} \ln(aa') H^2 \square 
+ \frac{11}{2} \ln(aa') H^3 \frac{\partial_{0}}{a} 
- 6\ln(aa')  H^4  - \frac32 \ln(aa') H^2 \frac{\nabla^2}{a^2} \Bigr) \delta^4 (x-x') + \mathcal{O}(D-4)\;,
\label{Totalregk}
\nonumber\\
\eea
where the coefficients $d_i$ are 
\bea
d_1 
&=& 0\;, \\
d_2 
&=& \frac{H^{D-4}}{(4 \pi)^{\frac{D}{2}}} \Biggl\{
\frac{\Gamma(\frac{D}{2})}{(D-4)(D-3)}\biggl[\frac{(D^5+3D^4-14D^3+82D^2-60D+24)}{2^2(D-2)(D-1)^3}\biggr] \nonumber\\
& &\hspace{1.5cm} + \frac{\Gamma(D-1)}{\Gamma(\frac{D}{2})} \biggl[-2 + [ \frac14  D(D-5) - (\frac{D-1}{D-3}) ] \pi \cot(\frac{\pi D}{2})\biggr]\Biggr\}\;,\nonumber\\
&=& \frac{H^{D-4}}{(4 \pi)^{\frac{D}{2}}} \Biggl\{ 
\frac{4D^8-64D^7+396D^6-1239D^5+2111D^4-1886D^3+742D^2+44D-72}{2^3(D-4)(D-3)(D-1)^3} -4
+ \frac{205}{18} \gamma  + \mathcal{O}(D-4) \Biggr\}\;, \nonumber\\ \\
d_3 
&=& \frac{H^{D-4}}{(4 \pi)^{\frac{D}{2}}} \Biggl\{
\frac{\Gamma(\frac{D}{2})}{(D-4)(D-3)}\biggl[\frac{(-15D^6+130D^5-406D^4+434D^3+372D^2-1112D+608)}{2^3(D-2)(D-1)}\biggr] \nonumber\\
& &\hspace{1.5cm} + \frac{\Gamma(D-1)}{\Gamma(\frac{D}{2})} \biggl[  -\frac92 + [ \frac14 (D-2)(3D-8) (\frac{D-1}{D-3}) ] \pi \cot(\frac{\pi D}{2}) \biggr] 
\Biggr\}\;,\nonumber\\
&=& \frac{H^{D-4}}{(4 \pi)^{\frac{D}{2}}} \Biggl\{ 
\frac{9D^6-150D^5+914D^4-2774D^3+4596D^2-3960D-1376}{2^4(D-4)(D-3)(D-1)} -9 + \frac{98}{3} \gamma  +  \mathcal{O}(D-4) \Biggr\}\;, \\
d_4 
&=& \frac{H^{D-4}}{(4 \pi)^{\frac{D}{2}}} \Biggl\{
\frac{\Gamma(\frac{D}{2})}{(D-4)(D-3)}\biggl[-\frac{D(D^3-14D^2 +28D -20)}{2^2(D-2)(D-1)}\biggr]\nonumber\\
& &\hspace{1.5cm} + \frac{\Gamma(D-1)}{\Gamma(\frac{D}{2})} \biggl[\frac52 - [\frac14 D(D-5)- (\frac{D-1}{D-3}) ] \pi \cot(\frac{\pi D}{2})\biggr] 
\Biggr\}\;, 
\nonumber\\
&=& \frac{H^{D-4}}{(4 \pi)^{\frac{D}{2}}} \Biggl\{ \frac{-4D^6 +56D^5 -281D^4 +638D^3-608D^2+108D+96}{2^3(D-4)(D-3)(D-1)} + 5
- \frac{65}{3} \gamma  + \mathcal{O}(D-4) \Biggr\}\;, \\
d_5 
&=& 0\;.
\eea
Here $\gamma$ is the Euler's constant which arises from the expansion of the Gamma function.

%%%%%%%%%%%%%%%%%%%%%%%%%%%%%%%%%%%%%%%%%%%%%%%%%%%%%%%%%%%%%%%%
\subsubsection{For 3-point cross part}\label{regcross}
%%%%%%%%%%%%%%%%%%%%%%%%%%%%%%%%%%%%%%%%%%%%%%%%%%%%%%%%%%%%%%%%

To illustrate how we put the divergences in the forms of the counterterms, we take the term with the external operator $\beta_1$ in Eq. \eqref{beta1} coming from the 3-point kinetic-conformal cross interactions as an example:
\bea\label{beta1y^(2-D)}
\lefteqn{\beta_{1} \Biggl\{-\frac{(D-1) H^{2D-4}\Gamma(\frac{D}2 -1)^2}{2\pi^D} \Bigl( \frac1{y}\Bigr)^{D-2} -\frac{3H^4}{2\pi ^4} \Bigl(\frac{1}{y}\Bigr) \Biggr\}\;, 
}
\nonumber \\ 
& & \hspace{0cm} = -\frac{(D-1) H^{2D-4}\Gamma(\frac{D}2 -1)^2}{2\pi^D} (aa')^{\frac{D}{2}+1} H^2 \square \Bigl(\frac1{y}\Bigr)^{D-2}
\;-\; (aa')^{\frac{D}2+1} \square \frac{3H^4}{2\pi^4} \Bigl(\frac1{y} \Bigr)\;,
\nn 
\\ 
& & \hspace{0cm} = -\frac{i H^{D-4}}{(4\pi)^\frac{D}2} \frac{\Gamma(\frac{D}2-1)}{(D-4)(D-3)} 
\Biggl\{16(D-1) H^2 a^2 \square + 4 D(D-2)(D-1) H^4 a +16(D-2)(D-1)H^3 a \partial_{0}\Biggr\} \delta^D (x-x')\;
\nonumber\\
& & \hspace{0.5cm} + \frac{H^{4}}{(4\pi)^4} \Biggl\{ 24 (aa')^{3}\square^2 \frac {\ln(x)}{x}  -  24  (aa')^{3} H^{2} \square (\frac{2\ln(x)}{x} \;+\; \frac{3}{x} ) \Biggr\}\;.
\eea
Similarly, the terms with the remaining ten external operators can be localized into the desired form of external operators acting on $\delta^D(x-x')$. 
We give the results in TABLE~\ref{ct1}, TABLE~\ref{ct2} and TABLE~\ref{ct3}.
The terms in the last line of Eq. \eqref{beta1y^(2-D)} are nonlocal and finite for $D=4$ dimension.
We denote the function $y(x;x')$ by $x \equiv \frac{y}{4}$ and take the $D=4$ limit for the finite nonlocal pieces.
Applying the same procedure to the terms with the remaining ten external operators,
we obtain the rest of nonlocal finite terms. 
We add these newly found finite terms to the finite terms in TABLES  ~\ref{cta} - ~\ref{ctz} and
report the summation of the two (i.e., all the finite terms from the 3-point cross interactions) in TABLE ~\ref{ct4}.

\begin{table}[htbp]
\begin{center}
\caption{Contributions to 3-point cross counterterms. All terms are multiplied by $\frac{i \tilde{\kappa}^2 H^{D-4}}{(4\pi)^{D/2}} \frac{\Gamma(\frac{D}2)}{(D-4)(D-3)}$.}
\begin{ruledtabular}
\begin{tabular}{lcl}
${\rm External\; operators}$&&${\rm Coef.\;of\;}H^{4}a^{2}\delta^D(x-x')$ \\[1ex]
\hline\\
$\beta_{1}$&& $-8D(D-1)$ \\[2ex]
$\beta_{2}$&& $- \frac {D^5-6D^4+2D^3+16D^2+8D}{2(D-2)}$\\[2ex]
$\gamma_{1}$&& $-(D^4+6D^3-58D^2-173D+344)$ \\[2ex]
$\gamma_{2}$&& $\frac{D^5-4D^4+10D^3-254D^2+764D-616}{2(D-2)}$\\[2ex]
$\gamma_{3}$&&$\frac{D^5+24D^4-253D^3+752D^2-852D+324}{(D-2)}$ \\[2ex]
\hline\\
${\rm Total}$&&$\frac{20D^4-187D^3+696D^2-1180D+704}{(D-2)}$ \\[1ex]
\end{tabular}\label{ct1}
\end{ruledtabular}
\end{center}
\end{table}

\begin{table}[htbp] 
\begin{center}
\caption{Contributions to 3-point cross counterterms. All terms are multiplied by $\frac{i \tilde{\kappa}^2 H^{D-4}}{(4\pi)^{D/2}}
\frac{\Gamma(\frac{D}2)}{(D-4)(D-3)}$.}
\begin{ruledtabular}
\begin{tabular}{lcc}
${\rm External}$&${\rm Coef.\;of}$&${\rm Coef.\;of}$\\[1ex]
${\rm operators}$&$a^{2}\square^2\delta^D(x-x')$&$H^{2}a^{2}\square\delta^D(x-x')$\\[1ex]
\hline\\
$\beta_{1}$&  & $-\frac{32}{(D-2)}$  \\[2ex]
$\beta_{2}$& $-\frac{8}{(D-2)}$ & $-\frac{2D^3-6D^2-36D+96}{(D-2)}$ \\[2ex]
$\gamma_{1}$&  & $-\frac{4(D^2+3D+1)}{(D-2)}$ \\[2ex]
$\gamma_{2}$&  & $\frac{2(D^2+4D+2)}{(D-2)}$  \\[2ex]
$\gamma_{3}$&  & $\frac{4(D^2+4D-10)}{(D-2)}$ \\[2ex]
\hline\\
${\rm Total}$&$-\frac{8}{(D-2)}$&$\frac{2D^3-8D^2-48D+168}{(D-2)}$ \\[1ex]
\end{tabular}\label{ct2}
\end{ruledtabular}
\end{center}
\end{table}

\begin{table}[htbp] 
\begin{center}
\caption{Contributions to 3-point cross counterterms. All terms are multiplied by $\frac{i \tilde{\kappa}^2 H^{D-4}}{(4\pi)^{D/2}}
\frac{\Gamma(\frac{D}2)}{(D-4)(D-3)}$.}
\begin{ruledtabular}
\begin{tabular}{lcc}
${\rm External}$&${\rm Coef.\;of}$&${\rm Coef.\;of}$\\[1ex]
${\rm operators}$&$H^{2}\nabla^2 \delta^D(x-x')$&$Ha \partial_{0} \square \delta^D(x-x')$\\[1ex]
\hline\\
$\beta_{2}$ &$-\frac{8D(D-2)+32}{(D-2)}$ & $-16$\\[2ex]
$\epsilon_{1}$&  $-4(D-6)$ & \\[2ex]
$\epsilon_{2}$&  $-\frac{2D(D-8)}{(D-1)}$ & \\[2ex]
$\epsilon_{3}$&  $-\frac{4(D^3-3D^2+22D-16)}{(D-3)(D-1)}$ & \\[2ex]
\hline\\
${\rm Total}$& $ -\frac{(D-2)(9D^4-71D^3+256D^2-400D+184)}{2(D-3)(D-1)}$ & $-16$ \\[1ex]
\end{tabular}\label{ct3}
\end{ruledtabular}
\end{center}
\end{table}

%%%Table X replaced
\begin{table}[htbp]
\begin{center}
\caption{All finite nonlocal 3-point cross contributions with $x \equiv \frac{y}{4}$, where $y(x;x')$ is defined in the equation (\ref{ydef}).}
\begin{ruledtabular}
\begin{tabular*}{1.5\textwidth}{ll}
$\phantom{ssssssssss}\rm External \; operators$&$\rm Coef. \; of \; \frac{\tilde{\kappa}^2 H^4}{(4\pi)^4}$\\[1ex]
\hline\\
$\phantom{ssssssssss}(aa')^3 \square^3/H^2$&$\frac{2\ln{x}}{x}\phantom{ssssssssssssssssssssss}$\\[2ex]
$\phantom{ssssssssss}(aa')^3 \square^2$&$-\frac{27\ln{x}}{x} $\\[2ex]
$\phantom{ssssssssss}(aa')^3 H^2\square$&$-\frac{378\ln{x}}{x} - \frac{59}{x}$\\[2ex]
$\phantom{ssssssssss}(aa')^3 H^4$&$\frac{608\ln{x}}{x} - \frac{436}{x}$\\[2ex]
$\phantom{ssssssssss}(aa')^2(a^2+a'^2)\square^3/H^2$&$\frac{\ln{x}}{x}$\\[2ex]
$\phantom{ssssssssss}(aa')^2(a^2+a'^2)\square^2$&$-\frac{28\ln{x}}{x} + \frac{1}{x}$\\[2ex]
$\phantom{ssssssssss}(aa')^2(a^2+a'^2)H^2\square$&$\frac{178\ln{x}}{x} - \frac{26}{x}$\\[2ex]
$\phantom{ssssssssss}(aa')^2(a^2+a'^2)H^4$&$-\frac{252\ln{x}}{x} - \frac{66}{x}$\\[2ex]
$\phantom{ssssssssss}(aa')^2 H^2 \nabla^2$&$-24\ln(x) - \frac{142\ln{x}}{3x} + \frac{508}{3x}$\\[2ex]
$\phantom{ssssssssss}(aa')^2 \nabla^2 \square$&$\frac{76\ln{x}}{3x}$\\[2ex]
$\phantom{ssssssssss}(aa')(a^2+a'^2) H^2 \nabla^2$&$-\frac{24\ln{x}}{x} + \frac{58}{x}$\\[2ex]
$\phantom{ssssssssss}(aa')(a^2+a'^2)\nabla^2\square$&$\frac{12\ln{x}}{x} + \frac{4}{x}$\\[2ex]
$\phantom{ssssssssss}(aa')\nabla^4$&$\frac{4}{x}$\\[2ex]
\end{tabular*}\label{ct4}
\end{ruledtabular}
\end{center}
\end{table}

The regulated self-mass-squared from the kinetic-conformal cross interactions consists of two finite parts: 
\begin{itemize}
\item{the local 3-point cross contributions from TABLES ~\ref{ct1} - ~\ref{ct3} 
and the equations \eqref{3-ptcrossloc} and \eqref{3-ptcrosslog},}
\item{the nonlocal contributions from the 3-point cross interactions in TABLE ~\ref{ct4},}
\end{itemize}
and we can write it as
\bea
-iM^2_{\rm regcross}(x;x') \;=\; i \kappa^2 a^2 \Bigl( d_1 \square^2 + d_2 H^2 \square + d_3 H^4 + d_4 H^2 \frac{\nabla^2}{a^2} + d_5 H \frac{\partial_{0} \square}{a} \Bigr) \delta^D(x-x') + {\rm Table ~\ref{ct4}} + \mathcal{O}(D-4)\;.
\label{Totalregc}
\nonumber\\
\eea
Here $\tilde{\kappa} = \frac{D-2}{8(D-1)}\kappa$ is taken. 
The multiplier $(\frac{D-2}{8(D-1)})^2$ is inserted in the coefficients $d_i$. Thus the coefficients $d_i$ become

\bea
d_1 
&=& \frac{H^{D-4}}{(4\pi)^\frac{D}{2}} \Biggl\{\frac{\Gamma(\frac{D}{2})}{(D-4)(D-3)} \;\biggl[-\frac{(D-2)}{2^3(D-1)^2}\biggr]\Biggr\}\;, \nonumber\\
&=& \frac{H^{D-4}}{(4\pi)^\frac{D}{2}} \Biggl\{-\frac{(D-2)^2}{2^4(D-4)(D-3)(D-1)^2} + \frac{1}{72}\gamma +\mathcal{O}(D-4)\Biggr\}\;, \\
d_2 
&=& \frac{H^{D-4}}{(4\pi)^\frac{D}{2}} \Biggl\{\frac{\Gamma(\frac{D}{2})}{(D-4)(D-3)}\;\biggl[\frac{(D-2)(D^3-4D^2-24D+84)}{2^5(D-1)^2} \biggr]\Biggr\}\;, \nonumber\\
&=& \frac{H^{D-4}}{(4\pi)^\frac{D}{2}} \Biggl\{\frac{(D-2)^2(D^3-4D^2-24D+84)}{2^6(D-4)(D-3)(D-1)^2} + \frac{1}{24} \gamma +  \mathcal{O}(D-4) \Biggr\}\;, \\
d_3 
&=& \frac{H^{D-4}}{(4\pi)^\frac{D}{2}} \Biggl\{\frac{\Gamma(\frac{D}{2})}{(D-4)(D-3)} \;\biggl[\frac{(D-2)(20D^4-187D^3+696D^2-1180D+704)}{2^6(D-1)^2}\biggl]\Biggr\}\;, \nonumber\\
&=& \frac{H^{D-4}}{(4\pi)^\frac{D}{2}}  \Biggl\{\frac{(D-2)^2(20D^4-187D^3+696D^2-1180D+704)}{2^7(D-4)(D-3)(D-1)^2} - \frac{17}{9} \gamma +  \mathcal{O}(D-4) \Biggr\}\;, \\
d_4 
&=& \frac{H^{D-4}}{(4\pi)^\frac{D}{2}} \Biggl\{\frac{\Gamma(\frac{D}{2})}{(D-4)(D-3)} \;\biggl[-\frac{(D-2)^3(9D^4 -71D^3 +256D^2 -400D +184)}{2^7(D-1)^2}\biggr]\Biggr\}\;, \nonumber\\
&=& \frac{H^{D-4}}{(4\pi)^\frac{D}{2}} \Biggl\{-\frac{(D-2)^4(9D^4 -71D^3 +256D^2 -400D +184)}{2^8(D-4)(D-3)(D-1)^2} + \frac{55}{108} \gamma +  \mathcal{O}(D-4) \Biggr\}\;, \\
d_5 
&=& \frac{H^{D-4}}{(4\pi)^\frac{D}{2}} \Biggl\{\frac{\Gamma(\frac{D}{2})}{(D-4)(D-3)} \;\biggl[-\frac{(D-2)^2}{2^2(D-1)^2}\biggr]\Biggr\}\;, \nonumber\\
&=& \frac{H^{D-4}}{(4\pi)^\frac{D}{2}} \Biggl\{-\frac{(D-2)^3}{2^3(D-4)(D-3)(D-1)^2} + \frac{1}{18} \gamma +  \mathcal{O}(D-4) \Biggr\}\;.
\eea

%%%%%%%%%%%%%%%%%%%%%%%%%%%%%%%%%%%%%%%%%%%%%%%%%%%%%%%%%%%%%%%%
\subsection{Renormalization} 
%%%%%%%%%%%%%%%%%%%%%%%%%%%%%%%%%%%%%%%%%%%%%%%%%%%%%%%%%%%%%%%%

Finally, we sum the contributions from kinetic and cross interactions:
\bea
\lefteqn{-iM^2_{\rm reg}(x;x') \;=\; -iM^2_{\rm regK}(x;x') \;+\; -iM^2_{\rm regcross}(x;x') \;,
}
\nonumber\\
& &\hspace{0cm}=\; i \kappa^2 a^2 \Bigl( d_1 \square^2 + d_2 H^2 \square + d_3 H^4 + d_4 H^2 \frac{\nabla^2}{a^2} + d_5 H \frac{\partial_{0} \square}{a}\Bigr) \delta^D(x-x') + {\rm Table ~\ref{kt4}} + (\frac{8(D-1)}{D-2})^2{\rm Table ~\ref{ct4}}
\nonumber\\
& &\hspace{0.5cm} +\; i \kappa^2 \frac{1}{(4\pi)^2}\; a^2 \Bigl( \frac{11}{4} \ln(aa') H^2 \square 
+ \frac{11}{2} \ln(aa') H^3 \frac{\partial_{0}}{a} 
- 6\ln(aa') H^4 - \frac32 \ln(aa') H^2 \frac{\nabla^2}{a^2} \Bigr) \delta^4 (x-x') 
+ \mathcal{O}(D-4)\;.
\nonumber\\
\label{Totalregkc}
\eea
The resulting total coefficients $d_i$ are
\bea
d_1 
&=& \frac{H^{D-4}}{(4\pi)^\frac{D}{2}} \Biggl\{ \frac{\Gamma(\frac{D}{2})}{(D-4)(D-3)} \;\biggl[-\frac{(D-2)}{2^3(D-1)^2}\biggr] \Biggr\}\;,\nonumber\\
&=& \frac{H^{D-4}}{(4\pi)^\frac{D}{2}} \Biggl\{-\frac{(D-2)^2}{2^4(D-4)(D-3)(D-1)^2} + \frac{1}{72} \gamma +  \mathcal{O}(D-4) \Biggr\}\;,\label{d1-final} \\
d_2 
&=& \frac{H^{D-4}}{(4 \pi)^{\frac{D}{2}}} \Biggr\{
\frac{\Gamma(\frac{D}{2})}{(D-4)(D-3)} \biggl[\frac{D^6 -D^5 +28D^4 +56D^3 +60D^2 +288D -144}{2^5 (D-2)(D-1)^3}\biggr] 
\nonumber\\
& &\hspace{1.5cm}+\frac{\Gamma(D-1)}{\Gamma(\frac{D}{2})} \biggl[ -2 + [ \frac14  D(D-5) - (\frac{D-1}{D-3}) ] \pi \cot(\frac{\pi D}{2}) \biggr] 
\Biggr\}\;,
\nonumber\\
&=& \frac{H^{D-4}}{(4 \pi)^{\frac{D}{2}}} \Biggr\{\frac{D^6 -D^5 +28D^4 +56D^3 +60D^2 +288D -144}{2^6(D-4)(D-3)(D-1)^3}
- 4  + \frac{823}{72} \gamma  + \mathcal{O}(D-4) \Biggr\}\;, \\
d_3 
&=& \frac{H^{D-4}}{(4 \pi )^{\frac{D}{2}}} 
\Biggl\{ -\frac{\Gamma(\frac{D}{2})}{(D-4)(D-3)} \biggl[\frac{100D^7 -853D^6 +2230D^5 +1040D^4 -17136D^3 +35824D^2 -31648D +10496}{2^6 (D-2)(D-1)^2}\biggr] 
\nonumber\\
& &\hspace{1.5cm}+\frac{\Gamma(D-1)}{\Gamma(\frac{D}{2})} \biggl[ -\frac92 +  [ \frac14 (D-2)(3D-8) (\frac{D-1}{D-3}) ] \pi \cot(\frac{\pi D}{2})\biggr] \Biggr\}\;,
\nonumber\\
&=& \frac{H^{D-4}}{(4 \pi )^{\frac{D}{2}}} 
\Biggl\{ - \frac{100D^7 -853D^6 +2230D^5 +1040D^4 -17136D^3 +35824D^2 -31648D +10496}{2^7 (D-4)(D-3)(D-1)^2} 
\nonumber\\
& & \hspace{1.5cm}-9 + \frac{1159}{36} \gamma  +  \mathcal{O}(D-4) \Biggr\}\;, \\
d_4 
&=& \frac{H^{D-4}}{(4 \pi)^{\frac{D}{2}}} \Biggl\{
\frac{\Gamma(\frac{D}{2})}{(D-4)(D-3)}\biggl[\frac{-9D^8+143D^7-1040D^6+4408D^5-11464D^4+19056D^3-19776D^2+11648D-2944}{2^7(D-2)(D-1)^2}\biggr]
\nonumber\\
& & \hspace{1.5cm}+\frac{\Gamma(D-1)}{\Gamma(\frac{D}{2})}\biggl[ \frac52 - [ \frac14 D(D-5)- (\frac{D-1}{D-3}) ] \pi \cot(\frac{\pi D}{2}) \biggr] \Biggr\}\;,
\nonumber\\
&=& \frac{H^{D-4}}{(4 \pi)^{\frac{D}{2}}} \Biggl\{
\frac{-9D^8+143D^7-1040D^6+4408D^5-11464D^4+19056D^3-19776D^2+11648D-2944}{2^8(D-4)(D-3)(D-1)^2}
\nonumber\\
& & \hspace{1.5cm} +5 - \frac{2285}{108} \gamma  + \mathcal{O}(D-4) \Biggr\}\;, \\
d_5 
&=& \frac{H^{D-4}}{(4\pi)^\frac{D}{2}} \Biggl\{\frac{\Gamma(\frac{D}{2})}{(D-4)(D-3)} \;\biggl[-\frac{(D-2)^2}{2^2(D-1)^2}\biggr]\Biggr\}\;, 
\nonumber\\
&=& \frac{H^{D-4}}{(4\pi)^\frac{D}{2}} \Biggl\{-\frac{(D-2)^3}{2^3(D-4)(D-3)(D-1)^2} + \frac{1}{18} \gamma +  \mathcal{O}(D-4) \Biggr\}\;.\label{d5-final}
\eea
Now we choose the coefficients of the counterterm vertices (\ref{alp1} - \ref{alp3}) and \eqref{alp4} as 
\bea
c_i &=& -d_i + \Delta c_i   \quad \mbox{for } ~ i =1, 3, 4 \;, \\ %\mbox{ and } 5
c_2 &=& -d_2 + (D-2)d_1 + \Delta c_2\;,
\eea
where $\Delta c_i$ is the arbitrary finite terms which remain after the cancellation of the divergent parts. 
Note that the coefficients $d_1$ and $d_5$ have the relation
\be
d_5 = 2 (D-2) \; d_1 \;,
\ee 
which makes them simultaneously cancelled by subtracting the counterterm vertex \eqref{alp1}. 
We finally obtain the final renormalized self-mass-squared from kinetic and kinetic-conformal cross parts at one loop order as
\bea
\lefteqn{
-iM^2_{\rm ren}(x;x') = i \kappa^2 a^2 \Bigl( \Delta c_1 \square^2 + \Delta c_2 H^2 \square + \Delta c_3 H^4 + \Delta c_4 H^2 \frac{\nabla^2}{a^2} + %\Delta c_5 
4 \Delta c_1 H \frac{\partial_{0} \square}{a} \Bigr) \delta^4(x-x')
}
\nonumber \\
& & + i \kappa^2 \frac{1}{(4\pi)^2}\; a^2 \Bigl( \frac{11}{4} \ln(aa') H^2 \square 
+ \frac{11}{2} \ln(aa') H^3 \frac{\partial_{0}}{a} 
- 6\ln(aa') H^4 - \frac32 \ln(aa') H^2 \frac{\nabla^2}{a^2} \Bigr) \delta^4 (x-x') 
\nonumber\\
& & + {\rm Table ~\ref{kt4}} + (\frac{8(D-1)}{D-2})^2 {\rm Table ~\ref{ct4}} 
+ \mathcal{O}(D-4)\;.
\label{Totalrenkc}
\eea

%%%%%%%%%%%%%%%%%%%%%%%%%%%%%%%%%%%%%%%%%%%%%%%%%%%%%%%%%%%%%%%%
\section{Discussion}\label{dis}
%%%%%%%%%%%%%%%%%%%%%%%%%%%%%%%%%%%%%%%%%%%%%%%%%%%%%%%%%%%%%%%%

We have evaluated the kinetic and kinetic-conformal cross parts of the CC scalar self-mass-squared at one loop order in the cosmological patch of de Sitter space. 
The CC scalar has the technical advantage that its propagator with the conformally rescaled metric looks the same as the MMC scalar propagator in flat space,
\be
i\Delta^{\rm flat} = \frac{\Gamma(\frac{D}{2}-1)}{4\pi^{D/2}}\frac{1}{\Delta x^{D-2}}\;,
\label{flat-scalar-propagator}
\ee
where $\Delta x^2$ is the Poincar\'e length function defined as
\be
\Delta x^2  \equiv \Vert \vec{x} - \vec{x}' \Vert^2 - \Bigl(\vert \eta - \eta'\vert - i \delta\Bigr)^2 \;.
\ee
We have first dimensionally regulated the divergences, renormalized the result by subtracting the four possible BPHZ counterterms and then taken the unregulated limit of $D=4$.  
The final result is given in Eq. \eqref{Totalrenkc}
with the finite nonlocal contributions in TABLE ~\ref{kt4} and TABLE ~\ref{ct4}.
Adding our previous result from the conformal-conformal part, Eq. (129) of Ref. \cite{kahya3} to Eq. \eqref{Totalrenkc}
completes the full renormalized CC scalar self-mass-squared at one loop order.

The point of this computation is to examine the effects of inflationary produced gravitons on the CC scalar mode functions.  
In a subsequent paper \cite{BKP17-modeftn} we will solve the quantum corrected CC scalar field equation \eqref{linear_eq} (with the full self-mass-squared) to obtain one loop corrections to the CC scalar mode functions. 
It is worthwhile at this point commenting 
that the Schwinger-Keldysh formalism \cite{Sch, KTM, LVK, Jrd, CHU, CVE, FW, SW11} is necessary when studying quantum responses in a time-dependent background such as de Sitter. Applying the formalism amounts to replacing the in-out self-mass-squared by the ones from the Schwinger-Keldysh formalism,    
\be
a^4 \square \phi(x) - \frac16 R \phi(x) - \int_{\eta_i}^0 \! d\eta'\!\int\!d^3x' \,\Bigl\{M^2_{\scriptscriptstyle ++}(x;x') + M^2_{\scriptscriptstyle +-}(x;x')\Bigr\} \phi(x') = 0 \; . \qquad \label{linear_eq_ren}
\ee
At one loop order, $-iM^2_{\scriptscriptstyle ++}(x;x')$ agrees with the in-out self-mass-squared $-iM^2(x;x')$. 
The $+-$ self-mass-squared  $-iM^2_{\scriptscriptstyle +-}(x;x')$ can be obtained by replacing $y(x;x')$ everywhere with  
\be
y(x;x') \rightarrow y_{\scriptscriptstyle +-}(x;x') \equiv a(\eta) a(\eta')H^2 \Bigl[ \Vert \vec{x}\!-\! \vec{x}' \Vert^2 - (\eta\!-\! \eta' \!+\! i \delta )^2 \Bigr]\;. \label{y+-} \qquad
\ee

It should be noted that the CC scalar field interacts with gravitons both through a kinetic term and the conformal coupling term. The case with a kinetic term only, also known as minimal coupling, was investigated in a previous work \cite{kahya, kahya2}. When the conformal coupling is added, the 4-point interaction part simply adds the contributions from the kinetic and conformal coupling terms. However, for the 3-point interaction part which has two vertices, there are three possibilities: (i) both vertices are kinetic, (ii) both vertices are conformal (iii) one vertex is kinetic and the other vertex is conformal. The case (ii) was studied in a previous paper \cite{kahya3} and the cases (i) and (iii) were investigated in the present paper. The difference between the computations of the case (i) in Ref. \cite{kahya} and in this paper is that in the former, the MMC scalar propagator was used (since there was no conformal coupling term); and in the latter we have used the CC scalar propagator.     

We also would like to comment on a way to check the accuracy of our computation.  
We note that the most singular part of the graviton propagator in de Sitter background agrees with the conformally coupled scalar propagator and it can be compared with the graviton propagator in flat space 
\be 
i[_{\mu\nu}\Delta^{\rm flat}_{\rho\sigma}](x;x') =\Bigl[ 2\eta_{\mu(\rho} \eta_{\sigma)\nu} - \frac{2}{D-2} \eta_{\mu\nu} \eta_{\rho\sigma} \Bigr] \frac{\Gamma(\frac{D}{2}-1)}{4\pi^{D/2}}\frac{1}{\Delta x^{D-2}}\;, 
\ee 
Here note that the scalar propagator multiplying the tensor factor is the flat space scalar propagator \eqref{flat-scalar-propagator}.
%%%
As the counterterm analysis in Eq. \eqref{alp1} shows, the coefficients of the divergent terms $d_1$ of $a^2 \Box^2 \delta^{D}(x-x')$ and  $d_5$ of $H a \partial_0  \Box \delta^{D}(x-x')$ should obey the relation $d_5 = 2(D-2)d_1$. And that is exactly what we get in Eqs. \eqref{d1-final} and \eqref{d5-final}, so these divergent terms can be removed by the relevant counterterms.
%%%%
Also, as can be seen in TABLE~\ref{kt3}, the total coefficients for the contributions coming from   the terms $\nabla^2 \partial_{0} \delta^D(x-x')$ and $\nabla^2 \square \delta^D(x-x')$ to 3-point kinetic counterterms add up to zero. There is no other reason why these relations are satisfied, but they turn out to satiate (through explicit computations) the conditions so that our counterterms respect the symmetries that are not broken by the gauge fixing term \eqref{gf}.    

Finally, the CC scalar self-mass squared is ready to be employed in the CC scalar effective field equation so that one can study how the inflationary produced gravitons affect the CC scalar mode functions. In a subsequent paper \cite{BKP17-modeftn} we will examine whether the mode functions get a secular growth effect in late times or not. If it does, it might leave an observational signature such as a correction to the scalar power spectrum. We will soon report whether this is the case or not.  

\vskip 1cm
\centerline{\bf Acknowledgements}
\vskip 0.3cm

We are very grateful to Richard P. Woodard for helpful comments and discussions.
SB and EOK acknowledge support from Tubitak Grant Number:112T817.
EOK acknowledges TUBA-GEBIP 2015 awards programme.

%%%%%%%%%%%%%%%%%%%%%%%%%%%%%%%%%%%%%%%%%%%%%%%%%%%%%%%%%%%%%%%%

\newpage

%%%%%%%%%%%%%%%%%%%%%%%%%%%%%%%%%%%%%%%%%%%%%%%%%%%%%%%%%%%%%%%%
\appendix
%%%%%%%%%%%%%%%%%%%%%%%%%%%%%%%%%%%%%%%%%%%%%%%%%%%%%%%%%%%%%%%%

%%%%%%%%%%%%%%%%%%%%%%%%%%%%%%%%%%%%%%%%%%%%%%%%%%%%%%%%%%%%%%%%
\section{Tables for the coefficient functions of external operators for 3-point kinetic part}
\label{app1}
%%%%%%%%%%%%%%%%%%%%%%%%%%%%%%%%%%%%%%%%%%%%%%%%%%%%%%%%%%%%%%%%
\begin{table}[htbp]
\caption{For $\alpha \equiv (aa')^{\frac{D}{2}+1} \square^2$ type terms.}
\begin{ruledtabular}
\begin{tabular}{|l|l|}
\hline
\multicolumn{2}{|c|}{}\\
  \multicolumn{2}{|c|}{$ f_{\alpha}(y)=C_{A}\;\;f_{\alpha(A)}(y)\;+\;C_{C}\;\;f_{\alpha(C)}(y),$}\\
  \multicolumn{2}{|c|}{}\\
\hline \hline
$f_{\alpha(A)}(y)$&$ 2 I^2[A F'']\phantom{sssssssssssssssssssssssssssssssssssssssssssssssss}$\\
\hline \hline
$f_{\alpha(C)}(y)$&$ 2 I^2[C F''] $\\
\hline \hline
\hline
\multicolumn{2}{|c|}{}\\
\multicolumn{2}{|c|}{${\rm Total\;for\;}f_{\alpha}(y)$}\\
\multicolumn{2}{|c|}{}\\
\hline \hline
\multicolumn{2}{|c|}{}\\
\multicolumn{2}{|c|}{$-\frac{D\;H^{2D-4}\pi^{-D}\;\Gamma{(\frac{D}{2}-1)^2}}{64(D-1)}\;(\frac{1}{y})^{D-2}.$}\\
\multicolumn{2}{|c|}{}\\
\end{tabular}\label{kta}
\end{ruledtabular}
\end{table}
%%%%%%%%%%%%%%%%%%%%%%%%%%%%%%%%%%%%%%%%%%%%%%%%%%%%%%%%%%%%%%%%
\begin{table}[htbp] 
\caption{For $\beta_2 \equiv  (aa')^{\frac{D}{2}} (a^2+a'^2) H^2 \square$ type terms.}
\begin{ruledtabular}
\begin{tabular}{|l|l|}
\hline
\multicolumn{2}{|c|}{}\\
\multicolumn{2}{|c|}{$f_{\beta_2}(y)=C_{A}\;\;f_{\beta_2(A)}(y)\;+\;f_{\beta_2(A)}(y)\;+\;C_{C}\;\;f_{\beta_2(C)}(y),$}\\
\multicolumn{2}{|c|}{}\\
\hline \hline
$f_{\beta_2(A)}(y)$&$-(\frac{D}{2} -1)\;\frac{D}{4} [AF] + (\frac{D}{2} -1) \;\frac12 (D-1) I[A' F] + I[A'F'] -\frac12 (D-1) I^2 [A'F']\phantom{ssssssssssssssssssssssssss}$\\
\hline \hline
$f_{\beta_2(B)}(y)$&$\phantom{sssssssssssssssssssssssss}-(D-1)I [B' F']$\\
\hline \hline
$f_{\beta_2(C)}(y)$&$-(\frac{D}{2} -1) \;\frac{D}{4} \; [CF] + (\frac{D}{2} -1) \frac12 (D-1) I[C' F] + I[C'F'] -\frac12 (D-1) I^2 [C'F']$\\
\hline \hline
\hline
\multicolumn{2}{|c|}{}\\
\multicolumn{2}{|c|}{${\rm Total\;for\;}f_{\beta_2}(y)$}\\
\multicolumn{2}{|c|}{}\\
\hline \hline
\multicolumn{2}{|c|}{}\\
\multicolumn{2}{|c|}{$\frac{(D-2)^2 D\;H^{2D-4}\pi^{-D}\;\Gamma{(\frac{D}{2}-1)^2}}{64(D-1)}\;(\frac1{y})^{D-1}$}\\ 
\multicolumn{2}{|c|}{}\\
\multicolumn{2}{|c|}{
$-\frac{(D-2)\;(D(D-4)-4)\;H^{2D-4} \pi^{-D}\;\Gamma{(\frac{D}{2} -1)^2}}{512}\; (\frac1{y})^{D-2}.$}\\
\multicolumn{2}{|c|}{}\\
\end{tabular}\label{ktb2}
\end{ruledtabular}
\end{table}
%%%%%%%%%%%%%%%%%%%%%%%%%%%%%%%%%%%%%%%%%%%%%%%%%%%%%%%%%%%%%%%%
\begin{table}[htbp] 
\begin{center}
\caption{For $\gamma_1 \equiv (aa')^{\frac{D}{2}+1} H^4 $ type terms.}
\begin{ruledtabular}
\begin{tabular}{|l|l|}
\hline
\multicolumn{2}{|c|}{}\\
\multicolumn{2}{|c|}{$f_{\gamma_1}(y)=C_{A}\;\;f_{\gamma_1(A)}(y)\;+\;f_{\gamma_1(B)}(y)\;+\;C_{C}\;\;f_{\gamma_1(C)}(y),$}\\
\multicolumn{2}{|c|}{}\\
\hline \hline
$f_{\gamma_1(A)}(y)$&$-(\frac{D}{2}-1)\;(\frac{D}{2})\;(D^2-4D+4) [AF] + (\frac{D}{2}-1)\;(D-1)\;(D^2-4D+4)  I[A'F]$\\
&$+(\;((\frac{D}{2}-1)^2-1)(2-y) - 2(D-1)\; ) I[A'F'] + (\frac{D}{2}-1) \;(D-1) \; I^{2}[A'F'] + (D-1) y I[A'F']\phantom{sssssssssssssss}$\\
&$-(D-1)\;\frac14 \;(D^2-4D+4) I^2[A'F'] $\\
\hline \hline
$f_{\gamma_1(B)}(y)$&$-(D-1)\;\frac14 \;(D^2-4D+4) I[B'F']$\\
\hline \hline
$f_{\gamma_1(C)}(y)$&$-(\frac{D}{2}-1)\;(\frac{D}{2})\;(D^2-4D+4) [CF] + (\frac{D}{2}-1)\;(D-1)\;(D^2-4D+4)  I[C'F]$\\
&$+(\;((\frac{D}{2}-1)^2-1)(2-y) - 2(D-1)\; ) I[C'F'] + (\frac{D}{2}-1) \;(D-1) \; I^{2}[C'F'] + (D-1) y I[C'F']$\\
&$-(D-1)\;\frac14 \;(D^2-4D+4) I^2[C'F'] $ \\
\hline \hline 
\hline
\multicolumn{2}{|c|}{}\\
\multicolumn{2}{|c|}{${\rm Total\;for\;}f_{\gamma_1}(y)$}\\
\multicolumn{2}{|c|}{}\\
\hline \hline
\multicolumn{2}{|c|}{}\\
\multicolumn{2}{|c|}{$\frac{(D-4)\; H^{2D-4} \pi^{-D} \;\Gamma {(\frac{D}{2}-1)}^2}{8(D-1)} \;(\frac{1}{y})^{D-1}$}\\
\multicolumn{2}{|c|}{}\\
\multicolumn{2}{|c|}{$\frac{(D-2)(D^5-9D^4+16D^3+84D^2-136D+32)\;H^{2D-4}\pi^{-D}\;\Gamma{(\frac{D}{2}-1)}^2}{1024 (D-1)}\;(\frac{1}{y})^{D-2}\;-\;\frac{3H^4}{128\pi ^4} \;\frac{1}{y}.$}\\
\multicolumn{2}{|c|}{}\\
\end{tabular}\label{ktg1}
\end{ruledtabular}
\end{center}
\end{table}
%%%%%%%%%%%%%%%%%%%%%%%%%%%%%%%%%%%%%%%%%%%%%%%%%%%%%%%%%%%%%%%%
\begin{table}[htbp] 
\begin{center}
\caption{For $\gamma_2 \equiv (aa')^{{D}{2}}(a^2+a'^2) H^4 $ type terms.}
\begin{ruledtabular}
\begin{tabular}{|l|l|}
\hline
\multicolumn{2}{|c|}{}\\
\multicolumn{2}{|c|}{$f_{\gamma_2}(y)= C_{A}\;\;f_{\gamma_2(A)}(y)
\;+\;f_{\gamma_2(B)}(y) + C_{C}\;\;f_{\gamma_2(C)}(y)$}\\
\multicolumn{2}{|c|}{}\\
\hline \hline
$f_{\gamma_2(A)}(y)$& $-(\frac{D}{2}-1)\;(\frac{D}{4}) y [AF]' + (\frac{D}{2}-1) \;(D-1) \frac12 y[A'F]\phantom{ssssssssssssssssssssssssssssssssssssssssssssssssssssssss}$\\
&$- \frac12 y (\;(4y-y^2) [A'F']' + D (2-y) [A'F'] \;) + (2-y) \frac12 y [A'F'] -(D-1) y I[A'F']$ \\
\hline \hline
$f_{\gamma_2(B)}(y)$& $-(D-1) y [B'F'] \phantom{sssssssssssssssssssssssssssssssssssssssssssss}$\\
\hline \hline
$f_{\gamma_2(C)}(y)$& $-(\frac{D}{2}-1)\;(\frac{D}{4}) y [CF]' + (\frac{D}{2}-1) \;(D-1) \frac12 y[C'F]$\\
&
$-\frac12 y (\;(4y-y^2) [C'F']' + D (2-y) [C'F'] \;) + (2-y) \frac12 y [C'F'] -(D-1) y I[C'F']$ \\
\hline \hline
\hline
\multicolumn{2}{|c|}{}\\
\multicolumn{2}{|c|}{${\rm Total\;for\;}f_{\gamma_2}(y)$}\\
\multicolumn{2}{|c|}{}\\
\hline \hline
\multicolumn{2}{|c|}{}\\
\multicolumn{2}{|c|}{$-\frac{(D-2)^2 D\;H^{2D-4}\pi^{-D} \;\Gamma{(\frac{D}{2}-1)^2} }{32} \;(\frac{1}{y})^{D-1}$}\\ 
\multicolumn{2}{|c|}{}\\
\multicolumn{2}{|c|}{$-\frac{(\frac{D}{2}-1)^2(D^2-17D+12)\; H^{2D-4} \pi^{-D}\;\Gamma{(\frac{D}{2}-1)^2}}
{64} \;(\frac{1}{y})^{D-2}\;+\;\frac{3H^4}{256 \pi^4}\;\frac{1}{y}.$}\\ 
\multicolumn{2}{|c|}{}\\
\end{tabular}\label{ktg2}
\end{ruledtabular}
\end{center}
\end{table}
%%%%%%%%%%%%%%%%%%%%%%%%%%%%%%%%%%%%%%%%%%%%%%%%%%%%%%%%%%%%%%%%
\begin{table}[htbp]
\begin{center}
\caption{For $\gamma_3 \equiv (aa')^{\frac{D}{2}}(a+a')^2 H^4 $ type terms.}
\begin{ruledtabular}
\begin{tabular}{|l|l|}
\hline
\multicolumn{2}{|c|}{}\\
\multicolumn{2}{|c|}{$f_{\gamma_3}(y)=C_{A}\;\;f_{\gamma_3(A)}(y)\;+\;f_{\gamma_3(B)}(y)\;+\;C_{C}\;\;f_{\gamma_3(C)}(y),$}\\
\multicolumn{2}{|c|}{}\\
\hline \hline
$f_{\gamma_3(A)}(y)$& $-(\frac{D}{2}-1)\;(\frac{D}{4})\;((D-1)y[AF]' + y^2[AF]'') + (\frac{D}{2}-1)(D-1)\; \frac12 ((D-1)y[A'F]+y^2[A'F]') -2 y [A'F']\phantom{ssssssss}$\\
&$+(2-y) \frac12 ((D-1)y[A'F'] + y^2 [A'F']') -(D-1) \frac12 ((D-1) yI[A'F'] + y^2 [A'F'])$\\
\hline \hline
$f_{\gamma_3(B)}(y)$& $-(D-1)\;(\;(D-1)y [B'F'] +y^2[B'F']'\;)$\\
\hline \hline
$f_{\gamma_3(C)}(y)$& $-(\frac{D}{2}-1)\;(\frac{D}{4})\;((D-1)y[CF]' + y^2[CF]'') + (\frac{D}{2}-1)(D-1)\; \frac12 ((D-1)y[C'F]+y^2[C'F]') -2 y [C'F']$\\
&$+(2-y) \frac12 ((D-1)y[C'F'] + y^2 [C'F']') -(D-1) \frac12 ((D-1) yI[C'F'] + y^2 [C'F'])$\\
\hline \hline
\hline
\multicolumn{2}{|c|}{}\\
\multicolumn{2}{|c|}{${\rm Total\;for\;}f_{\gamma_3}(y)$}\\
\multicolumn{2}{|c|}{}\\
\hline \hline
\multicolumn{2}{|c|}{}\\
\multicolumn{2}{|c|}{$\frac{(D-2)^2(D+2)\;H^{2D-4} \pi^{-D}\;\Gamma{(\frac{D}{2}-1)^2}}{64}\;(\frac{1}{y})^{D-1}$}\\ 
\multicolumn{2}{|c|}{}\\
\multicolumn{2}{|c|}{$-\frac{(\frac{D}{2}-1)^2(2D^3-8D^2+9D+14)\;H^{2D-4} \pi^{-D}\;\Gamma{(\frac{D}{2}-1)^2}}{64}\;(\frac{1}{y})^{D-2}\;+\;\frac{3H^4}{256 \pi^4} \;\frac{1}{y}.$}\\ 
\multicolumn{2}{|c|}{}\\
\end{tabular}\label{ktg3}
\end{ruledtabular}
\end{center}
\end{table}
%%%%%%%%%%%%%%%%%%%%%%%%%%%%%%%%%%%%%%%%%%%%%%%%%%%%%%%%%%%%%%%%
\begin{table}[htbp] 
\begin{center}
\caption{For $\delta \equiv (aa')^{\frac{D}{2}-1}(a^2+a'^2) \nabla^2 \square $ type terms.}
\begin{ruledtabular}
\begin{tabular}{|l|c|}
\hline
\multicolumn{2}{|c|}{}\\
\multicolumn{2}{|c|}{$f_{\delta}(y)=C_{A}\;\;f_{\delta (A)}(y)\;+\;f_{\delta (B)}(y)\;+\;C_{C}\;\;f_{\delta (C)}(y),$}\\
\multicolumn{2}{|c|}{}\\
\hline \hline
$f_{\delta (A)}(y)$&$(\frac{D}{2} -1)\;\frac14 I^2[AF'] - I^2[AF''] + \frac14 I^3[A'F'] + C_{A}^{-1} \;I^2[AF'']\phantom{sssssssssssssssssssssssssss}$\\
\hline \hline
$f_{\delta (B)}(y)$&$-\frac12 I^2[BF'']-I^2[BF'']\phantom{sssssssssssssssssssssssssssssssssssssssssssssss}$\\
\hline \hline
$f_{\delta (C)}(y)$&$(\frac{D}{2} -1) \; \frac14 I^2[CF'] - I^2[CF''] + \frac14 I^3[C'F']\phantom{sssssssssssssssssssssssssssssssssssssssss}$\\
\hline \hline
\hline
\multicolumn{2}{|c|}{}\\
\multicolumn{2}{|c|}{${\rm Total\;for\;}f_{\delta}(y)$}\\
\multicolumn{2}{|c|}{}\\
\hline \hline
\multicolumn{2}{|c|}{}\\
\multicolumn{2}{|c|}{$\frac{D\;H^{2D-4} \pi^{-D}\; \Gamma{(\frac{D}{2}-1)^2}}{128(D-1)}\;(\frac{1}{y})^{D-2}\;+\;\frac{H^4}{96\pi^4}\;\frac1{y}.$}\\
\multicolumn{2}{|c|}{}\\
\end{tabular}\label{ktd}
\end{ruledtabular}
\end{center}
\end{table}
%%%%%%%%%%%%%%%%%%%%%%%%%%%%%%%%%%%%%%%%%%%%%%%%%%%%%%%%%%%%%%%%
\begin{table}[htbp] 
\begin{center}
\caption{For $\epsilon_1 \equiv (aa')^{\frac{D}{2}} H^2 \nabla^2 $ type terms.}
\begin{ruledtabular}
\begin{tabular}{|l|l|}
\hline
\multicolumn{2}{|c|}{}\\
\multicolumn{2}{|c|}{$f_{\epsilon_1}(y)=C_{A}\;\;f_{\epsilon_1(A)}(y)\;+\;f_{\epsilon_1(B)}(y)
\;+\;C_{C}\;\;f_{\epsilon_1(C)}(y),$}\\
\multicolumn{2}{|c|}{}\\
\hline \hline
$f_{\epsilon_1(A)}(y)$& 
$ -C_{A}^{-1}\; 2(D-1) I[A'F'] -(\frac{D}{2}-1)\; \frac{D}{4} (\frac12 I[AF] - [AF])$\\
& $+ (\frac{D}{2}-1)\frac18 (D^2-4D+4) I^2[AF'] + (\frac{D}{2}-1) \;(D-1) (\frac12 I^2[A'F] - I[A'F]) $\\
& $+((\frac{D}{2}-1)^2 - 1 -(\frac{D}{2} -1)) \frac12 I^3[A'F'] - \frac12 I^2[yA'F'] + 2I^2[A'F'] + (-2I[A'F'] + I^2[y [A'F']']) \phantom{sssssssssssssssss}$\\
& $-(D-1) (\frac12 I^3[A'F'] - I^2[A'F']) - C_{A}^{-1} \;(\;(\frac{D}{2} -1)(D-1) I[AF'] + (D-1) I^2[A'F']\;)$\\
\hline \hline
$f_{\epsilon_1(B)}(y)$&
$ -(\frac{D}{2}-1)^2 I^2[BF''] - 2(D-1) (\frac12 I^2[B'F'] - I[B'F']) + (\frac{D}{2} -1)\; I[BF'] $\\
&$+ (\frac{D}{2}) I^2[B'F'] - (\frac{D}{2} -1)^2 [BF] + 2(D-1) I[B'F']$\\
\hline \hline
$f_{\epsilon_1(C)}(y)$& $-(\frac{D}{2}-1)\; \frac{D}{4} (\frac12 I[CF] - [CF])$\\
& $+ (\frac{D}{2}-1)\frac18 (D^2-4D+4) I^2[CF'] + (\frac{D}{2}-1) \;(D-1) (\frac12 I^2[C'F] - I[C'F]) $\\
& $+((\frac{D}{2}-1)^2 - 1 -(\frac{D}{2} -1)) \frac12 I^3[C'F'] - \frac12 I^2[yC'F'] + 2I^2[C'F'] + (-2I[C'F'] + I^2[y [C'F']']) $\\
& $-(D-1) (\frac12 I^3[C'F'] - I^2[C'F'])$\\
\hline \hline
\hline
\multicolumn{2}{|c|}{}\\
\multicolumn{2}{|c|}{${\rm Total\;for\;}f_{\epsilon_1}(y)$}\\
\multicolumn{2}{|c|}{}\\
\hline \hline
\multicolumn{2}{|c|}{}\\
\multicolumn{2}{|c|}{$-\frac{(D-2)^2 D \;H^{2D-4} \pi^{-D}\;\Gamma{(\frac{D}{2}-1)^2}}{32(D-1)}\;(\frac{1}{y})^{D-1}$}\\
\multicolumn{2}{|c|}{}\\
\multicolumn{2}{|c|}{
$-\frac{(D-2)(D^3-16D+16) \;H^{2D-4} \pi^{-D}\;\Gamma{(\frac{D}{2}-1)^2}}{256(D-1)}\;(\frac{1}{y})^{D-2} \;-\;\frac{H^4}{192\pi^4}\;\frac1{y}.$}\\ 
\multicolumn{2}{|c|}{}\\
\end{tabular}\label{kte1}
\end{ruledtabular}
\end{center}
\end{table}
%%%%%%%%%%%%%%%%%%%%%%%%%%%%%%%%%%%%%%%%%%%%%%%%%%%%%%%%%%%%%%%%
\begin{table}[htbp] 
\begin{center}
\caption{For $\epsilon_2 \equiv (aa')^{\frac{D}{2}-1}(a^2 + a^{\prime 2}) H^2 \nabla^2 $ type terms.}
\begin{ruledtabular}
\begin{tabular}{|l|c|}
\hline
\multicolumn{2}{|c|}{}\\
\multicolumn{2}{|c|}{$f_{\epsilon_2}(y)=C_{A}\;\;f_{\epsilon_2(A)}(y)\;+\;f_{\epsilon_2(B)}(y)\;+\;C_{C}\;\;f_{\epsilon_2(C)}(y),$}\\
\multicolumn{2}{|c|}{}\\
\hline \hline
$f_{\epsilon_2(A)}(y)$& $ (\frac{D}{2}-1)\;\frac14 yI[AF'] + \frac14 I^2[AF'] - C_{A}^{-1} \; (\frac{D}{2}) ((\frac{D}{2}-1)I^[AF']+I^2[A'F'])\phantom{sssssssssssssssssssssssssssssssssssssss}$\\
\hline \hline
$f_{\epsilon_2(B)}(y)$& $ -\frac12 yI[BF''] - (I[BF'] + I[yBF'']) +(\frac{D}{2}) I^2[B'F'] -\frac12 (\;(4y-y^2)[BF''] + D(2-y) [BF']\;)-\frac12 y[BF]\phantom{ssssssss}$\\
\hline \hline
$f_{\epsilon_2(C)}(y)$& $ (\frac{D}{2}-1)\;\frac14 yI[CF'] + \frac14 I^2[CF'] \phantom{sssssssssssssssssssssssssssssssssssssssssssssssssssssssssssssssssssssssssss}$\\
\hline \hline
\hline
\multicolumn{2}{|c|}{}\\
\multicolumn{2}{|c|}{${\rm Total\;for\;}f_{\epsilon_2}(y)$}\\
\multicolumn{2}{|c|}{}\\
\hline \hline
\multicolumn{2}{|c|}{}\\
\multicolumn{2}{|c|}{$-\frac{(D-2)(D(2D-7)+4)\;H^{2D-4} \pi^{-D}\;\Gamma{(\frac{D}{2}-1)^2}}{128(D-1)}  \;(\frac{1}{y})^{D-2}\;-\;\frac{H^4}{96 \pi^4}\;\frac{1}{y}.$}\\ 
\multicolumn{2}{|c|}{}\\
\end{tabular}\label{kte2}
\end{ruledtabular}
\end{center}
\end{table}
%%%%%%%%%%%%%%%%%%%%%%%%%%%%%%%%%%%%%%%%%%%%%%%%%%%%%%%%%%%%%%%%
\begin{table}[htbp] 
\begin{center}
\caption{For $\epsilon_3 \equiv (aa')^{\frac{D}{2}-1}(a + a')^2 H^2 \nabla^2 $ type terms.}
\begin{ruledtabular}
\begin{tabular}{|l|c|}
\hline
\multicolumn{2}{|c|}{}\\
\multicolumn{2}{|c|}{$f_{\epsilon_3}(y)=C_{A}\;\;f_{\epsilon_3(A)}(y)\;+\;f_{\epsilon_3(B)}(y)\;+\;C_{C}\;\;f_{\epsilon_3(C)}(y),$}\\
\multicolumn{2}{|c|}{}\\
\hline \hline
$f_{\epsilon_3(A)}(y)$& $(\frac{D}{2}-1) \frac12 (\frac12 (D-1) y I[AF'] + \frac12 y^2 [AF'])+ \frac12 \;(\frac12 (D-1) y I^[A'F'] +\frac12 y^2 I[A'F']) \phantom{sssssssssssssssssssssssssssss}$\\
&$- C_{A}^{-1} \;(\; (\frac{D}{2}-1) y[AF'] + y I[A'F']\;)\phantom{sssssssssssssssssssssssssssssssssssssssssssssssssssssssssssssssssssss}$\\
\hline \hline
$f_{\epsilon_3(B)}(y)$& $-\frac12 ((D-1)yI[BF''] + y^2 [BF'']) - (\frac{D}{2}-1) \;(I[BF'] + I[yBF''])
-y[BF'] -y^2 [BF''] +yI[B'F']\phantom{ssssssssssss}$\\
&$ -\frac12 ((D-1)y[BF']+y^2[BF''])\phantom{sssssssssssssssssssssssssssssssssssssssssssssssssssssssssssssssssssssssss}$\\
\hline \hline
$f_{\epsilon_3(C)}(y)$& $(\frac{D}{2}-1) \frac12 (\frac12 (D-1) y I[AF'] +\frac12 y^2 [AF']) + \frac12 (\frac12 (D-1) yI[A'F'] +\frac12 y^2 I[A'F']) \phantom{ssssssssssssssssssssssssssss}$\\
\hline \hline
\hline
\multicolumn{2}{|c|}{}\\
\multicolumn{2}{|c|}{${\rm Total\;for\;}f_{\epsilon_3}(y)$}\\
\multicolumn{2}{|c|}{}\\
\hline \hline
\multicolumn{2}{|c|}{}\\
\multicolumn{2}{|c|}{$-\frac{(D-2)^2 \;H^{2D-4} \pi^{-D}\;\Gamma{(\frac{D}{2}-1)^2}}{64}  \;(\frac{1}{y})^{D-1} \;-\;\frac{5H^4}{384 \pi^4}\;\frac{1}{y}.$}\\ 
\multicolumn{2}{|c|}{}\\
\end{tabular}\label{kte3}
\end{ruledtabular}
\end{center}
\end{table}
%%%%%%%%%%%%%%%%%%%%%%%%%%%%%%%%%%%%%%%%%%%%%%%%%%%%%%%%%%%%%%%%
\begin{table}[htbp] 
\begin{center}
\caption{For $\zeta \equiv (aa')^{\frac{D}{2}-1}\nabla^4 $ type terms.}
\begin{ruledtabular}
\begin{tabular}{|l|c|}
\hline
\multicolumn{2}{|c|}{}\\
\multicolumn{2}{|c|}{$f_{\zeta}(y)=C_{A}\;\;f_{\zeta (A)}(y)\;+\;f_{\zeta (B)} (y)\;+\;C_{C}\;\;f_{\zeta (C)}(y),$}\\
\multicolumn{2}{|c|}{}\\
\hline \hline
$f_{\zeta (A)}(y)$&$ -C_{A}^{-1} \; I^2[AF''] + (\frac{D}{2} -1) \frac12 (\frac12 I^3[AF'] - I^2[AF'])
+I^2[AF''] -\frac12 I^3[A'F'] + (\frac{D}{2} -1) \frac12 I^2[AF']\phantom{ssssssssssssss}$\\
& $+ C_{A}^{-1} \;(-2 I^2[AF''] + \frac12 I^3[A'F'])\phantom{ssssssssssssssssssssssssssssssssssssssssssssssssssssssssssssssssssssss}$\\
\hline \hline
$f_{\zeta (B)}(y)$&$ I^2[BF''] - (\frac{D}{2} -1) \frac12 I^2[BF'] + (2I^2[BF'']-\frac12 I^3[B'F']) + I^2[BF'']\phantom{sssssssssssssssssssssssssssssssssssss}$\\
\hline \hline
$f_{\zeta (C)}(y)$&$ (\frac{D}{2} -1) \frac12 (\frac12 I^3[CF'] - I^2[CF'])
+I^2[CF''] -\frac12 I^3[C'F'] + (\frac{D}{2} -1) \frac12 I^2[CF'] \phantom{sssssssssssssssssssssssssssssss}$\\
\hline \hline
\hline
\multicolumn{2}{|c|}{}\\
\multicolumn{2}{|c|}{${\rm Total\;for\;}f_{\zeta}(y)$}\\
\multicolumn{2}{|c|}{}\\
\hline \hline
\multicolumn{2}{|c|}{}\\
\multicolumn{2}{|c|}{$-\frac{H^4}{48\pi^4}\;\frac{1}{y}.$}\\
\multicolumn{2}{|c|}{}\\
\end{tabular}\label{ktz}
\end{ruledtabular}
\end{center}
\end{table}
%%%%%%%%%%%%%%%%%%%%%%%%%%%%%%%%%%%%%%%%%%%%%%%%%%%%%%%%%%%%%%%%

\newpage

\clearpage

\newpage

%%%%%%%%%%%%%%%%%%%%%%%%%%%%%%%%%%%%%%%%%%%%%%%%%%%%%%%%%%%%%%%%
\section{Tables for the coefficient functions of external operators for 3-point cross part}\label{app2}
%%%%%%%%%%%%%%%%%%%%%%%%%%%%%%%%%%%%%%%%%%%%%%%%%%%%%%%%%%%%%%%%
\begin{table}[htbp]
\caption{For $\alpha \equiv (aa')^{\frac{D}{2}+1} \square^2$ type terms.}
\begin{ruledtabular}
\begin{tabular}{|l|l|}
\hline
\multicolumn{2}{|c|}{}\\
  \multicolumn{2}{|c|}{$ f_{\alpha}(y) =  C_{4} \;\;f_{\alpha(4A)} (y) - C_{4} \;\;f_{\alpha(4C)} (y),$}\\
  \multicolumn{2}{|c|}{}\\
\hline \hline
$f_{\alpha(4A)}(y)$&$ 2 I^3[F A^{(3)}] - 2 I^2[F A^{(2)}]\phantom{sssssssssssssssssssssssssssssssssssssssssssssssss}$\\
\hline \hline
$f_{\alpha(4C)}(y)$&$ 2 I^3[F C^{(3)}] - 2 I^2[F C^{(2)}]$\\
\hline \hline
\hline
\multicolumn{2}{|c|}{}\\
\multicolumn{2}{|c|}{${\rm Total\;for\;}f_{\alpha}(y)$}\\
\multicolumn{2}{|c|}{}\\
\hline \hline
%\multicolumn{2}{|c|}{}\\
\multicolumn{2}{|c|}{$0.$}\\
%\multicolumn{2}{|c|}{}\\
\end{tabular}\label{cta}
\end{ruledtabular}
\end{table}

%%%%%%%%%%%%%%%%%%%%%%%%%%%%%%%%%%%%%%%%%%%%%%%%%%%%%%%%%%%%%%%%

\begin{table}[htbp]
\caption{For $\beta_1 \equiv (aa')^{\frac{D}{2}+1} H^2 \square$ type terms.}
\begin{ruledtabular}
\begin{tabular}{|l|l|}
\hline
\multicolumn{2}{|c|}{}\\
\multicolumn{2}{|c|}{$f_{\beta_1}(y) = C_{4}\;\;f_{\beta_1(4A)} (y) - C_{4}\;\;f_{\beta_1(4C)} (y),$}\\
\multicolumn{2}{|c|}{}\\
\hline \hline
$f_{\beta_1(4A)}(y)$&$2(D-2)(D-1)\;(I^3 [F A^{(3)}]-I^2 [F A^{(2)}]) \;+8\;I[F'A']\phantom{sssssssssssssssssssssssss}$\\
\hline \hline
$f_{\beta_1(4C)}(y)$&$2(D-2)(D-1)\;(I^3 [F C^{(3)}]-I^2 [F C^{(2)}]) \;+8\;I[F'C']$\\
\hline \hline
\hline
\multicolumn{2}{|c|}{}\\
\multicolumn{2}{|c|}{${\rm Total\;for\;}f_{\beta_1}(y)$}\\
\multicolumn{2}{|c|}{}\\
\hline \hline
\multicolumn{2}{|c|}{}\\
\multicolumn{2}{|c|}{$-\frac{(D-1)\;H^{2D-4}\pi^{-D}\;\Gamma{(\frac{D}{2}-1)^2}}{2} \; (\frac1{y})^{D-2}\;-\;\frac{3H^4}{2\pi^4}\;\frac{1}{y}.$}\\
\multicolumn{2}{|c|}{}\\
\end{tabular}\label{ctb1}
\end{ruledtabular}
\end{table}

%%%%%%%%%%%%%%%%%%%%%%%%%%%%%%%%%%%%%%%%%%%%%%%%%%%%%%%%%%%%%%%%

\begin{table}[htbp] 
\caption{For $\beta_2 \equiv  (aa')^{\frac{D}{2}} (a^2+a'^2) H^2 \square$ type terms.}
\begin{ruledtabular}
\begin{tabular}{|l|l|}
\hline
\multicolumn{2}{|c|}{}\\
\multicolumn{2}{|c|}{$f_{\beta_2}(y) = - C_{2}\;\;f_{\beta_2(3A)} (y) + C_{4}\;\;f_{\beta_2(4A)} (y) + C_{3}\;\;f_{\beta_2(3C)} (y) 
 - C_{4}\;\;f_{\beta_2(4C)} (y),$}\\
\multicolumn{2}{|c|}{}\\
\hline \hline
$f_{\beta_2(3A)}(y)$&$\phantom{ssssssssssss}-2(D-1)\;\;I [F' A']$\\
\hline \hline
$f_{\beta_2(4A)}(y)$&$-(D-2)^2\;\;(I^3 [F A^{(3)}] - I^2 [F A^{(2)}]) - 2\;\;I[F'A'] - \frac12 (D-2)^2\;\;[F A ]\phantom{sssssssssssssssssssss}$\\
\hline \hline
$f_{\beta_2(3C)}(y)$&$\phantom{ssssssssssss}-2(D-1)\;\;I [F' C']$\\
\hline \hline
$f_{\beta_2(4C)}(y)$&$-(D-2)^2\;\;(I^3 [F C^{(3)}] - I^2 [F C^{(2)}]) - 2\;\;I[F'C'] - \frac12 (D-2)^2\;\;[F C ]\phantom{sssssssssssssssssssss}$\\
\hline \hline
\hline
\multicolumn{2}{|c|}{}\\
\multicolumn{2}{|c|}{${\rm Total\;for\;}f_{\beta_2}(y)$}\\
\multicolumn{2}{|c|}{}\\
\hline \hline
\multicolumn{2}{|c|}{}\\
\multicolumn{2}{|c|}{$-\frac{(D-2)^2\;H^{2D-4}\pi^{-D}\Gamma{(\frac{D}{2}-1)^2}}{8} \;(\frac1{y})^{D-1}$}\\
\multicolumn{2}{|c|}{}\\
\multicolumn{2}{|c|}{
$-\frac{(\frac{D}{2}-1)\;\frac{D}{2}\;(D-1) \;H^{2D-4} \pi^{-D}\;\Gamma{(\frac{D}{2} -1)^2}}{16}\;(\frac1{y})^{D-2}\;+\;0\;\frac{1}{y}.$}\\
\multicolumn{2}{|c|}{}\\
\end{tabular}\label{ctb2}
\end{ruledtabular}
\end{table}

%%%%%%%%%%%%%%%%%%%%%%%%%%%%%%%%%%%%%%%%%%%%%%%%%%%%%%%%%%%%%%%%

\begin{table}[htbp] 
\begin{center}
\caption{For $\gamma_1 \equiv (aa')^{\frac{D}{2}+1} H^4 $ type terms.}
\begin{ruledtabular}
\begin{tabular}{|l|l|}
\hline
\multicolumn{2}{|c|}{}\\
\multicolumn{2}{|c|}{$f_{\gamma_1}(y) = -C_{1}\;\;f_{\gamma_1(2A)} (y) - C_{2}\;\;f_{\gamma_1(3A)}(y) + C_{4}\;\;f_{\gamma_1(4A)}(y) + C_{1} \;\;f_{\gamma_1(B)} (y) + C_{3}\;\;f_{\gamma_1(3C)} (y) - C_{4}\;\;f_{\gamma_1(4C)} (y), $}\\
\multicolumn{2}{|c|}{}\\
\hline \hline
$f_{\gamma_1(2A)}(y)$& $-16(D-1)(yI[F'A'']) -48(D-1) I^2[F'A''] + 16(D-1) I[yF'A''] $\\
\hline \hline
$f_{\gamma_1(3A)}(y)$& $-(D-4)(D-1)^2 I[F'A'] + 2(D-1)(2D-17)I^2 [F'A''] + (D-2)(D-1)^2 I^2[F''A']$\\
& $ - 2(D-1)(D+5)(yI[F'A'']) + 8D(D-1) I[F'A''] $\\
\hline \hline
& $-\frac{D}{2} (D-2) [(4y-y^2)FA'' + D(2-y) FA'] + \frac{D}{4} (D-2)^2(D-1) [FA] - \frac{D}{4} (D-2)^2 (D-1) I[F'A]$\\
& $+\frac12 (D-2)(D-1) [(4y-y^2)FA'' + D(2-y) FA'] -\frac{1}{2} (D-2)^2 (y[FA'] + y I[FA''-F''A]) $\\
&$-\frac14 (D-2)^3 (D-1) ([FA] + I^2[FA'' - F''A]) + (D-2)(D-1)^2 I[F'A'] $\\
&$+\frac12 (D-2)(D-1) I[(4y-y^2) F'A'' + D(2-y) F'A'] - \frac12 (D-2)^2 (y[FA]' -y I[FA''- F''A]) $\\
&$-\frac14 (D-2)^3 (D-1) ([FA]-I^2[FA''-F''A]) + (D-2)^2 (2-y) I[F'A'] - (D-2)^2(D-1) I^2[F'A']$\\
&$+6(D-2)(D-1) I^2[F'A''] + (D-2)(D-1)^2 I^2[F''A']$\\
$f_{\gamma_1(4A)}(y)$& $-(D-2)^3 (D-1) (I^3[FA^{(3)}]-I^2[FA^{(2)}]) +2(D-2)^2(yI^2[FA^{(3)}] - yI[FA^{(2)}])
-2(D-1) (I^3[F'A''] - I^2[F'A'])$\\
&$+8(yI[F'A''] - y[F'A']) -2(D-1) (1-y) (I^2 [F'A''] - I[F'A']) + (4y^3-16y^2) [F'A']' + D(4y-8) (y[F'A'])$ \\
&$+2(D-2)(D-1) (yI[F'A']) -2(D-2)(D-1) I^2[F'A'] -6(D-2)(D-1) I[F'A'] +2(D-1)(D+3) I^3[F'A'']$ \\
&$-24(y^2+1) I[F'A'']-8(D-1) I[yI[F'A'']] + 8(y^2[F'A'']) -2(D-1)^2(yI^2[F'A'']) - (14D-86) (yI[F'A'']) $\\
&$+ (D-1)(12D-80) I^2[F'A''] + 24(D-1) (yI[F'A'']) - 24(D-1) I[F'A''] -24(D-1) I^2[F'A'']$\\
\hline \hline
& $-60(D-1) I^2[F'B''] -4(D-1) (yI[F'B''] +4(D-1) I[yF'B''] $\\
$f_{\gamma_1(B)}(y)$&$ -(D-1)(D(D-1)-4) I[F'B'] + 2(D-1) (2D-23) I^2 [F'B''] + (D-2)(D-1)^2 I^2[F'' B'] $\\
& $+2(D^2-1) (yI[F'B''])+8D(D-1) I[F'B''] +2(D-2)(D-1) I[F'B']$\\
\hline \hline
$f_{\gamma_1(3C)}(y)$&  $-(D-4)(D-1)^2 I[F'C'] + 2(D-1)(2D-17)I^2 [F'C''] + (D-2)(D-1)^2 I^2[F''C']$\\
& $ - 2(D-1)(D+5)(yI[F'C'']) + 8D(D-1) I[F'C''] $\\
\hline \hline
& $-\frac{D}{2} (D-2) [(4y-y^2)FC'' + D(2-y) FC'] + \frac{D}{4} (D-2)^2(D-1) [FC] - \frac{D}{4} (D-2)^2 (D-1) I[F'C]$\\
& $+\frac12 (D-2)(D-1) [(4y-y^2)FC'' + D(2-y) FC'] -\frac{1}{2} (D-2)^2 (y[FC'] + y I[FC''-F''C]) $\\
&$-\frac14 (D-2)^3 (D-1) ([FC] + I^2[FC'' - F''C]) + (D-2)(D-1)^2 I[F'C'] $\\
&$+\frac12 (D-2)(D-1) I[(4y-y^2) F'C'' + D(2-y) F'C'] - \frac12 (D-2)^2 (y[FC]' -y I[FC''- F''C]) $\\
&$-\frac14 (D-2)^3 (D-1) ([FC]-I^2[FC''-F''C]) + (D-2)^2 (2-y) I[F'C'] - (D-2)^2(D-1) I^2[F'C']$\\
&$+6(D-2)(D-1) I^2[F'C''] + (D-2)(D-1)^2 I^2[F''C']$\\
$f_{\gamma_1(4C)}(y)$& $-(D-2)^3 (D-1) (I^3[FC^{(3)}]-I^2[FC^{(2)}]) +2(D-2)^2(yI^2[FC^{(3)}] - yI[FC^{(2)}])
-2(D-1) (I^3[F'C''] - I^2[F'C'])$\\
&$+8(yI[F'C''] - y[F'C']) -2(D-1) (1-y) (I^2 [F'C''] - I[F'C']) + (4y^3-16y^2) [F'C']' + D(4y-8) (y[F'C'])$ \\
&$+2(D-2)(D-1) (yI[F'C']) -2(D-2)(D-1) I^2[F'C'] -6(D-2)(D-1) I[F'C'] +2(D-1)(D+3) I^3[F'C'']$ \\
&$-24(y^2+1) I[F'C'']-8(D-1) I[yI[F'C'']] + 8(y^2[F'C'']) -2(D-1)^2(yI^2[F'C'']) - (14D-86) (yI[F'C'']) $\\
&$+ (D-1)(12D-80) I^2[F'C''] + 24(D-1) (yI[F'C'']) - 24(D-1) I[F'C''] -24(D-1) I^2[F'C'']$\\
\hline \hline 
\hline
\multicolumn{2}{|c|}{}\\
\multicolumn{2}{|c|}{${\rm Total\;for\;}f_{\gamma_1}(y)$}\\
\multicolumn{2}{|c|}{}\\
\hline \hline
\multicolumn{2}{|c|}{}\\
\multicolumn{2}{|c|}{$-\frac{(D^2+3D+1)(\frac{D}{2}-1)^2 \;H^{2D-4} \pi^{-D} \;\Gamma {(\frac{D}{2}-1)}^2 }{2}\;(\frac{1}{y})^{D-1}$}\\
\multicolumn{2}{|c|}{}\\
\multicolumn{2}{|c|}{$-\frac{(D-2)(D^4+5D^3-57D^2-162D+348)\; H^{2D-4} \pi^{-D}\;\Gamma {(\frac{D}{2}-1)}^2}{64} \;(\frac{1}{y})^{D-2}\;+\;
\frac{3H^4}{32\pi^4}\;\frac{1}{y}.$}\\
\multicolumn{2}{|c|}{}\\
\end{tabular}\label{ctg1}
\end{ruledtabular}
\end{center}
\end{table}

%%%%%%%%%%%%%%%%%%%%%%%%%%%%%%%%%%%%%%%%%%%%%%%%%%%%%%%%%%%%%%%%
\clearpage
%%%%%%%%%%%%%%%%%%%%%%%%%%%%%%%%%%%%%%%%%%%%%%%%%%%%%%%%%%%%%%%%

\begin{table}[htbp] 
\begin{center}
\caption{For $\gamma_2 \equiv (aa')^{\frac{D}{2}}(a^2+a'^2) H^4 $ type terms.}
\begin{ruledtabular}
\begin{tabular}{|l|l|}
\hline
\multicolumn{2}{|c|}{}\\
\multicolumn{2}{|c|}{$f_{\gamma_2}(y)=-C_{1}\;\;f_{\gamma_2(2A)}(y)
- C_{2}\;\;f_{\gamma_2(3A)}(y) + C_{4}\;\;f_{\gamma_2(4A)}(y) 
+ C_{3}\;\;f_{\gamma_2(3C)}(y) - C_{4}\;\;f_{\gamma_2(4C)}(y),$}\\
\multicolumn{2}{|c|}{}\\
\hline \hline
$f_{\gamma_2(2A)}(y)$& $-8(D-1) y I[F' A''] + 8(D-1) I^2 [F'A''] + 8(D-1) I[yF'A'']$ \\
\hline \hline
$f_{\gamma_2(3A)}(y)$& $-6(D-1) y I[F' A''] + 8(D-1) I^2 [F'A''] +\frac12 (D^2-2)(D-1) I[F'A']$\\
\hline \hline
$f_{\gamma_2(4A)}(y)$& $\frac{D}{2} (D-2)(D-1) I[F'A'] -(D-2)^2 y[FA]' + 4(yI[F'A''] -y[F'A'])-14(y I[F'A'']) + 16(D-1) I^2[F'A''] $\\
&$+4(y^2 [F'A'']) + 4 (D-1) (yI[F'A''])$\\
\hline \hline
$f_{\gamma_2(3C)}(y)$&  $-6(D-1) y I[F' C''] + 8(D-1) I^2 [F'C''] +\frac12 (D^2-2)(D-1) I[F'C']$\\
\hline \hline
$f_{\gamma_2(4C)}(y)$& $\frac{D}{2} (D-2)(D-1) I[F'C'] -(D-2)^2 y[FC]' + 4(yI[F'C''] -y[F'C'])-14(y I[F'C'']) + 16(D-1) I^2[F'C''] $\\
&$+4(y^2 [F'C'']) + 4 (D-1) (yI[F'C''])$\\
\hline \hline
\hline
\multicolumn{2}{|c|}{}\\
\multicolumn{2}{|c|}{${\rm Total\;for\;}f_{\gamma_2}(y)$}\\
\multicolumn{2}{|c|}{}\\
\hline \hline
\multicolumn{2}{|c|}{}\\
\multicolumn{2}{|c|}{$\frac{(D^2+6D-2)(\frac{D}{2}-1)^2\;H^{2D-4} \pi^{-D}\;\Gamma{(\frac{D}{2}-1)^2}}{8} \;(\frac{1}{y})^{D-1}$}\\ 
\multicolumn{2}{|c|}{}\\
\multicolumn{2}{|c|}{$+\frac{(D-2)(D^4-3D^3+4D^2-212D-304)\;H^{2D-4} \pi^{-D}\;\Gamma{(\frac{D}{2}-1)^2}}{256}  \;(\frac{1}{y})^{D-2}.$}\\ 
\multicolumn{2}{|c|}{}\\
\end{tabular}\label{ctg2}
\end{ruledtabular}
\end{center}
\end{table}

%%%%%%%%%%%%%%%%%%%%%%%%%%%%%%%%%%%%%%%%%%%%%%%%%%%%%%%%%%%%%%%%

\begin{table}[htbp]
\begin{center}
\caption{For $\gamma_3 \equiv (aa')^{\frac{D}{2}}(a+a')^2 H^4 $ type terms.}
\begin{ruledtabular}
\begin{tabular}{|l|l|}
\hline
\multicolumn{2}{|c|}{}\\
\multicolumn{2}{|c|}{$f_{\gamma_3}(y) = -C_{2}\;\;f_{\gamma_3(3A)}(y) + C_{4}\;\;f_{\gamma_3(4A)}(y) 
+ C_{1}\;\;f_{\gamma_3(B)}(y) + C_{3}\;\;f_{\gamma_3(3C)}(y) - C_{4}\;\;f_{\gamma_3(4C)}(y)  , $}\\
\multicolumn{2}{|c|}{}\\
\hline \hline
$f_{\gamma_3(3A)}(y)$& $-(D-4)(D-1) (yI[F'A']) + (D-2)(D-1) (I[F'A'] + I[yF''A']) $\\
& $-2(D-1) (yI[F'A''] + y^2I[F'A'']) + 4(D-1) I[yF'A'']\phantom{ssssssssssssssssssssssssssssss}$\\
\hline \hline
& $ -\frac14 (D-2)^2 (y^2 [FA]'' + y^2 [FA'' - F''A]) - \frac14 (D-3)^2 (y [FA]' + y I[FA'' - F''A]) + (D-2)(D-1) (y [F'A'])$\\
& $+\frac14 (D-2)^2(D-1) (yI[FA''-F''A] - y [FA]') + (D-2)(D-1) I[(yF')'A']$\\
$f_{\gamma_3(4A)}(y)$& $ -D(D-2)^2 (yI^2[FA^{(3)}] -y I[FA^{(2)}]) - (D-2)^2 (y^2 I[FA^{(3)}] - y^2 [FA^{(2)}])$\\
& $ - 2(yI[F'A'']-y[F'A']) - 2(D-1) (y[F'A']) + 4(D-1) I[yF'A''] -12 (y[F'A'']) + 2 (y^2[F'A''])$\\
\hline \hline
$f_{\gamma_3(B)}(y)$& $ 4(D-1) (y[F'B''])$\\
\hline \hline
$f_{\gamma_3(3C)}(y)$& $-(D-4)(D-1) (yI[F'C']) + (D-2)(D-1) (I[F'C'] + I[yF''C']) $\\
&$-2(D-1) (yI[F'C''] + y^2I[F'C'']) + 4(D-1) I[yF'C'']\phantom{ssssssssssssssssssssssssssssss}$\\
\hline \hline
& $ -\frac14 (D-2)^2 (y^2 [FC]'' + y^2 [FC'' - F''C]) - \frac14 (D-3)^2 (y [FC]' + y I[FC'' - F''C]) + (D-2)(D-1) (y [F'C'])$\\
& $+\frac14 (D-2)^2(D-1) (yI[FC''-F''C] - y [FC]') + (D-2)(D-1) I[(yF')'C']$\\
$f_{\gamma_3(4C)}(y)$& $ -D(D-2)^2 (yI^2[FC^{(3)}] -y I[FC^{(2)}]) - (D-2)^2 (y^2 I[FC^{(3)}] - y^2 [FC^{(2)}])$\\
& $ - 2(yI[F'C'']-y[F'C']) - 2(D-1) (y[F'C']) + 4(D-1) I[yF'C''] -12 (y[F'C'']) + 2 (y^2[F'C''])$\\
\hline \hline
\hline
\multicolumn{2}{|c|}{}\\
\multicolumn{2}{|c|}{${\rm Total\;for\;}f_{\gamma_3}(y)$}\\
\multicolumn{2}{|c|}{}\\
\hline \hline
\multicolumn{2}{|c|}{}\\
\multicolumn{2}{|c|}{$\frac{(D^2+4D-10)(\frac{D}{2}-1)^2 \;H^{2D-4} \pi^{-D}\;\Gamma{(\frac{D}{2}-1)^2}}{8} \;(\frac{1}{y})^{D-1}$}\\ 
\multicolumn{2}{|c|}{}\\
\multicolumn{2}{|c|}{$+\frac{(D-2)(D^4+25D^3-204D^2+372D-192)\;H^{2D-4} \pi^{-D}\;\Gamma{(\frac{D}{2}-1)^2}}{256}  \;(\frac{1}{y})^{D-2}.$}\\ 
\multicolumn{2}{|c|}{}\\
\end{tabular}\label{ctg3}
\end{ruledtabular}
\end{center}
\end{table}

%%%%%%%%%%%%%%%%%%%%%%%%%%%%%%%%%%%%%%%%%%%%%%%%%%%%%%%%%%%%%%%%
\clearpage
%%%%%%%%%%%%%%%%%%%%%%%%%%%%%%%%%%%%%%%%%%%%%%%%%%%%%%%%%%%%%%%%

\begin{table}[htbp]
\caption{For $\delta \equiv (aa')^{\frac{D}{2}-1}(a^2+a'^2) H^2 \nabla^2 \square $ type terms.}
\begin{ruledtabular}
\begin{tabular}{|l|l|}
\hline
\multicolumn{2}{|c|}{}\\
\multicolumn{2}{|c|}{$f_{\delta}(y) = -C_{1}\;\;f_{\delta (2A)}(y) -C_{2}\;\;f_{\delta (3A)}(y)+ C_{4}\;\;f_{\delta (4A)}(y) + C_{1}\;\;f_{\delta (B)}(y) + C_{3}\;\;f_{\delta (3C)}(y) - C_{4}\;\;f_{\delta (4C)}(y) , $}\\
\multicolumn{2}{|c|}{}\\
\hline \hline
$f_{\delta (2A)}(y)$&$\phantom{ssssssssssssssss}I^3 [ F A^{(3)}]- I^2[ F A^{(2)}]$\\
\hline \hline
$f_{\delta (3A)}(y)$&$\phantom{ssssssssssssssss}I^3 [ F A^{(3)}]- I^2[ F A^{(2)}]$\\
\hline \hline
$f_{\delta (4A)}(y)$&$\phantom{ssssssssssssss}2(I^3 [ F A^{(3)}]- I^2[ F A^{(2)}])$\\
\hline \hline
$f_{\delta (B)}(y)$&$\phantom{ssssssssssssss}2(I^3 [ F B^{(3)}]- I^2[ F B^{(2)}])$\\
\hline \hline
$f_{\delta (3C)}(y)$&$\phantom{ssssssssssssssss}I^3 [ F C^{(3)}]- I^2[ F C^{(2)}]$\\
\hline \hline
$f_{\delta (4C)}(y)$&$\phantom{ssssssssssssss}2(I^3 [ F C^{(3)}]- I^2[ F C^{(2)}])$\\
\hline \hline
\hline
\multicolumn{2}{|c|}{}\\
\multicolumn{2}{|c|}{${\rm Total\;for\;}f_{\delta}(y)$}\\
\multicolumn{2}{|c|}{}\\
\hline \hline 
\multicolumn{2}{|c|}{}\\
\multicolumn{2}{|c|}{$-\frac{H^4}{16\pi^4}\;\frac{1}{y}.$}\\
\multicolumn{2}{|c|}{}\\
\end{tabular}\label{ctd}
\end{ruledtabular}
\end{table}

%%%%%%%%%%%%%%%%%%%%%%%%%%%%%%%%%%%%%%%%%%%%%%%%%%%%%%%%%%%%%%%

\begin{table}[htbp] 
\begin{center}
\caption{For $\epsilon_1 \equiv (aa')^{\frac{D}{2}} H^2 \nabla^2 $ type terms.}
\begin{ruledtabular}
\begin{tabular}{|l|l|}
\hline
\multicolumn{2}{|c|}{}\\
\multicolumn{2}{|c|}{$f_{\epsilon_1}(y) = C_{1} \;\;f_{\epsilon_1(1A)}(y) - C_{2} \;\;f_{\epsilon_1(2A)}(y)
+ C_{2} \;\;f_{\epsilon_1(3A)}(y) + C_{4} \;\;f_{\epsilon_1(4A)}(y) - C_{1} \;\;f_{\epsilon_1(7A)}(y)$}\\
\multicolumn{2}{|c|}{$+ C_{1}\;\;f_{\epsilon_1(B)} (y)$}\\ 
\multicolumn{2}{|c|}{$+ C_{3} \;\;f_{\epsilon_1(3C)}(y) - C_{4} \;\;f_{\epsilon_1(4C)}(y),$}\\
\multicolumn{2}{|c|}{}\\
\hline \hline
$f_{\epsilon_1(1A)}(y)$& $ 4(D-1) I[F'A'] - 4(D+1) I^{2}[F'A''] $\\
\hline \hline
$f_{\epsilon_1(2A)}(y)$& $4 (yI[F'A']) + 2(D-1) I^2[F'A'] - 4I[F'A'] -8I^2[yF'A''] -4(D-5) I^3[F'A''] + 4I^2[F'A''] $\\ 
&$+(D-2)(D-1) (I^{3}[FA^{(3)}] - I^{2}[FA^{(2)}])$\\
\hline \hline
$f_{\epsilon_1(3A)}(y)$& $ (\frac{D}{2}-1) (D-1) I^2[FA''] + \frac12 (D-4)(D-1) I^2[F'A'] +  \frac12 (D(D-7)+26) I^{3} [F'A''] - \frac12 (D-2)(D-1) I^{3}[F''A'] $\\
&$ + 4(D-1) I[F'A'] + (D-1) I^2 [yF'A''] - 4(D-1)I^2[F'A''] + (D-2)(D-1) (I^3[FA^{(3)}]-I^2[FA^{(2)}])$\\
\hline \hline
& $ -\frac{D}{8}(D-2)^2 I^2[FA'] - \frac14 (D-2) I[(4y-y^2) [FA''] + D(2-y)[FA']] + \frac14 (D-2)^3 I[FA] + (D-2)^2 [FA] $\\
& $ +\frac12 (D-2)(D-1) (I^2[FA''] - I^2 [F'A'])  -\frac14 (D-2) I^2[(4y-y^2)F'A'' + D(2-y) F'A']$\\
& $ -\frac12 (D-1)^2 I^3[F'A'] + \frac12 (D-2)(D-5) I^3[F'A''] - \frac12 (D-2)(D-1)I^3[F''A'] $\\
$f_{\epsilon_1(4A)}(y)$& $+4(D-1) I[y FA'' - yI[FA'']] -(D-2) I[y^2 FA'' - y^2 I[FA'']] - 2D(D-2) (I^3[FA^{(3)}] - I^2[FA^{(2)}]) $\\
& $ +D(D-2) I[yI^2[FA^{(3)}]-yI[FA^{(2)}]] + 2(D-2)(2D-3) (I^3[FA^{(3)}] - I^2[FA^{(2)}]) $\\
& $ +\frac12 (D-3)^3 (I^4[FA^{(3)}] - I^3[FA^{(2)}]) - I[yI[F'A''] - y[F'A']] + (I^3[F'A'']-I^2[F'A']) -(D-2) I^2[yF'A'] $\\
& $ +3(D-2) I^2[F'A'] + 4I[F'A'] - 2I^4 [F'A''] + 7I^3[F'A''] + 2 I^2 [y^2F'A''] + (D-1) I^2[y I[F'A'']] $\\
& $ -12(D-3) I^3 [F'A''] + I^2 [yF'A''] - 12 I^2[yF'A''] + 12 I^2 [F'A'']$\\
\hline \hline
$f_{\epsilon_1(7A)}(y)$& $ 4I[F'A'] -4(D+1) I^2[F'A'']$\\
\hline \hline
& $\frac12 (D-2)(D-1) I^2[FB''] + 3 I^2[F'B'] -4(D-7) I^3[F'B''] - 2 I^2[yF'B''] + 4 I^2[F'B'']$\\
$f_{\epsilon_1(B)}(y)$& $\frac12(D-1)(D+2) I^2 [F'B'] + \frac14 (2(D-2)(D-5)+64) I^3[F'B''] - \frac12 (D-2)(D-1) I^3[F''B'] $\\
& $+(D-1) I^2[yF'B''] - 4(D-1) I^2[F'B''] + (D-2) I^2[F'B']$\\
&$+ 4(yI^2[FB^{(3)}]-yI[FB^{(2)}]) + 2(D-2)(D-1)(I^{3}[FB^{(3)}]-I^{2}[FB^{(2)}])$\\
\hline \hline
$f_{\epsilon_1(3C)}(y)$& $ (\frac{D}{2}-1) (D-1) I^2[FC''] + \frac12 (D-4)(D-1) I^2[F'C'] +  \frac12 (D(D-7)+26) I^{3} [F'C''] - \frac12 (D-2)(D-1) I^{3}[F''C'] $\\
&$+ 4(D-1) I[F'C'] + (D-1) I^2 [yF'C''] - 4(D-1)I^2[F'C''] + (D-2)(D-1) (I^3[FC^{(3)}]-I^2[FC^{(2)}])$\\
\hline \hline
& $ -\frac{D}{8}(D-2)^2 I^2[FC'] - \frac14 (D-2) I[(4y-y^2) [FC''] + D(2-y)[FC']] + \frac14 (D-2)^3 I[FC] + (D-2)^2 [FC] $\\
& $ +\frac12 (D-2)(D-1) (I^2[FC''] - I^2 [F'C'])-\frac14 (D-2) I^2[(4y-y^2)F'C'' + D(2-y) F'C'] $\\
& $ -\frac12 (D-1)^2 I^3[F'C'] + \frac12 (D-2)(D-5) I^3[F'C''] - \frac12 (D-2)(D-1)I^3[F''C'] $\\
$f_{\epsilon_1(4C)}(y)$& $+4(D-1) I[y FC'' - yI[FC'']] -(D-2) I[y^2 FC'' - y^2 I[FC'']] - 2D(D-2) (I^3[FC^{(3)}] - I^2[FC^{(2)}]) $\\
& $ +D(D-2) I[yI^2[FC^{(3)}]-yI[FC^{(2)}]] + 2(D-2)(2D-3) (I^3[FC^{(3)}] - I^2[FC^{(2)}]) $\\
& $ +\frac12 (D-3)^3 (I^4[FC^{(3)}] - I^3[FC^{(2)}]) - I[yI[F'A''] - y[F'C']] + (I^3[F'C'']-I^2[F'C']) -(D-2) I^2[yF'C'] $\\
& $ +3(D-2) I^2[F'C'] + 4I[F'C'] - 2I^4 [F'C''] + 7I^3[F'C''] + 2 I^2 [y^2F'C''] + (D-1) I^2[y I[F'C'']] $\\
& $ -12(D-3) I^3 [F'C''] + I^2 [yF'C''] - 12 I^2[yF'C''] + 12 I^2 [F'C'']$\\
\hline \hline
\hline
\multicolumn{2}{|c|}{}\\
\multicolumn{2}{|c|}{${\rm Total\;for\;}f_{\epsilon_1}(y)$}\\
\multicolumn{2}{|c|}{}\\
\hline \hline
\multicolumn{2}{|c|}{}\\
\multicolumn{2}{|c|}{$-\frac{(D-2)(D-6)\;H^{2D-4} \pi^{-D} \;\Gamma{(\frac{D}{2}-1)^2}}{16}\;(\frac{1}{y})^{D-2}.$}\\ 
\multicolumn{2}{|c|}{}\\
\end{tabular}\label{cte1}
\end{ruledtabular}
\end{center}
\end{table}

%%%%%%%%%%%%%%%%%%%%%%%%%%%%%%%%%%%%%%%%%%%%%%%%%%%%%%%%%%%%%%%%

\begin{table}[htbp] 
\begin{center}
\caption{For $\epsilon_2 \equiv (aa')^{\frac{D}{2}-1}(a^2 + a^{\prime 2}) H^2 \nabla^2 $ type terms.}
\begin{ruledtabular}
\begin{tabular}{|l|c|}
\hline
\multicolumn{2}{|c|}{}\\
\multicolumn{2}{|c|}{$f_{\epsilon_2}(y) = -C_{1}\;\;f_{\epsilon_2(2A)}(y) - C_{2}\;\;f_{\epsilon_2(3A)}(y) 
+ C_{4}\;\;f_{\epsilon_2(4A)}(y) + C_{1} \;\;f_{\epsilon_2(B)}(y) + C_{3} \;\;f_{\epsilon_2(3C)}(y) 
- C_{4}\;\;f_{\epsilon_2(4C)}(y),$}\\
\multicolumn{2}{|c|}{}\\
\hline \hline
$f_{\epsilon_2(2A)}(y)$& $ 2(yI[F'A']) - 2 I^2[F'A'] - 6 I^3[F'A''] - 4 I^2[yF'A''] \phantom{ssssssssssssssssssssssssss}$\\
\hline \hline
$f_{\epsilon_2(3A)}(y)$& $ \frac{D}{2} (\frac{D}{2}-1) I^2[F A''] - 6 I^3[F'A'']
\phantom{ssssssssssssssssssssssssss}$\\
\hline \hline
$f_{\epsilon_2(4A)}(y)$& $ \frac{D}{4} (D-2) I^2[FA''] - 2 (I^3[F'A''] - I^2[F'A']) + 2 I^2[yF'A'' + y F''A'] - 4 I[F'A']-10 I^2[F'A''] - 2 I^2 [y F'A'']$\\
\hline \hline
$f_{\epsilon_2(B)}(y)$& $ 4 (y I^2[FB^{(3)}] - y I[FB^{(2)}]) \phantom{ssssssssssssssssssssssssss}$\\
\hline \hline
$f_{\epsilon_2(3C)}(y)$& $ \frac{D}{2} (\frac{D}{2}-1) I^2[F C''] - 6 I^3[F'C'']
\phantom{ssssssssssssssssssssssssss}$\\
\hline \hline
$f_{\epsilon_2(4C)}(y)$& $ \frac{D}{4} (D-2) I^2[FC''] - 2 (I^3[F'C''] - I^2[F'C']) + 2 I^2[yF'C'' + y F''C'] - 4 I[F'C'] -10 I^2[F'C''] - 2 I^2 [y F'C'']$\\
\hline \hline
\hline
\multicolumn{2}{|c|}{}\\
\multicolumn{2}{|c|}{${\rm Total\;for\;}f_{\epsilon_2}(y)$}\\
\multicolumn{2}{|c|}{}\\
\hline \hline
\multicolumn{2}{|c|}{}\\
\multicolumn{2}{|c|}{$-\frac{D(D-8)(D-2)\;H^{2D-4}\pi^{-D}\;\Gamma{(\frac{D}{2}-1)^2}}{64(D-1)}\;(\frac{1}{y})^{D-2}\;+\;\frac{3H^4}{4\pi^4}\;\frac{1}{y}.$}\\ 
\multicolumn{2}{|c|}{}\\
\end{tabular}\label{cte2}
\end{ruledtabular}
\end{center}
\end{table}

%%%%%%%%%%%%%%%%%%%%%%%%%%%%%%%%%%%%%%%%%%%%%%%%%%%%%%%%%%%%%%%%

\begin{table}[htbp] 
\begin{center}
\caption{For $\epsilon_3 \equiv (aa')^{\frac{D}{2}-1}(a + a')^2 H^2 \nabla^2 $ type terms.}
\begin{ruledtabular}
\begin{tabular}{|l|c|}
\hline
\multicolumn{2}{|c|}{}\\
\multicolumn{2}{|c|}{$f_{\epsilon_3}(y) = -C_{1}\;\;f_{\epsilon_3(2A)}(y) - C_{2} \;\;f_{\epsilon_3(3A)}(y)
+ C_{1} \;\;f_{\epsilon_3(B)}(y) + C_{3}\;\;f_{\epsilon_3(3C)}(y) - C_{4} \;\;f_{\epsilon_3(4C)} (y),$}\\
\multicolumn{2}{|c|}{}\\
\hline \hline
$f_{\epsilon_3(2A)}(y)$& $(D-2) (y I^2[FA^{(3)}] - y I[FA^{(2)}] )\phantom{ssssssssssssssssssssssssssssssssssssssssss}$\\
\hline \hline
$f_{\epsilon_3(3A)}(y)$& $(\frac{D}{2}-1) (y I[FA''] )
+ ((\frac{D}{2}-1)-2) I^2[y F'A''] + (D-2) (yI^2[FA^{(3)}] - yI[FA^{(2)}])
\phantom{ssssssssssssssssssssssssssssssss}$\\
\hline \hline
$f_{\epsilon_3(4A)}(y)$& $\frac12 (D-2) (yI[FA'']) + \frac12 (D-2) I^2[yF'A''] + 2(D-2) (yI^2[FA^{(3)}] - yI[FA^{(2)}]) - 2 I^2[yF'A'']$\\
\hline \hline
$f_{\epsilon_3(B)}(y)$& $2(y^2 I[FB^{(3)}] - y^2 [FB^{(2)}]) + 2(D-2) (yI^2[FB^{(3)}] - yI[FB^{(2)}])
\phantom{ssssssssssssssssssssssssssssssss}$\\
\hline \hline
$f_{\epsilon_3(3C)}(y)$& $(\frac{D}{2}-1) (y I[FC''] ) 
+ ((\frac{D}{2}-1)-2) I^2[y F'C''] + (D-2) (yI^2[FC^{(3)}] - yI[FC^{(2)}])
\phantom{ssssssssssssssssssssssssssssssss}$\\
\hline \hline
$f_{\epsilon_3(4C)}(y)$& $\frac12 (D-2) (yI[FC'']) + \frac12 (D-2) I^2[yF'C''] + 2(D-2) (yI^2[FC^{(3)}] - yI[FC^{(2)}]) - 2 I^2[yF'C'']$ \\
\hline \hline
\hline
\multicolumn{2}{|c|}{}\\
\multicolumn{2}{|c|}{${\rm Total\;for\;}f_{\epsilon_3}(y)$}\\
\multicolumn{2}{|c|}{}\\
\hline \hline
\multicolumn{2}{|c|}{}\\
\multicolumn{2}{|c|}{$-\frac{(D-2)(D^3-3D^2+22D-16)\;H^{2D-4} \pi^{-D} \;\Gamma{(\frac{D}{2}-1)^2}}{64(D-3)(D-1)} \;(\frac{1}{y})^{D-2}\;+\;\frac{9H^4}{8\pi^4}\;\frac{1}{y}.$}\\ 
\multicolumn{2}{|c|}{}\\
\end{tabular}\label{cte3}
\end{ruledtabular}
\end{center}
\end{table}

%%%%%%%%%%%%%%%%%%%%%%%%%%%%%%%%%%%%%%%%%%%%%%%%%%%%%%%%%%%%%%%%

\begin{table}[htbp] 
\begin{center}
\caption{For $\zeta \equiv (aa')^{\frac{D}{2}-1}\nabla^4 $ type terms.}
\begin{ruledtabular}
\begin{tabular}{|l|c|}
\hline
\multicolumn{2}{|c|}{}\\
\multicolumn{2}{|c|}{$f_{\zeta}(y) = C_{1}\;\;f_{\zeta (1A)} (y) - C_{1}\;\;f_{\zeta (2A)} (y)  - C_{2}\;\;f_{\zeta (3A)} (y) + C_{4}\;\;f_{\delta (4A)}(y) - C_{1} \;\;f_{\zeta (7A)} (y)$}\\
\multicolumn{2}{|c|}{$+ C_{1}\;\;f_{\zeta (B)} (y) $}\\
\multicolumn{2}{|c|}{
$+ C_{3}\;\;f_{\zeta (3C)} (y) -C_{4}\;\;f_{\zeta (4C)}(y),$}\\
\multicolumn{2}{|c|}{}\\
\hline \hline
$f_{\zeta (1A)}(y)$&$ -2(I^3[F A^{(3)}] - I^2[F A^{(2)}]) \phantom{sssssssss}$\\
\hline \hline
$f_{\zeta (2A)}(y)$&$ -I^3[F' A'] + 2 I^4[F'A''] - \frac12 (D-2) (I^4 [F A^{(3)}] - I^3 [F A^{(2)}]) - 2 (I^{3}[F A^{(3)}]-I^{2}[F A^{(2)}])\phantom{sssssss}$\\
\hline \hline
$f_{\zeta (3A)}(y)$&$  -\frac14 (D-2) (I^4[F' A''] + I^3 [F A'']) + I^4 [F'A''] -\frac12 (D-2) (I^4 [F A^{(3)}] - I^3 [F A^{(2)}]) - 2 (I^3 [F A^{(3)}] - I^2 [F A^{(2)} ]) \phantom{sssssss}$\\
\hline \hline
$f_{\zeta (4A)}(y)$&$  -(D-2) (I^4[FA^{(3)}] - I^3 [F A^{(2)}]) -2 (I^3 [FA^{(2)}] -I^2 [FA^{(2)}])
+3I^4 [F'A''] - \frac14 (D-2) (I^4 [F' A''] + I^3 [F A''])  \phantom{sssssss}$\\
\hline \hline
$f_{\zeta (7A)}(y)$&$ -2(I^3[F A^{(3)}] - I^2[F A^{(2)}]) \phantom{sssssssss}$\\
\hline \hline
$f_{\zeta (B)}(y)$&$ -\frac14 (D-2) I^3[FB''] - \frac12 I^3[F'B'] + 2 I^4[F'B''] -\frac14 (D-2) I^4[F'B''] - \frac12 I^3 [F'B'] + I^4[F'B'']$\\
& $-(D-2) (I^4[FB^{(3)}] - I^3 [F B^{(2)}]) - 4(I^3 [F B^{(3)}] -I^2 [F B^{(2)}])\phantom{sss}$\\
\hline \hline
$f_{\zeta (3C)}(y)$&  $-\frac14 (D-2) (I^4[F' C''] + I^3 [F C'']) + I^4 [F'C''] -\frac12 (D-2) (I^4 [F C^{(3)}] - I^3 [F C^{(2)}]) - 2 (I^3 [F C^{(3)}] - I^2 [F C^{(2)} ]) \phantom{sssssss}$\\
\hline \hline
$f_{\zeta (4C)}(y)$&$  -(D-2) (I^4[FC^{(3)}] - I^3 [F C^{(2)}]) -2 (I^3 [F C^{(2)}] -I^2 [F C^{(2)}])
+3I^4 [F'C''] - \frac14 (D-2) (I^4 [F' C''] + I^3 [F C''])  \phantom{sssssss}$\\
\hline \hline
\hline
\multicolumn{2}{|c|}{}\\
\multicolumn{2}{|c|}{${\rm Total\;for\;}f_{\zeta}(y)$}\\
\multicolumn{2}{|c|}{}\\
\hline \hline
\multicolumn{2}{|c|}{}\\
\multicolumn{2}{|c|}{$-\frac{H^4}{16 \pi^4}\;\frac{1}{y}.$}\\
\multicolumn{2}{|c|}{}\\
\end{tabular}\label{ctz}
\end{ruledtabular}
\end{center}
\end{table}

%%%%%%%%%%%%%%%%%%%%%%%%%%%%%%%%%%%%%%%%%%%%%%%%%%%%%%%%%%%%%%%%

\end{document}